\documentclass[12pt]{article}
\usepackage{amsmath,amssymb,epsfig,epsf,psfrag}
\usepackage{graphicx,color}
\definecolor{DarkBlue}{rgb}{0.1,0.1,0.5}
\definecolor{Red}{rgb}{0.9,0.0,0.1}

\usepackage{setspace}
\begin{document}

\title{A geometric approach to maximum likelihood estimation of
the functional principal components from sparse longitudinal
data\\
}

\vskip.2in

\author{Jie Peng$^\ast$ and Debashis Paul\\
(Department of Statistics, University of California,
Davis)\\
$^\ast$ Correspondence author: email: jie@wald.ucdavis.edu }
\date{}

\maketitle
\begin{abstract}
In this paper, we consider the problem of estimating  the
eigenvalues and eigenfunctions of the covariance kernel (i.e., the
\textit{functional principal components}) from sparse and
irregularly observed longitudinal data. We approach this problem
through a maximum likelihood method assuming that the covariance
kernel is smooth and finite dimensional.  We exploit the
smoothness of the eigenfunctions to reduce dimensionality by
restricting them to a lower dimensional space of smooth functions.
The estimation scheme is developed based on a Newton-Raphson
procedure using the fact that the basis coefficients representing
the eigenfunctions lie on a \textit{Stiefel manifold}. We also
address the selection of the right number of basis functions, as
well as that of the dimension of the covariance kernel by a second
order approximation to the leave-one-curve-out cross-validation
score that is computationally very efficient. The effectiveness of
our procedure is demonstrated by simulation studies and an
application to a CD4 counts data set. In the simulation studies,
our method performs well on both estimation and model selection.
It also outperforms two existing approaches: one based on a local
polynomial smoothing of the empirical covariances, and another
using an EM algorithm.

\end{abstract}


\section{Introduction}\label{sec:intro}


In recent years there have been numerous works on data that may be
considered as noisy curves. When the individual observations can
be regarded as measurements on an interval, the data thus obtained
can be classified as functional data. For analysis of data arising
in various fields, such as longitudinal data analysis,
chemometrics, econometrics, etc. [Ferraty and Vieu (2006)], the
functional data analysis viewpoint is becoming increasingly
popular. Depending on how the individual curves are measured, one
can think of two different scenarios - (i) when the individual
curves are measured on a dense grid; and (ii) when the
measurements are observed on an irregular, and typically sparse
set of points on an interval. The first situation usually arises
when the data are recorded by some automated instrument, e.g. in
chemometrics, where the curves represent the spectra of certain
chemical substances. The second scenario is more typical in
longitudinal studies where the individual curves could represent
the level of concentration of some substance, and the measurements
on the subjects may be taken only at irregular time points.

In these settings, when the goal of analysis is  either data
compression, model building or studying covariate effects, one may
want to extract information about the mean, variability,
correlation structure, etc. In the first scenario, i.e., data on a
regular grid, as long as the individual curves are smooth, the
measurement noise level is low, and the grid is dense enough, one
can essentially treat the data to be on a continuum, and employ
techniques similar to the ones used in classical multivariate
analysis. However, the irregular nature of data in the second
scenario, and the associated measurement noise require a different
treatment.

The main goal of this paper is the estimation of the functional
principal components from sparse, irregularly, observed functional
data (scenario (ii)). The eigenfunctions give a nice basis for
representing functional data, and hence are very useful in
problems related to model building and prediction for functional
data [see e.g. Cardot, Ferraty and Sarda (1999), Hall and Horowitz
(2007), Cai and Hall (2006)]. Ramsay and Silverman (2005) and
Ferraty and Vieu (2006) give an extensive survey of the
applications of {\it functional principal components analysis}
(FPCA).

The focus throughout this paper thus is in the estimation of
covariance kernel of the underlying process. Covariance is a
positive semidefinite operator. The space of covariance operators
is a nonlinear manifold. Thus, from statistical as well as
aesthetic point of view, it is important that any estimator of the
covariance is also positive semidefinite. Moreover, Smith (2005)
gives a compelling argument in favor of utilizing the intrinsic
geometry of the parameter space in the context of estimating
covariance matrix in a multivariate Gaussian setting. He obtains
Cram\'{e}r-Rao bounds for the risk, that are described in terms of
intrinsic gradient and Hessian of the log-likelihood function.
This work brings out important features of the estimators that are
not obtained through the usual Euclidean viewpoint.
It also provides a strong motivation for a likelihood-based
approach that respects the intrinsic geometry of the parameter
space. In this paper, we shall adopt a \textit{restricted maximum
likelihood} approach and explicitly utilize the intrinsic geometry
of the parameter space when fitting the maximum likelihood
estimator.

Now we shall give an outline of the model for the sparse
functional data.
Suppose that we observe $n$ independent realizations of an
$L^2$-stochastic process $\{X(t): t \in [0,1]\}$ at a sequence of
points on the interval $[0,1]$ (or, more generally, on an interval
$[a,b]$), with additive measurement noise. That is, the observed
data $\{Y_{ij} : 1\leq j \leq m_i; 1 \leq i \leq n\}$ can be
modeled as :
\begin{equation}\label{eq:model}
Y_{ij} = X_i(T_{ij}) + \sigma \varepsilon_{ij},
\end{equation}
where $\{\varepsilon_{ij}\}$ are i.i.d. with mean 0 and variance
1. Since $X(t)$ is an $L^2$ stochastic process, by Mercer's
theorem [cf. Ash (1972)] there exists a positive semi-definite
kernel $C(\cdot,\cdot)$ such that $Cov(X(s),X(t)) = C(s,t)$ and
each $X_i(t)$ has the following a.s. representation in terms of
the eigenfunctions of the kernel $C(\cdot,\cdot)$ :
\begin{equation}\label{eq:Karhunen}
X_i(t) = \mu(t) + \sum_{\nu=1}^\infty \sqrt{\lambda_\nu}
\psi_\nu(t) \xi_{i\nu},
\end{equation}
where $\mu(\cdot) = \mathbb{E}(X(\cdot))$ is the mean function;
$\lambda_1 \geq \lambda_2 \geq \ldots \geq 0$ are the eigenvalues
of $C(\cdot,\cdot)$; $\psi_\nu(\cdot)$ are the corresponding
orthonormal eigenfunctions; and the random variables
$\{\xi_{i\nu}:\nu \geq 1\}$, for each $i$, are uncorrelated with
zero mean and unit variance. In the observed data model
(\ref{eq:model}), we assume that $\{T_{ij}:j=1,\ldots,m_i\}$ are
randomly sampled from a continuous distribution. In the problems
we shall be interested in, the number of measurements $m_i$ is
typically small.


Our estimation procedure is based on the assumption that the
covariance kernel $C$ is of finite rank, say $r$; and the
representation of the eigenfunctions $\{\psi_\nu\}_{\nu =1}^r$ in
a known, finite, basis of smooth functions. This results in an
orthogonality constraint on the matrix of basis coefficients, say
$B$, as described in Section \ref{sec:method}. Specifically, the
matrix $B$ lies in a \textit{Stiefel manifold}, that is the space
of real valued matrices with orthonormal columns. Our estimation
procedure involves maximization of the log-likelihood under the
working assumption of normality, satisfying the orthonormality
constraint on $B$. To implement this, we employ a Newton-Raphson
algorithm based on the work by Edelman, Arias and Smith (1998) for
optimization on a \textit{Stiefel manifold}, that utilizes its
intrinsic Riemannian geometric structure. As a remark, the
procedure we proposed here is intrinsically nonlinear. Linear
estimation procedures for covariance, that assume the basis
representation framework, have been studied by various authors
including Cardot (2000), Rice and Wu (2001), and Besse, Cardot and
Ferraty (1997).

At this point, we would like to mention the main contributions of
this paper. The approach based on utilizing the intrinsic geometry
of the parameter space in the context of covariance estimation
using a maximum likelihood approach is new. The resulting
estimation procedure can handle different regimes of sparsity of
data efficiently. Its implementation is computationally
challenging in the current context because of the irregular nature
of the measurements. The resulting estimator is very accurate,
based on the simulation studies we conducted.
The geometric viewpoint has further important implications.
Selection of an appropriate model in the context of longitudinal
data is of great importance, and devising a computationally
practical, yet effective, model selection procedure remains a
challenge [cf. Marron \textit{et al.} (2004), p. 620]. It is
well-known that the computational cost of the usual
leave-one-curve-out cross-validation score (for selecting the
dimension of the basis used in the representation) is prohibitive.
We utilize the geometry of the parameter space to derive an
approximation of the CV score that is computationally very
efficient. This is another main contribution of this paper.
Finally, our approach involves analysis of the gradient and
Hessian of the log-likelihood. A detailed asymptotic analysis of
any estimation procedure based on the likelihood necessarily
involves understanding of these objects and the geometry of the
parameter space. The work presented here serves as a first step in
this direction.

Before ending this section, we give a brief overview of the
existing literature on FPCA. The idea of maximizing the restricted
likelihood is in the same framework as that studied by James,
Hastie and Sugar (2000), who propose an EM algorithm to maximize
the log-likelihood. However, there are important differences
between the proposed approach and the EM approach. First, the EM
algorithm results in an estimator not necessarily satisfying the
orthonormality constraints, that is, being outside the parameter
space, which is corrected through an eigen-decomposition. But
nevertheless, this can lead to an inefficient estimator. Secondly,
the EM algorithm does not set the intrinsic gradient of the
log-likelihood to zero. Therefore it does not utilize the
redundancy in the optimization problem induced by the
orthonormality constraints. This could also result in a loss of
efficiency in estimation since the value of the objective function
may stabilize even though the optimal parameter value in the
restricted space may not have been attained. On the other hand,
our approach addresses the problem of finding the optimal
parameter value more directly. Moreover, the approximate CV score
proposed in Section \ref{sec:CV} rely heavily on the gradient of
the objective function being zero at the estimator, which is not
satisfied by the EM algorithm, but is one property of our proposed
estimator.

We already mentioned the basis representation approach. Another
approach to FPCA is through kernel smoothing. In this approach,
the $i$-th curve is pre-smoothed by taking weighted average of
$\{Y_{ij}\}_{j=1}^{m_i}$'s where the weights are evaluations of a
kernel centered at the time points $\{T_{ij}\}_{j=1}^{m_i}$.
Unfortunately, when the number of measurements is small, this
procedure results in a highly biased estimate of the covariance
kernel as demonstrated by Yao, M\"{u}ller and Wang (2005). These
authors thus propose to estimate the covariance by local
polynomial smoothing of the empirical covariances at observed
pairs of time points $\{(T_{ij},T_{ij'}): i=1,\cdots,n, 1 \leq j
\not=j' \leq m_i\}$. Hall, M\"{u}ller and Wang (2006) prove
optimality of this procedure under rather weak assumptions on the
process for optimal choice of bandwidths. Their work clearly
separates the problem of estimating the covariance and its
eigenfunctions, and identifies the latter as a one dimensional
nonparametric function estimation problem. In spite of its nice
asymptotic properties, there are some aspects of the local
polynomial method that are somewhat unsatisfactory. First, it does
not ensure a positive semi-definite estimate of the population
covariance kernel. A common practice to fix that is through
projection, however this can lead to an inefficient estimator.
Secondly, this procedure sometimes results in  a negative estimate
of the error variance $\sigma^2$. In contrast, the proposed
procedure gives positive semi-definite estimate as well as
positive estimate of $\sigma^2$.

The novelty of our work is the explicit utilization of the intrinsic
geometry of the parameter space, which results in more efficient
estimators. Moreover, this enables an efficient approximation of the
cross validation score. As far as we know, an efficient cross
validation based model selection procedure has not been discovered
for most of the existing procedures in this field, including the two
approaches we mentioned above. Simulation studies presented in
Section \ref{sec:simulation} indicate a significant improvement of
the proposed method over both EM (James \textit{et al.} (2000)) and
local polynomial (Yao \textit{et al.}(2005)) approaches, as well as
a satisfactory performance in model selection based on the
approximate CV score derived in Section \ref{sec:CV}. We also want
to emphasize that, the estimation procedure presented in this paper
should be regarded as a demonstration of the usefulness of the
geometric viewpoint while tackling a complex statistical problem.
Even though our focus throughout this paper remains on solving the
problem described above, the tools developed here can be easily
extended to other situations that involve matrix-valued parameters
with orthonormality constraints. Two such examples are discussed in
Section \ref{sec:discussion}.

The rest of the paper is organized as follows. In Section
\ref{sec:method}, we describe the restricted maximum likelihood
framework. In Section \ref{sec:computation}, we give an outline of
the Newton-Raphson algorithm for optimization on Stiefel
manifolds. In Section \ref{sec:CV}, we derive an approximation to
the leave-one-curve-out cross-validation score. Section
\ref{sec:simulation} is devoted to detailed simulation studies and
the comparison and discussion of the performance of various
procedures. In Section \ref{sec:real_data}, the proposed procedure
is illustrated through an application to a CD4 counts data set. In
Section \ref{sec:discussion}, we discuss some possible extensions
and future works. Technical details are given in the appendices.
Tables, Figures and supplementary material are attached at the
end.

\section{Restricted MLE framework}\label{sec:method}

We first describe the basis representation framework.
Under some weak conditions on the stochastic processes (like
$L^2$-differentiability of certain order, see, e.g. Ash (1972)),
the eigenfunctions have some degree of smoothness. This assumption
has been used in various studies, including Boente and Fraiman
(2000), Cardot (2000), James \textit{et al.} (2000), Yao
\textit{et al.} (2005, 2006), and Hall \textit{et al.} (2006).
Smoothness of the eigenfunctions means that they can be well
approximated in some stable basis for smooth function classes,
e.g. the B-spline basis [Chui (1987)].
If in addition, in model (\ref{eq:Karhunen}), we assume that
$\lambda_\nu = 0$ for $\nu > r$, for some $r \geq 1$, then we can
choose a finite set of linearly independent, $L^2$ functions
$\{\phi_1(\cdot),\ldots, \phi_M(\cdot)\}$ with $M \geq r$, such that
eigenfunctions can be modeled as $\psi_\nu(\cdot) = \sum_{k=1}^M
b_{k\nu} \phi_k(\cdot)$ for $\nu=1,\ldots,r$. Then, for every $t$,
\begin{equation}\label{eq:representation}
\boldsymbol{\psi}(t)^T := (\psi_1(t),\ldots,\psi_r(t)) =
(\phi_1(t),\ldots,\phi_M(t)) B
\end{equation}
for an $M \times r$ matrix $B=((b_{k\nu}))$ that satisfies the
constraint
\begin{equation}\label{eq:basis_repr_B_ortho}
B^T (\int \boldsymbol{\phi}(t) \boldsymbol{\phi}(t)^T dt) B = \int
\boldsymbol{\psi}(t) \boldsymbol{\psi}(t)^T dt = I_r,
\end{equation}
where $\boldsymbol{\phi}(\cdot) =
(\phi_1(\cdot),\ldots,\phi_M(\cdot))^T$. Since the $M \times M$
matrix $\int \boldsymbol{\phi}(t) \boldsymbol{\phi}(t)^T dt$ is
known and nonsingular, without loss of generality, hereafter we
assume $B^T B = I_r$, by orthonormalizing
$\{\phi_1(\cdot),\ldots,\phi_M(\cdot)\}$.


Here, we are assuming a reduced rank model for the covariance
kernel as in James \textit{et al.} (2000). This model can be
motivated as follows. Suppose that the covariance kernel $C(s,t)$
of the underlying process has the infinite Karhunen-Lo\'{e}ve
expansion:
\begin{equation}\label{eq:Karhunen_Loeve}
C(s,t) = \sum_{k=1}^\infty \lambda_k \psi_k(s) \psi_k(t),
\end{equation}
where $\lambda_1 \geq \lambda_2 \geq \cdots \geq 0$,
$\sum_{k=1}^\infty \lambda_k < \infty$, and
$\{\psi_k\}_{k=1}^\infty$ forms a complete orthonormal basis for
$L^2[0,1]$. The condition $\sum_{k=1}^\infty \lambda_k < \infty$
implies that $\lambda_k \to 0$ as $k \to \infty$. Also, the
orthonormality of the eigenfunctions $\{\psi_k\}$ implies that
$\psi_k$ typically gets more and more ``wiggly'' as $k$ increases,
at least for most reasonable processes with smooth covariance
kernel. Therefore, modeling the full covariance kernel remains a
challenge. However, one can truncate the series on the RHS of
(\ref{eq:Karhunen_Loeve}) at some finite $r \geq 1$ to get the
\textit{projected covariance kernel}
\begin{equation}\label{eq:covariance_projected}
C_{proj}^r(s,t) = \sum_{k=1}^r \lambda_k \psi_k(s) \psi_k(t).
\end{equation}
Note that $\parallel C - C_{proj}^r\parallel_F^2 =
\sum_{k=r+1}^\infty \lambda_k^2$. Thus, as long as  the
eigenvalues decay to zero fast, even with a relatively small $r$,
the approximation $C_{proj}^r$ only results in a small bias. This
motivates the choice of a finite rank model as described above.
Furthermore, the restriction to reduced rank model helps in
modeling the eigenfunctions as well. If $\psi_k$ for larger $k$
are more wiggly, it takes a lot more basis functions to represent
them well. On the other hand, for a model with $r$ relatively
small, we can get good approximations to the eigenfunctions with a
moderate number of basis functions.

Of course, in practice one could encounter situations for which
the projected kernel $C_{proj}^r$ is not a good approximation for
any small value of $r$. This will for example happen if the
eigenvalues are decaying slowly. Then in the modeling step one
needs to choose large $r$.  However under such situations, there
is an intrinsic instability of the estimates of the
eigenfunctions, as it is well known that, the estimation error
grows inversely with the gap between successive eigenvalues.
Moreover, it is harder to choose the sufficient rank $r$ by a
model selection procedure if it is too large. Appropriate
statistical methods to deal with such data still need to be
developed.

Finally, it is worthwhile to point out some advantages of the
reduced rank formulation over the mixed effects model by Rice and Wu
(2000), as also noted by James \textit{et al.} (2000). Notice that,
in the unconstrained mixed effects model, one needs to model the
covariance kernel using a full-rank representation. Thus if one uses
$M$ basis functions to represent it, there are $M(M+1)/2$ basis
coefficients of the covariance kernel that need to be estimated.
When the observations are sparse, this could lead to an
over-parametrization, and it will result in highly variable
estimates. Furthermore, if one uses a maximum likelihood approach,
the over-parametrization would cause a very rough likelihood
surface, with multiple local maxima. Therefore, restricting the rank
of the covariance kernel can also be viewed as a form of
regularization of the likelihood.


If one assumes Gaussianity of the processes, i.e., $\xi_{i\nu}
\stackrel{i.i.d.}{\sim} N(0,1)$ and $\varepsilon_{ij}
\stackrel{i.i.d.}{\sim} N(0,1)$, and they are independent, then
under the assumption (\ref{eq:representation}), the negative
log-likelihood of the data, conditional on $\{(m_i,
\{T_{ij}\}_{j=1}^{m_i})\}_{i=1}^n$ is given by
\begin{eqnarray}\label{eq:basis_repr_logL}
-\log L(B,\mathit{\Lambda},\sigma^2) &=& const.
+\frac{1}{2}\sum_{i=1}^n Tr [(\sigma^2 I_{m_i} + \mathit{\Phi}_i^T B
\mathit{\Lambda} B^T \mathit{\Phi}_i)^{-1}(\mathbf{Y}_i -
\boldsymbol{\mu}_i)
(\mathbf{Y}_i - \boldsymbol{\mu}_i)^T] \nonumber\\
&& +\frac{1}{2}\sum_{i=1}^n \log|\sigma^2 I_{m_i} +
\mathit{\Phi}_i^T B \mathit{\Lambda} B^T \mathit{\Phi}_i|,
\end{eqnarray}
where $\mathit{\Lambda}$ is the $r \times r$ diagonal matrix of
non-zero eigenvalues of $C(\cdot,\cdot)$, $\mathbf{Y}_i =
(Y_{i1},\ldots,Y_{im_i})^T$ and $\boldsymbol{\mu}_i =
(\mu(T_{i1}),\ldots,\mu(T_{im_i}))^T$ are $m_i \times 1$ vectors,
and $\mathit{\Phi}_i =
[\boldsymbol{\phi}(T_{i1}):\ldots:\boldsymbol{\phi}(T_{im_i})]$ is
an $M \times m_i$ matrix. One can immediately see that the
difficulty with the maximum likelihood approach mainly lies in the
irregularity of the form of the objective function
(\ref{eq:basis_repr_logL}), and the fact that the parameter $B$ has
orthonormal constraints (\ref{eq:basis_repr_B_ortho}). Moreover,
this is a non-convex optimization problem with respect to the
parameters.

We propose to directly minimize (\ref{eq:basis_repr_logL}) subject
to (\ref{eq:basis_repr_B_ortho}) by a Newton-Raphson algorithm on
the Stiefel manifold, whose general form has been developed in
Edelman \textit{et al.} (1998).
The proposed estimator is
$$
(\widehat B, \mathit{\widehat{\Lambda}}, \widehat \sigma^2) =
\arg\min_{B \in {\cal S}_{M,r},(\mathit{\Lambda},\sigma^2) \in
\Theta} -\log L(B,\mathit{\Lambda},\sigma^2),
$$
where $\Theta = \mathbb{R}_+^{r+1}$, and ${\cal S}_{M,r} := \{A
\in \mathbb{R}^{M\times r} : A^T A = I_r\}$ is the Stiefel
manifold of $M \times r$ real-valued matrices (with $r \leq M$)
with orthonormal columns.
The Newton-Raphson procedure involves computation of the intrinsic
gradient and Hessian of the objective function, and on
convergence, it sets the gradient to zero. Thus the proposed
estimator solves the score equation:
$$
\nabla_{(B,\Lambda,\sigma^2)}\log L(B,\Lambda,\sigma^2)=0.
$$
 We shall discuss the
details of this algorithm and its implementation in Section
\ref{sec:computation}.

It is important to note that, one does not need to assume
Gaussianity in order to carry out the proposed estimation as well as
model selection using the approximated CV score derived in Section
\ref{sec:CV}. This is because (\ref{eq:basis_repr_logL}) is a
\textit{bona fide} loss function. Thus the Gaussianity should be
viewed as a working assumption which gives the form of the loss
function. It is assumed throughout that (\ref{eq:basis_repr_logL})
is differentiable with respect to the eigenvalues and
eigenfunctions. This in turn depends on the assumption that all the
nonzero eigenvalues of the covariance kernel are distinct, since
multiplicity of eigenvalues results in the covariance kernel being
non-differentiable with respect to both eigenvalues and
eigenfunctions. It is  also worth pointing out that the M-step of
the EM algorithm in James \textit{et al.} (2000) does not utilize
the orthonormality constraint on $B$. This restriction can be
imposed, and the minimization of the corresponding objective
function can be carried out in a similar fashion as in the proposed
method. This may lead to an improvement in the performance of the EM
estimates.

\section{Newton-Raphson algorithm}\label{sec:computation}


In this section, we describe the Newton-Raphson algorithm for
minimising the loss function (\ref{eq:basis_repr_logL}). In a
seminal paper, Edelman \textit{et al.} (1999) derive Newton-Raphson
and conjugate gradient algorithms for optimising functions on
Stiefel and Grassman manifolds. As their counterparts in the
Euclidean space, these algorithms aim to set the gradient of the
objective function (viewed as a function on the manifold) to zero.
The algorithms involve the following steps : (i) update the
\textit{tangent vector} at the current parameter value; (ii) move
along the geodesic in the direction of the recently updated tangent
vector to a new point on the manifold.


In our setting, the objective is to minimise the loss function
(\ref{eq:basis_repr_logL}). For notational simplicity, drop the
irrelevant constants and re-write (\ref{eq:basis_repr_logL}) as
\begin{equation}\label{eq:objective}
F(B,\mathit{\Lambda},\sigma^2) := \sum_{i=1}^n \left[
F_i^1(B,\mathit{\Lambda},\sigma^2) +
F_i^2(B,\mathit{\Lambda},\sigma^2)\right],
\end{equation}
where
\begin{equation}\label{eq:F1_and_F2}
F_i^1(B,\mathit{\Lambda},\sigma^2) =
Tr[P_i^{-1}\widetilde{\mathbf{Y}}_i \widetilde{\mathbf{Y}}_i^T], \
F_i^2(B,\mathit{\Lambda},\sigma^2) = \log|P_i|,
\end{equation}
with $P_i=\sigma^2 I_{m_i} + \mathit{\Phi}_i^T B \mathit{\Lambda}
B^T \mathit{\Phi}_i$, $\widetilde{\mathbf{Y}}_i = \mathbf{Y}_i -
\boldsymbol{\mu}_i$, and $\mathit{\Phi}_i$ as defined in Section
\ref{sec:method}. Here we treat $\boldsymbol{\mu}_i$ as known, since
we propose to estimate it separately. The parameter spaces for
$\mathit{\Lambda}$  and $\sigma^2$ are positive cones in Euclidean
spaces and hence convex. The parameter space for $B$ is ${\cal
S}_{M,r}$, the Stiefel manifold of $M\times r$ matrices with
orthonormal columns.

We adopt a two-step procedure for updating the parameters. Each
Newton-Raphson updating step is broken into two parts - (a) an
update of $(\mathit{\Lambda}, \sigma^2)$, keeping $B$ at the
current value; and (b) an update of $B$, setting
$(\mathit{\Lambda}, \sigma^2)$ at the recently updated value.
Thus, the algorithm proceeds by starting at an initial estimate
and then cycling through these two steps iteratively till
convergence.


For now, assume that the orthonormal basis functions $\{\phi_k\}$
and dimensions $M$ and $r$ ($M \geq r$) are given. The choice of
these objects will be discussed later. Since $\lambda_k
> 0$ for all $k=1,\ldots,r$ and $\sigma^2 > 0$, it is convenient
to define $\boldsymbol{\zeta} = \log(\mathit{\Lambda})$, i.e.
$\zeta_k = \log \lambda_k$, and $\tau = \log \sigma^2$, and treat
$F$ as a function of $\boldsymbol{\zeta}$ and $\tau$. Note that
$\zeta_k$, $\tau$ can vary freely over $\mathbb{R}$. Then the
Newton-Raphson step for updating  $(\mathit{\Lambda}, \sigma^2)$
(or equivalently ($\boldsymbol{\zeta}, \tau$)) is straightforward.
In the rest of the paper, we treat $\boldsymbol{\zeta}$
interchangeably as an $r\times r$ matrix and as a $1\times r$
vector.

We then give an outline of the Newton-Raphson step for updating $B$.
This involves finding the \textit{intrinsic} gradient and Hessian of
$F$, while treating $\mathit{\Lambda}$ and $\sigma^2$ as fixed. The
key point is the fact that the gradient is a vector field acting on
the tangent space of the manifold ${\cal S}_{M,r}$, and the Hessian
is a bilinear operator acting on the same tangent space. Some facts
about the Stiefel manifold, the tangent space, and its canonical
metric, that are essential to describing and implementing the
algorithm, are given in Appendix A. Based on the notations used
there, we outline the Newton-Raphson algorithm for minimising an
arbitrary function $F(B)$ where $B \in {\cal S}_{M,r}$. For more
details, see Edelman et al. (1999). In the following, we use ${\cal
M}$ to denote ${\cal S}_{M,r}$. Let $\mathit{\Delta}$ denote an
element of the tangent space of ${\cal S}_{M,r}$ at the current
value $B$, denoted by $\mathit{\Delta} \in {\cal T}_B {\cal M}$. It
represents the direction in the tangent space in which a
Newton-Raphson step moves from the current $B$. Let $F_B$ denote the
usual Euclidean gradient, i.e., $F_B = ((\frac{\partial F}{\partial
b_{kl}}))$. For any $\mathit{\Delta} \in {\cal T}_B {\cal M}$,
$F_{BB}(\mathit{\Delta})$ is defined to be the element of ${\cal
T}_B {\cal M}$ satisfying
$$
\langle F_{BB}(\mathit{\Delta}), X \rangle_c =
\frac{\partial^2}{\partial s\partial t}
F(B+s\mathit{\Delta}+tX)\left|_{s,t=0}\right. ,~~\mbox{for all}~~ X
\in {\cal T}_B {\cal M},
$$
where  $\langle , \rangle_c$ denotes the canonical metric on the
Stiefel manifold ${\cal M}$. Also, let $H_F$ denote the Hessian
operator acting on the tangent space ${\cal T}_B{\cal M}$.

\medskip
\noindent{\bf Outline of Newton-Raphson algorithm on ${\cal
S}_{M,r}$:} Given $B \in {\cal S}_{M,r}$,
\begin{enumerate}
\item compute the intrinsic gradient  $\nabla F\left|_B\right.$ of
$F$ at $B$, given by $\nabla F\left|_B\right. = G := F_B - B F_B^T
B$;

\item compute the tangent vector $\mathit{\Delta} := - H_F^{-1}(G)$ at
$B$, by solving the linear system
\begin{eqnarray}
\hskip-.1in F_{BB}(\mathit{\Delta}) - B~\mbox{skew}(F_B^T
\mathit{\Delta}) - ~\mbox{skew}(\mathit{\Delta} F_B^T) B
- \frac{1}{2}\mathit{\Pi} \mathit{\Delta} B^T F_B \hskip-.1in &=& \hskip-.1in - G, \label{eq:NR_1}\\
B^T \mathit{\Delta} + \mathit{\Delta}^T B &=& 0, \label{eq:NR_2}
\end{eqnarray}
where $\mathit{\Pi} = I - BB^T$, and skew$(X) := (X-X^T)/2$;

\item
move from $B$ in the direction $\mathit{\Delta}$ to $B(1)$ along the
geodesics $B(t) = B M(t) + Q N(t)$, where
\begin{itemize}
\item[(i)] $QR = (I - BB^T)\mathit{\Delta}$ is the QR-decomposition, so that
$Q$ is $M\times r$ with orthonormal columns, and $R$ is $r\times r$,
upper triangular;

\item[(ii)]  $A = B^T \mathit{\Delta}$, and
$$
\begin{bmatrix}
M(t) \\
N(t)
\end{bmatrix} = \exp \left\{ t \begin{bmatrix} A & - R^T \\ R & 0 \end{bmatrix}
\right\} ~\begin{bmatrix} I_r \\ 0 \end{bmatrix};
$$
Note that the matrix within exponent is a skew-symmetric matrix and
so the exponential of that can be calculated using the singular
value decomposition.
\end{itemize}

\item
set $B = B(1)$, and repeat until convergence. This means that the
sup-norm of the gradient $G$ is less than a pre-specified tolerance
level.

\end{enumerate}

In the calculation of $F_B$ and $F_{BB}(\mathit{\Delta})$ for $F$
defined by (\ref{eq:objective}), complications associated with the
inversion of $m_i \times m_i$ matrices $P_i$ arise, since $m_i$'s
could vary from sample to sample. We avoid this by a suitable
utilisation of matrix inversion formulae that reduce the problem
to computing inverses of $r\times r$ matrices $Q_i$ instead
(Appendix B). Therefore the proposed procedure can also
efficiently handle the case of relatively dense measurements,
where $m_i$ could be much larger than $r$.  The formulae of these
quantities are given in equations (\ref{eq:F_1_grad_alt}) -
(\ref{eq:F_2_hessian_alt}) in Appendix B. In order to update the
tangent vector $\Delta$, in step 2 of the algorithm, we need to
solve the system of equations given by (\ref{eq:NR_1}) and
(\ref{eq:NR_2}). These are matrix equations and we propose to
solve them via vectorisation [cf. Muirhead (1982)]. This step
requires a considerable amount of computational effort since it
involves the inversion of an $Mr \times Mr$ matrix.


In order to apply the Newton-Raphson algorithm, we need to choose
a suitable basis for representing the eigenfunctions. In the
simulation studies presented in Section \ref{sec:simulation}, we
have used the (orthonormalised) cubic B-spline basis, with equally
spaced knots [Green and Silverman (1994), p. 157]. It is well
known that B-splines provide a flexible, localised and stable
basis for a wide class of smooth functions and are very easy to
compute [Chui (1987), de Boor (1978)]. Different choices of basis
functions are certainly possible, and can be implemented without
changing the structure of the algorithm. Besides the choice of the
basis, the number of basis functions $M$ and the dimension of the
process $r$ need to be specified. We treat the determination of
these two numbers as a model selection problem and discuss this in
Section \ref{sec:CV}.

Given a basis  $\{\phi_k(\cdot)\}$ and fixed $M, r $ with $M \geq
r$, an initial estimate of $\mathit{\Lambda}$ and $B$ can be
obtained by projecting an initial estimate of the covariance
kernel $\widehat{C}(\cdot,\cdot)$ onto the basis functions:
$\{\phi_1(\cdot),\cdots,\phi_M(\cdot)\}$, and then performing an
eigen-decomposition. In the simulation studies, the local
polynomial method and the EM algorithm discussed in Section
\ref{sec:intro} are used to obtain initial estimates of the
covariance kernel, as well as that of the noise variance
$\sigma^2$. The dependence of the proposed method on the initial
estimates is discussed in Section \ref{sec:simulation}.

\section{Approximate cross validation score}\label{sec:CV}


One of the key questions pertaining to nonparametric function
estimation is the issue of model selection. This, in our context
means selecting $r$, the number of nonzero eigenvalues, and the
basis for representing the eigenfunctions. Once we have a scheme
for choosing the basis, the second part of the problem boils down
to selecting $M$, the number of basis functions.  Various methods
for dealing with this include standard criteria like AIC, BIC,
multi-fold cross-validation and leave-one-curve-out
cross-validation.

In this paper, we propose to choose $(M, \ r)$ based on the
criterion of minimising an approximation of the leave-one-curve-out
cross-validation score
\begin{equation}\label{eq:CV_def}
CV := \sum_{i=1}^n
\ell_i(\mathbf{Y}_i,\mathbf{T}_i,\widehat\Psi^{(-i)}),
\end{equation}
where $\mathbf{T}_i = (T_{i1},\ldots,T_{im_i})$. Here $\Psi =
(B,\tau,\boldsymbol{\zeta})$, and $\widehat\Psi^{(-i)}$ is the
estimate of $\Psi$ based on the data excluding curve $i$. In this
paper
$$
\ell_i(\mathbf{Y}_i,\mathbf{T}_i,\Psi) = F_i^1 + F_i^2
$$
[cf. (\ref{eq:objective}) and (\ref{eq:F1_and_F2})]. Note that in
the Gaussian setting, $\ell_i$ is proportional to the negative
log-likelihood (up to an additive constant) of the $i$-th curve.
Therefore in that setting, CV defined through (\ref{eq:CV_def}) is
the \textit{empirical predictive Kullback-Leibler risk}. As
indicated in Section \ref{sec:intro}, the computational cost to
get $CV$ is prohibitive. This necessitates the use of efficient
approximations. Our method of approximation, which parallels the
approach taken by Burman (1990) in the context of fitting
generalized additive models, is based on the following
observations. The Newton-Raphson estimate $\widehat \Psi$, which
is based on the whole data, satisfies the equation
\begin{equation}\label{eq:Psi_hat_eqn}
\sum\limits_{i=1}^n
\nabla\ell_i(\mathbf{Y}_i,\mathbf{T}_i,\widehat\Psi) = 0.
\end{equation}
Also, for each $i$, the corresponding estimate $\widehat
\Psi^{(-i)}$ satisfies the equation
\begin{equation}\label{eq:Psi_hat_i_eqn}
\sum_{j\neq i}
\nabla\ell_j(\mathbf{Y}_j,\mathbf{T}_j,\widehat\Psi^{(-i)}) = 0.
\end{equation}
Here $\{ \ell_i\}_{i=1}^n$ are viewed as functions on the product
space $\widetilde{\cal M} = {\cal M} \times \mathbb{R}^{r+1}$, where
${\cal M}$ is the Stiefel manifold with the canonical metric, to be
denoted by $g \equiv \langle~, ~\rangle_c$. The parameter space
$\mathbb{R}^{r+1}$ refers to $\{(\tau,\underset{r \times
r}{\boldsymbol{\zeta}}) : \tau \in \mathbb{R}, \zeta_k \in
\mathbb{R},~k=1,\ldots,r\}$, with Euclidean metric. $\nabla \ell_i$
denotes the gradient of $\ell_i$ viewed as a vector field on the
product manifold.

The main idea for our approximation scheme is the observation that
for each $i=1,\ldots,n$, the ``leave curve $i$ out'' estimate
$\widehat\Psi^{(-i)}$ is a perturbation of the estimate $\widehat
\Psi$ based on the whole data. Thus, one can expand the left hand
side of (\ref{eq:Psi_hat_i_eqn}) around $\widehat\Psi$ to obtain an
approximation of $\widehat \Psi^{(-i)}$. Then we shall use this
approximation to get a second order approximation to the cross
validation score given by (\ref{eq:CV_def}).

We introduce some notations first. Let $\delta_\tau^i =
\widehat\tau^{(-i)} - \widehat\tau$, $\boldsymbol{\delta}_\zeta^i =
\widehat{\boldsymbol{\zeta}}^{(-i)} - \widehat{\boldsymbol{\zeta}}$
(a $1\times r$ vector), and $\mathit{\Delta}_i =
\overset{.}{\gamma}(0) \in {\cal T}_{\widehat{B}} {\cal M}$, with
$\gamma(t)$ a geodesic on $({\cal M},g)$ starting at $\gamma(0) =
\widehat B$, and ending at $\gamma(1) = \widehat B^{(-i)}$. Note
that,  $\mathit{\Delta}_i$ is an element of the tangent space at
$\widehat B$. Hereafter, we shall use $\ell_j(\widehat\Psi)$, and
$\ell_j(\widehat\Psi^{(-i)})$ to denote
$\ell_j(\mathbf{Y}_j,\mathbf{T}_j,\widehat\Psi)$ and
$\ell_j(\mathbf{Y}_j,\mathbf{T}_j,\widehat\Psi^{(-i)})$,
respectively, for $1\leq i,j \leq n$. Let $\nabla_B \ell_i$ and
$\nabla_B^2\ell_i$ denote gradient and Hessian of $\ell_i$ with
respect to $B$, and $\nabla_{(\tau,\zeta)} \ell_i$ and
$\nabla_{(\tau,\zeta)}^2\ell_i$ denote gradient and Hessian of
$\ell_i$ with respect to $(\tau,\boldsymbol{\zeta})$.
Since the parameter $(\tau,\boldsymbol{\zeta})$ lies in an
Euclidean space, $\nabla_{(\tau,\zeta)}\ell_i$ is an $(r+1)\times
1$ vector and $\nabla_{(\tau,\zeta)}^2\ell_i$ is an $(r+1)\times
(r+1)$ matrix. As mentioned before, $\nabla_B \ell_i$ is a tangent
vector and $\nabla_B^2\ell_i$ is a bilinear operator on the
tangent space $\mathcal{T}_B\mathcal{M}$ of the Stiefel manifold
at the point $B$. The Hessian $\nabla^2 \ell_i$ with respect to
$\Psi = (B,\tau,\boldsymbol{\zeta})$ can be approximated by
$$
\widetilde{\nabla}^2 \ell_i =
\begin{bmatrix}
\nabla_B^2\ell_i & 0 \\
0 & \nabla_{(\tau,\zeta)}^2\ell_i
\end{bmatrix},
$$
by ignoring the mixed-derivative terms
$\nabla_{(\tau,\zeta)}(\nabla_B\ell_i)$ and
$\nabla_B(\nabla_{(\tau,\zeta)}\ell_i)$. This approximation
simplifies the calculation considerably and allows us to treat the
terms involving approximation of $\widehat B^{(-i)}$ (keeping
$(\tau,\boldsymbol{\zeta})$ fixed at
$(\widehat\tau,\widehat{\boldsymbol{\zeta}})$) and that of
$(\widehat\tau^{(-i)},\widehat{\boldsymbol{\zeta}}^{(-i)})$ (keeping
$B$ fixed at $\widehat B$) separately. Thus, a second order Taylor
expansion of the CV score around $\widehat \Psi$ becomes
\begin{eqnarray}\label{eq:cv_MLE_approx_first}
CV &:=& \sum_{i=1}^n
\ell_i(\widehat\Psi^{(-i)})\nonumber\\
&\approx& \sum_{i=1}^n \ell_i(\widehat\Psi) + \Bigl[\sum_{i=1}^n
\langle \nabla_{(\tau,\zeta)} \ell_i(\widehat\Psi),
(\delta_\tau^i,\boldsymbol{\delta}_\zeta^i)^T\rangle +
\frac{1}{2}\sum_{i=1}^n \langle[\nabla_{(\tau,\zeta)}^2
\ell_i(\widehat\Psi)](\delta_\tau^i,\boldsymbol{\delta}_\zeta^i)^T,
(\delta_\tau^i,\boldsymbol{\delta}_\zeta^i)^T\rangle \Bigr] \nonumber\\
&& + \Bigl[\sum_{i=1}^n \langle \nabla_B \ell_i(\widehat\Psi),
\mathit{\Delta}_i\rangle_c + \frac{1}{2}\sum_{i=1}^n \nabla_B^2
\ell_i(\widehat\Psi)(\mathit{\Delta}_i,\mathit{\Delta}_i)\Bigr].
\end{eqnarray}
In order to get first order approximations to the second and third
terms in (\ref{eq:cv_MLE_approx_first}), we shall use equations
(\ref{eq:Psi_hat_eqn}) and (\ref{eq:Psi_hat_i_eqn}). These equations
separate into two sets of equations involving the gradients
$\nabla_{(\tau,\zeta)}\ell_i$ and $\nabla_B\ell_i$, respectively.
The treatment of the former does not require any extra concept
beyond regular matrix algebra, whereas the treatment of the latter
requires Riemannian geometric concepts. However, in terms of the
final form of the approximation, both expressions are very similar.
Denote the Hessian operator of $\sum_j \ell_j$ with respect to $B$
and $(\tau,\boldsymbol{\zeta})$ by $\mathbf{H}_B$ and
$\mathbf{H}_{(\tau,\zeta)}$, respectively. Then our final
approximation to the CV score is given by
\begin{eqnarray}\label{eq:cv_MLE_approx_final}
\widetilde{CV} &:=& \sum_{i=1}^n \ell_i(\widehat\Psi) + \sum_{i=1}^n
\langle \nabla_{(\tau,\zeta)}\ell_i(\widehat\Psi),
[\mathbf{H}_{(\tau,\zeta)}(\widehat\Psi)]^{-1}
\nabla_{(\tau,\zeta)}\ell_i(\widehat\Psi)\rangle\nonumber\\
&& + \sum_{i=1}^n \langle \nabla_B\ell_i(\widehat\Psi),
[\mathbf{H}_B(\widehat\Psi)]^{-1}
\nabla_B\ell_i(\widehat\Psi)\rangle_c\nonumber\\
&& + \frac{3}{2} \sum_{i=1}^n \langle \nabla_{(\tau,\zeta)}^2
\ell_i(\widehat\Psi) [\mathbf{H}_{(\tau,\zeta)}(\widehat\Psi)]^{-1}
\nabla_{(\tau,\zeta)}\ell_i(\widehat\Psi),[\mathbf{H}_{(\tau,\zeta)}(\widehat\Psi)]^{-1}
\nabla_{(\tau,\zeta)}\ell_i(\widehat\Psi)\rangle
\nonumber\\
&&+ \frac{3}{2} \sum_{i=1}^n \nabla_B^2 \ell_i(\widehat\Psi) (
[\mathbf{H}_B(\widehat\Psi)]^{-1} \nabla_B\ell_i(\widehat\Psi),
[\mathbf{H}_B(\widehat\Psi)]^{-1} \nabla_B\ell_i(\widehat\Psi)).
\end{eqnarray}
The details of this derivation are given in Appendix C. Observe
that, in order to obtain the estimate $\widehat\Psi$ using the
Newton-Raphson algorithm, we need to compute the objects
$\nabla_B\ell_i$, $\nabla_{(\tau,\zeta)}\ell_i$,
$\nabla_B^2\ell_i$, $\nabla_{(\tau,\zeta)}^2\ell_i$,
$\mathbf{H}_B$, and $\mathbf{H}_{(\tau,\zeta)}$ at each step.
Indeed, since the Newton-Raphson procedure aims to solve
(\ref{eq:Psi_hat_eqn}), whenever the procedure converges, we
immediately have these objects evaluated at $\widehat\Psi$.
Therefore, the additional computational cost for computing
$\widetilde{CV}$ is a negligible fraction of the cost of obtaining
the estimate $\widehat\Psi$. This provides huge computational
advantage in comparison to the usual leave-one-curve-out CV score
approach. We shall discuss the effectiveness of $\widetilde{CV}$
in model selection in Section \ref{sec:simulation}.

\vskip.15in\noindent{\bf Remark :} It is worth noting that this
approximation approach can be extended to other settings, for
example in nonparametric regression problems. In that context, the
approximation $\widetilde{CV}$ is different from the usual GCV
(\textit{generalized cross validation}) score. Indeed, in the
regression setting, GCV score is obtained by performing a first
order approximation to the usual leave-one-out CV score. In
contrast, our method relies on a second order approximation.



\section{Simulation}\label{sec:simulation}


In this section, we conduct two simulation studies. The first
study is focussed on the estimation accuracy of the proposed
method (henceforth, \texttt{Newton}) and comparing it with two
existing procedures: the local polynomial method (henceforth,
\texttt{loc}) [Yao \textit{et al.} (2005)], and the EM algorithm
(henceforth, \texttt{EM}) [James \textit{et al.} (2000)]. The
second study aims to illustrate the usefulness of the model
selection approach described in Section \ref{sec:CV}. All data are
generated under model (\ref{eq:model}) with Gaussian principal
component scores $\{\xi_{i\nu}\}$. For all settings, $\mu(t)
\equiv 0$, and its estimate $\widehat\mu(t)$, obtained by a local
linear smoothing, is subtracted from the observations before
estimating the other model parameters. The number of measurements
$m_i$ are i.i.d. $\sim {\rm uniform}\{2,\cdots,10\}$; the
measurement points for the ith subject $\{T_{ij}:
j=1,\cdots,m_i\}$ are i.i.d. $\sim {\rm uniform}[0,1]$. For
\texttt{Newton}, cubic $B$-splines with equally spaced knots are
used as basis functions. \texttt{loc} and \texttt{EM} are used to
obtain two different sets of initial estimates. The resulting
estimates by \texttt{Newton} are therefore denoted by
\texttt{New.loc} and \texttt{New.EM}, respectively. For
\texttt{EM}, only initial value of $\sigma$ is needed. Since the
result is rather robust to this choice [James \textit{et al.}
(2000)], it is set to be one. $B$-splines are used as basis
functions; for some cases natural splines are also considered. To
make a distinction, we use \texttt{EM.ns} to denote EM algorithm
that uses natural splines, and \texttt{New.EM.ns} to denote its
corresponding \texttt{Newton} method. For \texttt{loc}, bandwidths
are selected by
the \texttt{h.select()} function in the \texttt{R} package
\texttt{sm}, with \texttt{method="cv"}. Due to the limitation of
space, we only report simulation results in detail for selected
cases which we believe are representative. More results are given
as supplementary material (attached at the end of this paper).

\begin{figure}[h]
\caption{True and estimated eigenfunctions for \texttt{easy} with
$n=200$, $\sigma^2=1/16$, Gaussian noise: true eigenfunctions
(\texttt{Black}); Point-wise average of estimated eigenfunctions
by \texttt{New.EM} (\texttt{Red}); Point-wise $0.95$ and $0.05$
quantiles of estimated eigenfunctions by \texttt{New.EM}
(\texttt{Green})} \label{figure:eigenf_easy}
\begin{center}
\includegraphics[width=6in,height=6in, angle=270]{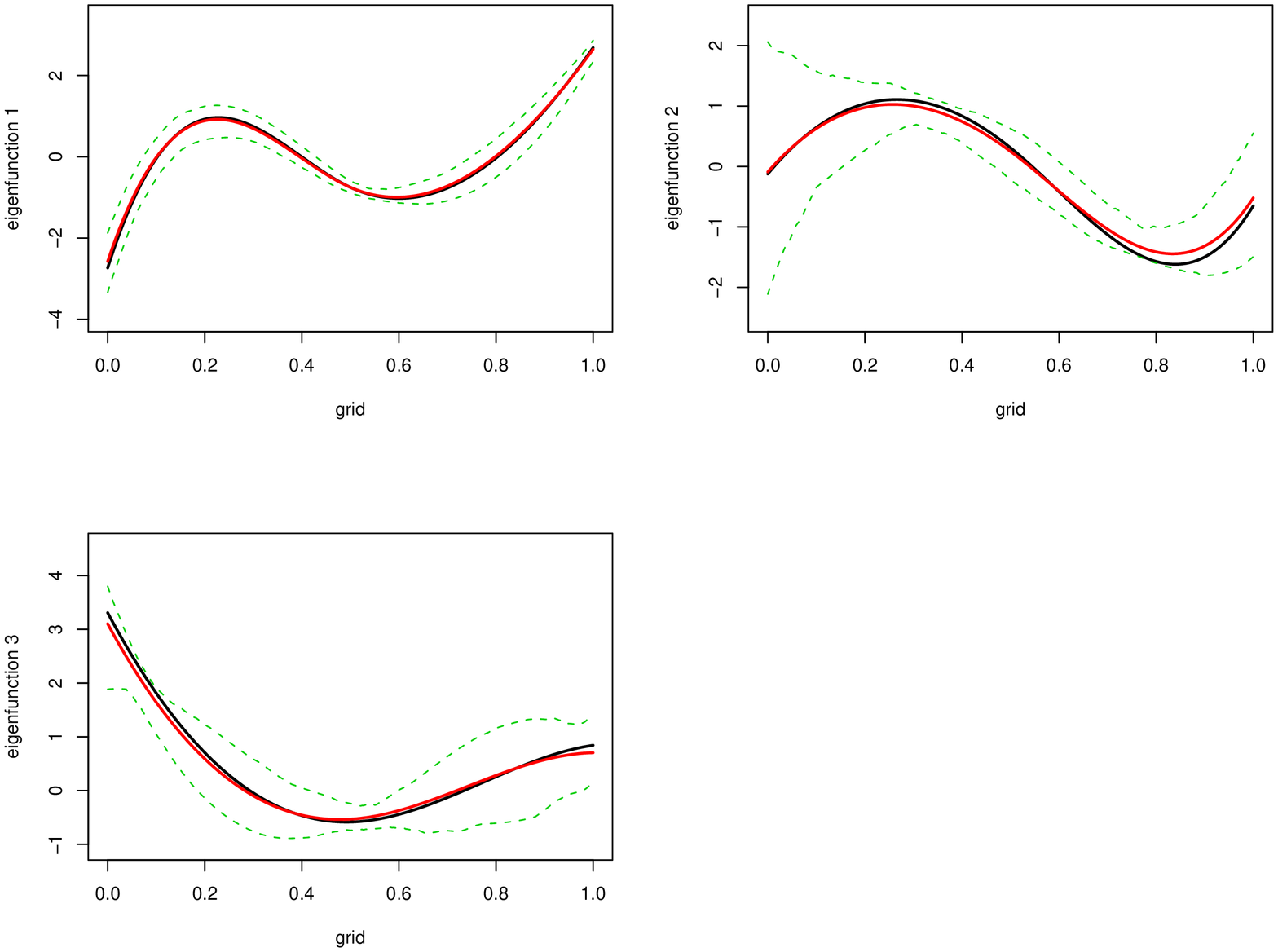}
\end{center}
\end{figure}

\begin{figure}[h]
\caption{True and estimated eigenfunctions for \texttt{practical}
with $n=500$, $\sigma^2=1/16$, Gaussian noise:  true
eigenfunctions (\texttt{Black}); Point-wise average of estimated
eigenfunctions by \texttt{New.EM} (\texttt{Red}); Point-wise
$0.95$ and $0.05$ quantiles of estimated eigenfunctions by
\texttt{New.EM} (\texttt{Green})} \label{figure:eigenf_prac}
\begin{center}
\includegraphics[width=6in,height=6in,angle=270]{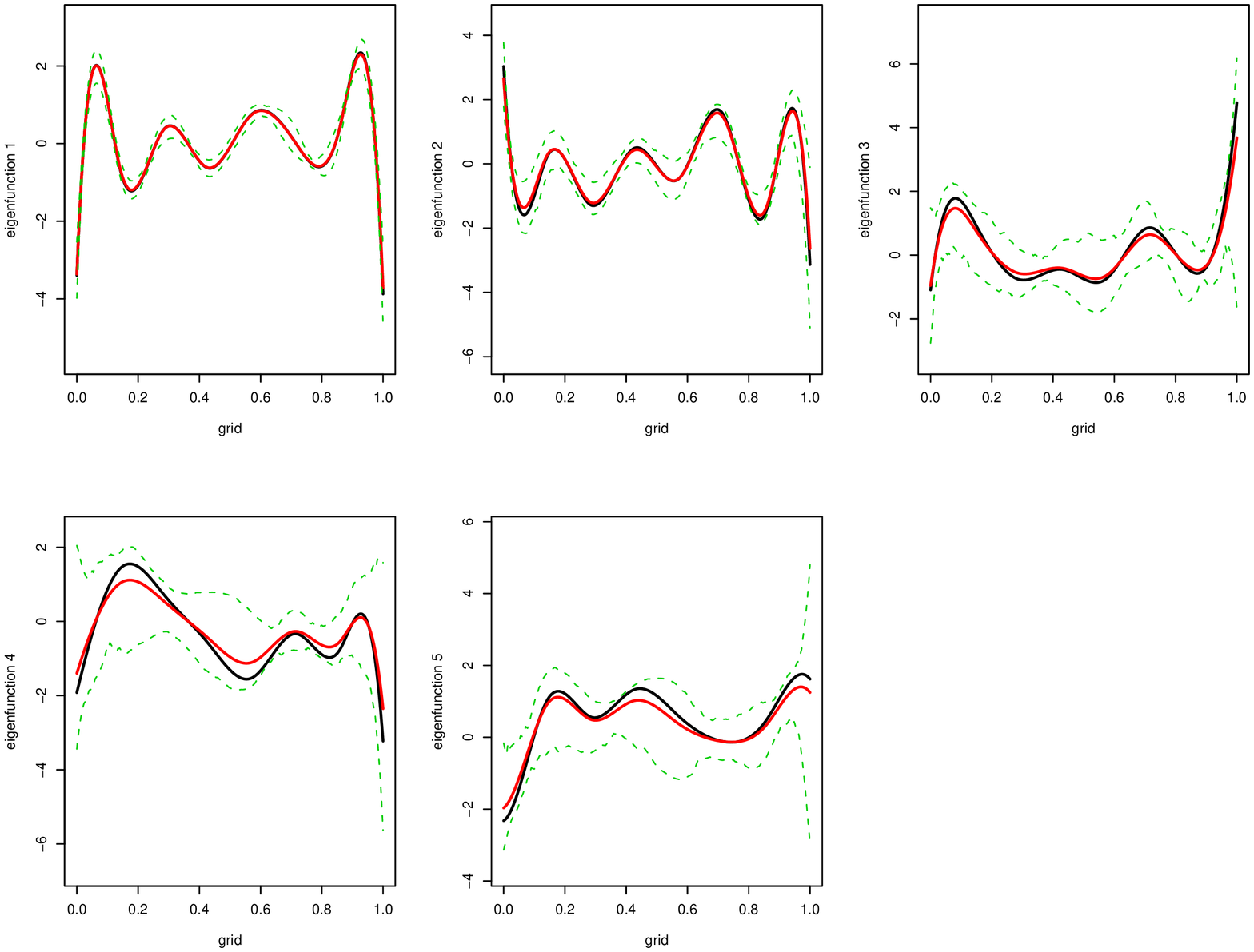}
\end{center}
\end{figure}

\begin{figure}[h]
\caption{True and estimated eigenfunctions for
\texttt{challenging} with $n=500$, $\sigma^2=1/16$, Gaussian
noise: true eigenfunctions (\texttt{Black}); Point-wise average of
estimated eigenfunctions by \texttt{New.EM} (\texttt{Red});
Point-wise $0.95$ and $0.05$ quantiles of estimated eigenfunctions
by \texttt{New.EM} (\texttt{Green})} \label{figure:eigenf_spike}
\begin{center}
\includegraphics[width=6in,height=6in,angle=270]{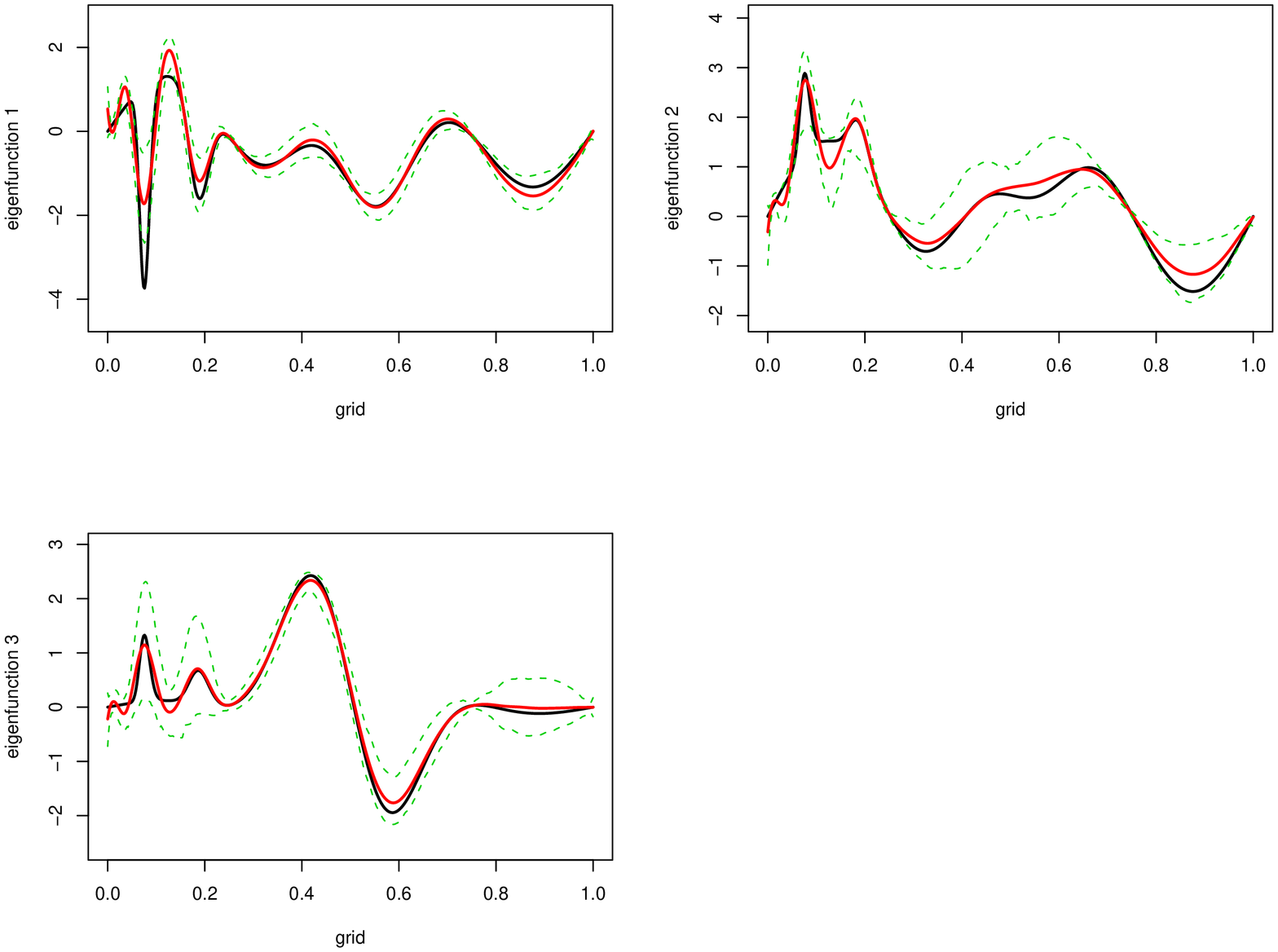}
\end{center}
\end{figure}

In the first study, data are generated under three different
settings (\texttt{easy}, \texttt{practical} and
\texttt{challenging}) with $100$ independent replicates for each
combination of parameters. The simulation scheme is summarised in
Table \ref{table:setting}. As can be seen, different sample sizes,
error variances and error distributions are considered. For the
\texttt{easy} and \texttt{practical} cases (Figures
\ref{figure:eigenf_easy} and \ref{figure:eigenf_prac},
respectively), eigenfunctions are represented by the cubic B-splines
with $M=5$ and $M=10$ equally spaces knots, respectively. For the
\texttt{challenging} case (Figure \ref{figure:eigenf_spike}), the
eigenfunctions are represented by three ``spike" functions and they
can not be represented exactly by cubic B-splines. {
\begin{table}
\scriptsize \centering \caption{Simulation Settings. Shown are the
parameters used in the first simulation study: nonzero eigenvalues
($\lambda_{\nu}$); basis for eigenfunctions ($\psi_{\nu}$); error
variances ($\sigma^2$); error distributions
($\mathcal{L}(\varepsilon)$); sample sizes ($n$).}
\label{table:setting}
\begin{tabular}{cccccc}\\\hline\hline
name& $\lambda_\nu$ & $\psi_\nu$ &$\sigma^2$
&$\mathcal{L}(\varepsilon)$& sample size $n$
\\\hline\hline \texttt{easy}& $(1:3)^{-0.6}$& $<5$ B-spline functions$>$
& $1/16,\ 1/8 $ &$N(0,1)$, $t_4$, $\exp(1)$ & $100, \ 200, \ 500$\\
\texttt{practical}& $(1:5)^{-0.6}$& $<10$ B-spline functions$>$&
$1/16,\ 1/8 $ & $N(0,1)$, $t_4$, $\exp(1)$ & $300, \ 500, \
1000$\\
\texttt{challenging}&$(1:3)^{-0.6}$& $<3$ spike functions$>$&
$1/16,\ 1/8$& $N(0,1)$ &$300, \ 500, \ 1000$
\\\hline\hline
\end{tabular}
\end{table}
}

In the first study,  the true $r$  is used by all three methods.
Note that, the estimation of covariance kernel by \texttt{loc} does
not rely on either $M$ or $r$. For a given $r$, the first $r$
eigenfunctions and eigenvalues of the estimated covariance
$\widehat{C}(\cdot,\cdot)$ (using the optimal choice of bandwidth)
are used. For \texttt{Newton} and \texttt{EM}, a number of different
values of $M$, including the truth, are used to fit the model. For
the \texttt{challenging} case, the ``true" $M$ means the $M$
resulting in least biased projection of the eigenfunctions onto the
B-spline basis, which is $30$. The selection of $(M, r)$ is
discussed in the second study. For \texttt{Newton}, we report the
number of converged replicates (cf. Section \ref{sec:computation})
for each combination of parameters and for each $M$ (Table
\ref{table:modsel_all}). As we shall see, lack of convergence of
\texttt{Newton} is primarily caused by poor initial estimates.
Therefore, it is fair to compare all three methods on the converged
replicates only. The performance of these three methods (based on
converged replicates only) is summarised in Tables
\ref{table:easy_200} to \ref{table:spike_500}. For the estimation of
eigenfunctions, mean integrated squared error (MISE) is used as a
measure of accuracy. We also report the standard deviations of the
integrated squared errors. To evaluate the estimation of eigenvalues
and error variance, mean squared error (MSE) is used as the measure.
Since these quantities are of different orders of magnitude, for the
ease of comparison, the MSEs are divided by the square of the
corresponding true values.

As can be seen from Tables \ref{table:easy_200} to
\ref{table:spike_500}, the MISE/MSE corresponding to
\texttt{Newton} (\texttt{New.loc}, \texttt{New.EM} and
\texttt{New.EM.ns}) shows a good risk behaviour under the true
$M$. The results under a larger $M$ are comparable  to that under
the true $M$. As expected, the performance under an inadequate $M$
is much worse, which reflects the lack of fit. To give a visual
illustration, in Figures \ref{figure:eigenf_easy} to
\ref{figure:eigenf_spike}, we plot the point-wise average of
estimated eigenfunctions by \texttt{New.EM} over all converged
replicates, as well as the point-wise $0.95$ and $0.05$ quantiles,
all under the true $M$ ($M=30$ for the \texttt{challenging} case).
As can be seen from these figures, the average is very close to
the truth, meaning only small biases, except for the
\texttt{challenging} case where $\psi_1$ is not estimated as
accurately mainly due to the intrinsic bias in the B-spline
representation. The width between two quantiles is fairly narrow
meaning small variations, except for occasional large variances at
the boundaries.

In comparison with \texttt{loc} and \texttt{EM}, \texttt{Newton}
generally performs better in terms of MISE for eigenfunctions
under an adequate $M$ ($\geq$ truth). The reduction in MISE varies
from $30\%$ to as high as $95\%$ compared to \texttt{loc}; and
$10\%$ to around $65\%$ compared to \texttt{EM} (except for the
first eigenfunction of the \texttt{challenging} case) (Tables
\ref{table:easy_200} to \ref{table:spike_500}, where the reduction
is always for \texttt{Newton} compared to its initial estimate).
Moreover, comparison of Table \ref{table:prac_500} with Table
\ref{table:prac_1000} shows greater improvement by \texttt{Newton}
with larger sample sizes. As is evident from the tables, there is
also a big improvement of \texttt{New.loc} over \texttt{loc} in
estimation of eigenvalues when $M$ is adequately large. The
reduction in MSEs varies from $30\%$ to as high as $90\%$ with the
exception for the last two eigenvalues of the \texttt{challenging}
case with $n=500, M=30$, where only a little improvement is
observed. In the \texttt{practical} case, there is also an
improvement by \texttt{New.EM} over \texttt{EM}, although the
reduction in MSEs is much less compared to the improvement over
\texttt{loc}. Moreover, under the \texttt{easy} and
\texttt{challenging} cases, small percentages of increase in MSEs
of eigenvalues by \texttt{New.EM} compared to \texttt{EM} are
sometimes observed. In terms of estimating the error variance
$\sigma^2$, \texttt{Newton} is much better than both \texttt{loc}
and \texttt{EM} in most of cases as long as $M$ is adequate. One
problem with \texttt{loc} is that, it gives highly variable, and
sometimes even negative, estimate of $\sigma^2$. For example, for
the easy case with $n=200$, $56$ out of $100$ replicates give a
negative estimate of $\sigma^2$ and for all the simulations we
have done, at least around $20\%$ replicates result in a negative
estimate (see numbers reported in Tables 10-28 in the
supplementary material). This fact is also reflected by the larger
MSE of $\widehat \sigma^2$ using \texttt{New.loc} than using
\texttt{New.EM}.

We observe that \texttt{New.loc} often suffers from lack of
convergence. This phenomenon is more pronounced for the two higher
dimensional cases: \texttt{practical} and \texttt{challenging}
(Table \ref{table:modsel_all}). This is mainly due to the poor
initial estimates by \texttt{loc}. For example, for the
\texttt{practical} case with $n=500$ and $M=10$ (Table
\ref{table:prac_500}), the MISE of the first eigenfunction by
\texttt{loc} is $0.434$, while that by \texttt{EM} is only
$0.054$. However, among the converged replicates, the performance
of \texttt{New.loc} is not much worse than that of
\texttt{New.EM}, especially for the leading eigenfunctions. In the
above case, the MISEs of the first eigenfunction by
\texttt{New.loc} and \texttt{New.EM} are $0.035$ and $0.036$,
respectively. \texttt{New.loc} does tend to give less accurate and
more variable estimates for eigenfunctions corresponding to
smaller eigenvalues. It is also noteworthy that, for the
\texttt{practical} case, \texttt{EM.ns} does not work very well
compared to \texttt{EM} at the true $M$ ($M=10$), but its
performance is much better for a larger $M$ (e.g., $M=20$). This
is because the actual eigenfunctions are represented by cubic
B-splines, thus the use of a natural spline basis could result in
a significant bias. However, among the converged replicates, the
performance of \texttt{New.EM} and \texttt{New.EM.ns} is rather
comparable. The main difference lies in the number of converged
replicates. For $n=500, M=10$, there are $93$ replicates
converging under \texttt{New.EM}, but only $60$ replicates
converging under \texttt{New.EM.ns} (Table
\ref{table:modsel_all}). In contrast, in the \texttt{challenging}
case, the difference between \texttt{EM} and \texttt{EM.ns} is
smaller, since now the biases resulting from representing the
eigenfunctions in the cubic B-spline basis and that in the natural
spline basis are more similar. We also study the impact of
increasing the error variance, as well as different error
distributions (see supplementary material: Tables 13-17, 20, 22-24
and 27 for detailed results). These simulations show that all
three methods are quite robust with respect to these two aspects.


In summary, we observe satisfactory performance of \texttt{Newton}
in terms of estimation, as well as improvements of \texttt{Newton}
over the two alternative methods, especially over \texttt{loc}.
These improvements pertain to both average and standard deviation of
the measures of accuracy, and they increase with the sample size. We
also want to point out that, for the Newton-Raphson algorithm, good
initial estimates are important mainly for the convergence of the
procedure. As long as the estimates converge, the difference in
performance is not very large for the estimation of eigenfunctions
and eigenvalues. We will discuss possible ways to improve
convergence at the end of this section.


As mentioned in Section \ref{sec:intro}, we shall use the
approximate cross validation score defined through
(\ref{eq:cv_MLE_approx_final}) as the criterion for selecting $M$
and $r$ for the \texttt{Newton} procedure. As can be seen from Table
\ref{table:modsel_all}, where $r$ is fixed at the true value, as
long as \texttt{Newton} converges for the correct model (i.e., true
$M$), it is selected almost all the time. Moreover, a model with
inadequate $M$ is not selected unless it is the only model under
which \texttt{Newton} converges. In order to study the selection of
$M$ and $r$ simultaneously, we conduct the second simulation study,
in which there are three leading eigenvalues ($1,  0.66, 0.52$), and
a fourth eigenvalue which is comparable to the error variance
($\lambda_4=0.07$). Additionally, there are 6 smaller eigenvalues
($9.47 \times 10^{-3}, 1.28 \times 10^{-3}, 1.74 \times 10^{-4},
2.35 \times 10^{-5}, 3.18 \times 10^{-6}, 4.30 \times 10^{-7}$).
Thus we refer $r=4$ as the adequate dimension. The corresponding
orthonormal eigenfunctions are represented in a cubic $B$-spline
basis with $M=10$ equally spaced knots. Data are generated with
sample size $n=500$ and Gaussian noises with $\sigma^2 = 1/16$. This
setting is referred as the \texttt{hybrid} case. We fit models with
$M=10,15,20,25$, and $r=2,\ldots,7$. In this setting, our aim is to
get an idea about the typical sizes (meaning $(M,r)$) of models
selected. At the same time, we want to see, whenever a larger than
adequate $r$ (i.e., $r=4$) is selected, whether small eigenvalues
are estimated to be small. This is important because if one uses
this estimation procedure for reducing dimensionality of the data,
for example by projecting the data onto the selected eigen-basis,
then ``spurious'' components should not have large weights in that
representation. Moreover, even if a larger $r$ is selected, as long
as the small or zero eigenvalues are estimated to be small, the
result is not going to be too misleading, in that, people can always
choose a smaller model based on the \textit{fraction of explained
variation (FEV)}.

In Table \ref{table:hybrid_modsel_500}, for both \texttt{New.EM} and
\texttt{New.loc}, there is a big drop in the number of converged
replicates from $r=5$ to $r=6$ and even bigger drop from $r=6$ to
$r=7$. Now the lack of convergence is a reflection of a combination
of poor initial estimates and larger than adequate $r$. The latter
is actually a safeguard against selecting unnecessarily large
models. Note that, under large $r$, the system under true parameters
is going to be (nearly) singular. In the case of \texttt{New.loc},
both factors apply whenever there is lack of convergence. In the
case of \texttt{New.EM}, the second factor is predominant. We find
that, for \texttt{New.EM}, $M=10$ and $r=5$ or $6$ are the preferred
models by the proposed approximate CV score; however, for
\texttt{New.loc}, $M=10$ and $r=3$ or $4$ are the ones selected most
often. The latter is mainly due to lack of convergence of
\texttt{New.loc} for  $r \geq 4$. Therefore, we will focus on the
results of \texttt{New.EM} hereafter.  We observe that, for models
with $r=5$ and $r=6$, the small eigenvalues (the fifth one and/or
the sixth one) are estimated to be reasonably small by
\texttt{New.EM} (data not shown). We then use the standard procedure
of FEV on the selected model to further prune down the value of r:
for every model $(M^*,r^*)$ selected by the CV criterion, we choose
the smallest index $\overline{r}$ for which the ratio
$\sum_{\nu=1}^{\overline{r}}\widehat\lambda_\nu/\sum_{\nu=1}^{r*}
\widehat \lambda_\nu$ exceeds a certain threshold $\kappa$. In this
study, we consider $\kappa = 0.995, 0.99, 0.95$. The results of the
model selection using this additional FEV criterion are reported in
Table \ref{table:hybrid_modsel_500} in the parentheses. As can be
seen, under $\kappa=0.995$, the most frequently selected models
become $M=10$ and $r=4$ or $r=5$ (the first number in the
parentheses). If we set $\kappa=0.99$, the most frequently selected
models become $M=10$ and $r=4$ (the second number in the
parentheses). Under $\kappa=0.95$, the preferred model becomes
$M=10$ and $r=3$. Note that, in \texttt{hybrid} case, the first
three eigenvalues are dominant and compared to the error variance
$\sigma^2$, the first four eigenvalues are not negligible.
Therefore, the additional FEV criterion gives very reasonable model
selection results. This is another indicator that in the models
selected by the approximate CV criterion, the small eigenvalues
indeed are estimated to be small.

In summary, the approximate CV score (\ref{eq:cv_MLE_approx_final})
is very effective in selecting the correct $M$-- the number of basis
functions needed to represent the eigenfunctions. It has the
tendency to select slightly larger than necessary $r$. However, in
those selected models, the \texttt{Newton} estimates of the small or
zero eigenvalues are quite small. Therefore, the model selection
results are not going to be very misleading and an additional  FEV
criterion  can be applied to select a smaller model (in terms of
$r$).

Finally, we want to discuss some practical aspects of the proposed
method. It is noted that, the $Mr \times Mr$ linear system
(\ref{eq:NR_1}) and (\ref{eq:NR_2}) is sometimes nearly singular,
causing \texttt{Newton} to terminate without the gradient
converging to zero (i.e., fail to converge). This phenomenon is
most obvious for \texttt{New.loc} due to the poor initial
estimates. The system becomes more stable as sample size $n$
becomes larger, as demonstrated by comparing the
\texttt{practical} case with $n=500$ to $n=1000$ in Table
\ref{table:modsel_all}. Combining the \texttt{Newton} results with
different initial estimates, for example \texttt{New.loc} and
\texttt{New.EM} (replace one by another if the first one fails to
converge), can improve convergence, and consequently the model
selection results (cf. \texttt{combine} and \texttt{combine.ns} in
Table \ref{table:modsel_all}). In addition, it is well known that
the initial steps of a Newton-Raphson algorithm are typically too
large [Boyd and Vandenberghe (2004)]. To avoid this, we suggest to
use smaller step sizes in the beginning. That is, in the
Newton-Raphson algorithm, instead of updating $B$ by $B(1)$, we
update $B$ by $B(\alpha)$ for some $0<\alpha<1$. We have already
incorporated this in our implementation. All codes for the
simulation studies are written in \texttt{R} language and running
under \texttt{R} version 2.4.0 on a machine with Pentium Duo core,
CPU 3.20 GHz and 3.50 GB RAM. The code for \texttt{EM} is kindly
provided by professor James at USC via personal communication. The
computational cost is summarised in Table \ref{table:comp} for two
settings. Note that, since \texttt{Newton} needs an initial
estimate, the computational cost reported there is the additional
time cost. As can be seen, \texttt{Newton} (together with
obtaining initial estimates and calculating the approximate CV
score) requires around $2.5$ times as much effort as \texttt{EM},
and sometimes more compared to \texttt{loc}. A more efficient
implementation of the estimation procedure is currently being
pursued.

\section{Application}
\label{sec:real_data}

As an application of our method to a real problem, we analyse the
data on \textit{CD4$+$ cell number count} collected as part of the
Multicenter AIDS Cohort Study (MACS) [Kaslow \textit{et al.}
(1987)]. The data is from Diggle, Heagerty, Liang and Zeger
(2002), and is downloadable at {\small
\texttt{http://www.maths.lancs.ac.uk/$\sim$diggle/lda/Datasets/lda.dat}}
. It consists of $2376$ measurements of CD4+ cell counts against
time since seroconversion (time when HIV becomes detectable which
is used as zero on the time line) for $369$ infected men enrolled
in the study. Five patients, for whom there was only one
measurement, were removed from our analysis. For the rest $364$
subjects, the number of measurements varies between $2$ and $12$,
with a median of $6$ and a standard deviation of $2.66$. The time
span of the study is about $8.5$ years (covering about three years
before seroconversion and 5.5 years after that). The goal of our
analysis is to understand the variability of CD4 counts as a
function of time since seroconversion. We expect that this will
provide useful insights into the dynamics of the process. This
data set has been analysed by many other authors using various
approaches, including varying coefficient models [Fan and Zhang
(2000), Wu and Chiang (2000)], functional principal component
approach [Yao \textit{et al.} (2005)] and parametric random
effects models [Diggle \textit{et al.} (2002)].

\begin{figure}[h]
\caption{\texttt{CD4 counts} data: estimated mean and
eigenfunctions. First panel: estimated mean function; Second
panel: estimated eigenfunctions by \texttt{New.EM}:
$\widehat{\psi}_1$=\texttt{Black},
$\widehat{\psi}_2$=\texttt{Red},
$\widehat{\psi}_3$=\texttt{Green},
$\widehat{\psi}_4$=\texttt{Blue}; Third to sixth panels: estimated
eigenfunctions by \texttt{loc} (\texttt{Blue}), \texttt{New.EM}
(\texttt{Red}), \texttt{EM} (\texttt{Green})}
\label{figure:cd4_eigenf}
\begin{center}
\includegraphics[width=6in,height=6in,angle=270]{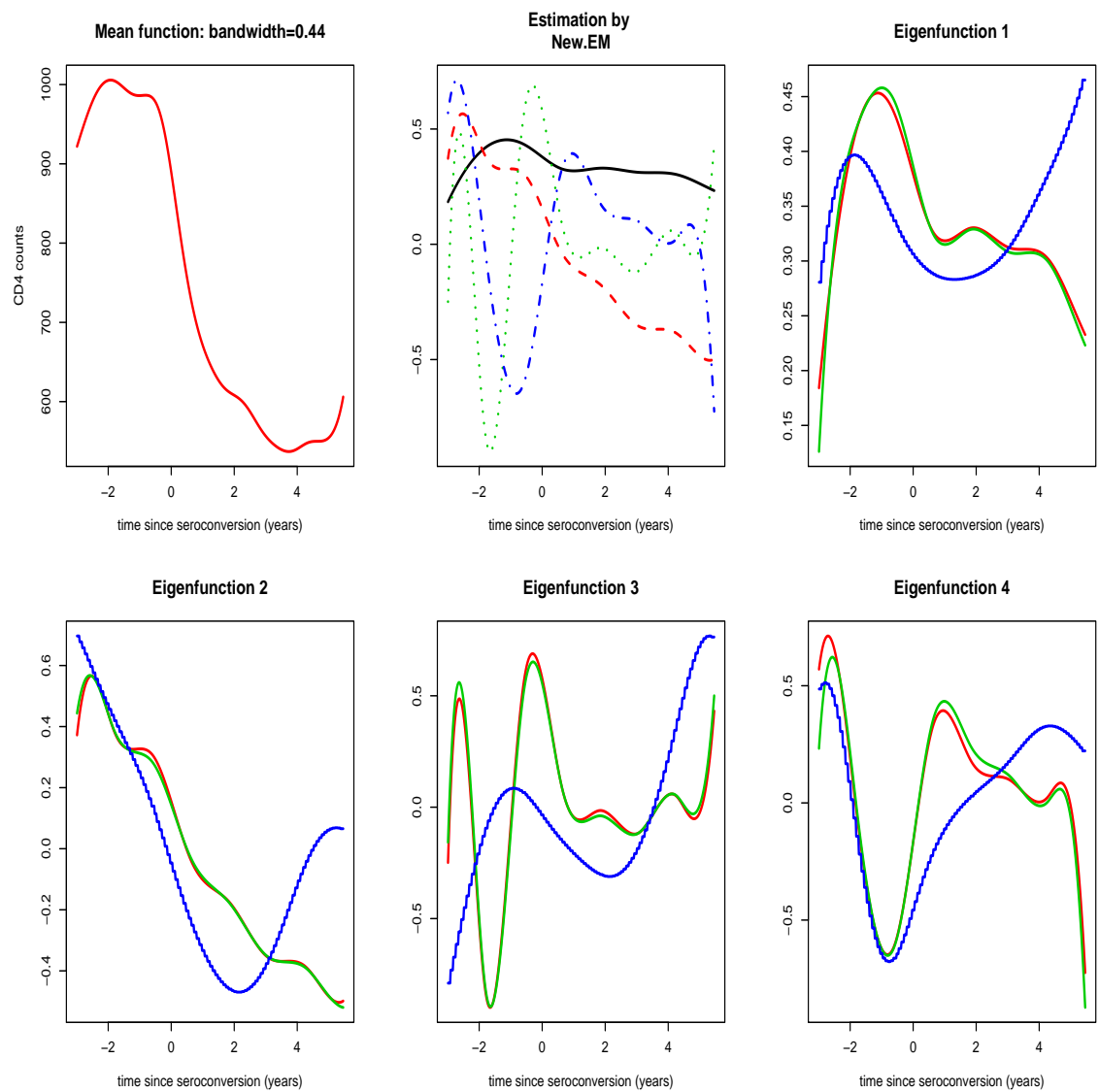}
\end{center}
\end{figure}

In our analysis, four methods: \texttt{EM}, \texttt{New.EM},
\texttt{loc} and \texttt{New.loc} are used. (The cubic B-spline
basis with equally spaced knots are used for \texttt{EM} and
\texttt{Newton}). Several different models, with $M$ taking values
$5, 10, 15, 20$, and $r$ taking values $2,\ldots,6$ are
considered. The approximate cross-validation criterion
$\widetilde{CV}$ is used for model selection. The model with
$M=10,  r=4$ results in the smallest score and thus is selected.
Figure \ref{figure:cd4_eigenf} shows the estimated eigenfunctions
under the selected model.
The estimates of the error variance and eigenvalues are given in
Table \ref{table:real_data}. Under the selected model,
\texttt{New.EM} and \texttt{EM} result in quite similar estimates
for both eigenvalues and eigenfunctions. On the other hand, the
estimates of \texttt{loc} are very different. For \texttt{loc},
$\widehat \lambda_1$ is much larger compared to that of
$\widehat\lambda_2$, whereas in the case of \texttt{New.EM} and
\texttt{EM}, they are of the same order. Since \texttt{New.loc}
fails to converge under the selected  model, its estimates are not
reliable and thus not reported here. Moreover, based on our
experience with the simulation studies, this might be an indicator
that the corresponding results by \texttt{loc} are not altogether
reliable either.
The estimated error variance is about $38,000$ by \texttt{New.EM}
and the results of \texttt{New.EM} suggest that there are $4$
non-negligible eigenvalues, two of which are large, and the other
two are relatively small.

Next, we give an interpretation of the shape of the mean function
and that of the eigenfunctions. The estimated mean function is
shown as the first panel of Figure \ref{figure:cd4_eigenf},
together with the optimal bandwidth by \texttt{h.select()}
function in R package \texttt{sm}. The shape of the mean function
reflects the fact that with the progression of the disease, the
CD4+ cell count tends to decrease. The eigenfunctions capture the
fluctuation of individual trajectories around the mean function.
The first eigenfunction is rather flat compared to the other three
eigenfunctions (Figure \ref{figure:cd4_eigenf}, panel two). This
means that it mainly captures the baseline variability in the CD4+
cell count from one subject to another. This is consistent with
the random effects model proposed in Diggle \textit{et al.} (2002)
(page 108-113). It is also noticeable that the second
eigenfunction has a shape similar to that of the mean function
(Figure \ref{figure:cd4_eigenf}, panel four). The shapes of the
first two eigenfunctions, and the fact that their corresponding
eigenvalues are relatively large, seem to indicate that a simple
linear dynamical model, with random initial conditions, may be
employed in studying the dynamics of CD4+ cell count. This
observation is also consistent with the implication by the
time-lagged graphs used in Diggle \textit{et al.} (2002) (Fig.
3.13, p. 47). The third and fourth eigenvalues are comparable in
magnitude to the error variance, and the corresponding
eigenfunctions have somewhat similar shapes. They correspond to
the contrast in variability between early and late stages of the
disease. Of course, there are a number of measured and unmeasured
covariates that are very likely to influence the dynamics of this
process. Thus a more elaborate model that incorporates covariate
effects should give a better interpretation of the eigenfunctions
corresponding to the smaller eigenvalues, and that of the
individual trajectories.

\section{Discussion}\label{sec:discussion}


In this paper, we presented a method that utilizes the intrinsic
geometry of the parameter space explicitly to obtain the estimate
in a non-regular problem, that of estimating eigenfunctions and
eigenvalues of the covariance kernel when the data are only
observed at sparse and irregular time points.  We did comparative
studies with two other estimation procedures by James \textit{et
al.} (2000) and Yao \textit{et al.} (2005). We presented a model
selection approach based on the minimization of an approximate
cross-validation score with respect to the model parameters. Based
on our simulation studies, we have found that the proposed
geometric approach works well for both estimation and model
selection. Moreover, its performance is in general better than
that of the other two methods. We also looked at a real-data
example to see how our method captures the variability in the
data. In the following, we briefly sketch some on-going work
relating to the problem studied in this paper.

There are a few aspects of the problem that can be investigated
further. One is the asymptotic behaviour of our estimates.
%
%
Asymptotic results have been established under very general
conditions for the local polynomial method in Hall \textit{et al.}
(2006). Based on the numerical comparisons, it is expected that
\texttt{Newton} with either \texttt{loc} or \texttt{EM} as initial
estimate, should have at least as good a risk behavior as that of
the local polynomial method.
A closely related problem  is the estimation of the eigenvalues
and eigenvectors of covariance matrix for high dimensional
multivariate Gaussian data, under the rank-restricted assumption.
It is known that, in this case the usual PCA estimates, i.e.,
eigenvalues and eigenvectors of the sample covariance matrix, are
the MLE's for their population counterparts [Muirhead (1982)]. In
Paul (2005), it has been shown that under the above setting, risk
of the PCA estimators of eigenvectors, measured under the squared
error loss, achieves the optimal nonparametric rate when the
dimension-to-sample size ratio converges to zero. Works currently
being pursued by the present authors indicate that the PCA (i.e.
restricted ML) estimators should also achieve the asymptotically
optimal risk. This is through an efficient score representation of
the PCA estimator that utilizes the intrinsic geometry of the
parameter space.
We also proved consistency, and obtain rates of convergence of the
proposed estimator for functional principal components, in a
regime of relatively dense measurements. We are currently working
on extending these results to sparse measurements case.
The key components of the asymptotic analysis of our estimator are
: (i) analysis of the expected loss function
(for the Gaussian model, this is the Kullback-Leibler
discrepancy); (ii) study of the Hessian of the loss function
(\textit{intrinsic information operator} in the Gaussian case).
The essential difficulty of the analysis in the current context
lies in the fact that the measurements are sparsely distributed in
time.
Regarding the rate of convergence, we do not expect the
distribution of noise to play any significant role and the
existence of enough moments should suffice.

Finally, we want to point out that, there are many statistical
problems with (part of) the parameters having orthornormality
constraints. Some examples include, extension of the current
framework to spatio-temporal data, inclusion of covariate effects
in FPCA, and problems involving orthonormality as natural
identifiability constraints. As long as we have  (i) explicit form
and smoothness of the loss function; (ii) the ability to compute
the intrinsic gradient and Hessian of the loss function, we can
adopt a similar approach, for both estimation and model selection.
Here we briefly discuss two examples which are closely related to
the problem studied in this paper.

The first example relates to an alternative to the
\textit{restricted maximum likelihood} approach pursued in this
paper. This involves representing the eigenfunctions in a
sufficiently rich class of basis functions (i.e., $M$ large), and
then adding a roughness penalty, e.g. $\kappa \sum_{\nu=1}^r \int
(\psi_\nu''(t))^2 dt$, for some $\kappa > 0$, to the negative
log-likelihood/loss function to control the degree of smoothness
of the eigenfunctions [cf. Green and Silverman (1994)]. If we use
the expansion $\psi_\nu(t) = \sum_{k=1}^M b_{k\nu}\phi_k(t)$, for
a known set of  orthonormal functions $\{\phi_1,\ldots,\phi_M\}$,
then the roughness  penalty can be expressed as $\kappa~ Tr(B^T
\mathbf{R}B)$, where  $\mathbf{R}$ is the $M \times M$ matrix
given by $\mathbf{R} = \int  (\Phi''(t)) (\Phi''(t))^T dt$, with
$\Phi(\cdot) =  (\phi_1(\cdot),\ldots,\phi_M(\cdot))^T$. Thus, the
penalized  log-likelihood is still a function of $B$ (and of
$\Lambda$ and  $\sigma^2$), where $B \in {\cal S}_{M,r}$.
Straightforward  algebra shows that the corresponding
\textit{penalized maximum likelihood estimate} can be obtained by
simple modifications of the proposed procedure.


Another problem relates to incorporation of covariate effects in the
analysis of longitudinal data.
For example, Cardot (2006) studies a model where the covariance of
$X(\cdot)$ conditioning on a covariate $W$ has the following
expansion : $C^w(s,t) := \mbox{Cov}(X(s),X(t)|W=w) = \sum_{\nu\geq
1} \lambda_\nu(w)\psi_\nu(s,w)\psi_\nu(t,w)$. He proposes a
kernel-based nonparametric approach for estimating the eigenvalues
and eigenfunctions (now dependent on $w$). In practice this method
would require dense measurements. A modification of our method can
easily handle the case, even for sparse measurements, when the
eigenvalues are considered to be simple parametric functions of
$w$, and eigenfunctions do not depend on $w$. For example, one
model is $\lambda_\nu(w) := \alpha_\nu e^{w^T  \beta_\nu}$,
$\nu=1,\ldots,r$, for some parameters $\beta_1,\ldots,\beta_r$,
assuming that $\alpha_\nu =0$ and $\beta_\nu=0$ for $\nu > r$.
This model captures the variability in amplitude of the
eigenfunctions in the individual sample curves as a function of
the covariate. In this setting we can express the conditional
likelihood of the data
$\{(\{Y_{ij}\}_{j=1}^{m_i},W_i):i=1,\ldots,n\}$ explicitly. Its
maximization under the restriction that the estimated
eigenfunctions represented by a set of smooth basis functions can
be carried out by a modification of the procedure proposed in this
paper.


\section*{Acknowledgements}
The authors thank G. James for providing the code for the EM
approach. We thank D. Temple Lang for helping in implementing the
proposed procedure in R and P. Burman for insightful discussions.

\section*{References}


\vskip.075in\noindent Ash, R. B. (1972)  \textit{Real Analysis and
Probability}, Academic Press.

\vskip.075in\noindent Besse, P., Cardot, H. and Ferraty, F. (1997)
Simultaneous nonparametric regression of unbalanced longitudinal
data. \textit{Computational Statistics and Data Analysis} {\bf 24},
255-270.


\vskip.075in\noindent Boente, G. and Fraiman, R. (2000)
Kernel-based functional principal components analysis.
\textit{Statistics and Probability Letters} {\bf 48}, 335-345.

\vskip.075in\noindent Boyd, S. and Vandenberghe, L. (2004)
\textit{Convex Optimization}. Cambridge University Press.

\vskip.075in\noindent Burman, P. (1990) Estimation of generalized
additive models. \textit{Journal of Multivariate Analysis} {\bf 32},
230-255.

\vskip.075in\noindent Cai, T. and Hall, P. (2006) Prediction in
functional linear regression. \textit{Annals of Statistics} {\bf
34}, 2159-2179.

\vskip.075in\noindent Cardot, H., Ferraty F. and Sarda P. (1999)
Functional Linear Model. \textit{Statistics and Probability Letters}
{\bf 45}, 11-22.

\vskip.075in\noindent Cardot, H. (2000) Nonparametric estimation of
smoothed principal components analysis of sampled noisy functions.
\textit{Journal of Nonparametric Statistics} {\bf 12}, 503-538.

\vskip.075in\noindent Cardot, H. (2006) Conditional functional
principal components analysis. \textit{Scandinavian Journal of
Statistics} {\bf 33}, 317-335.

\vskip.075in\noindent Chui, C. (1987) \textit{Multivariate Splines}.
SIAM.


\vskip.075in\noindent de Boor, C. (1978) \textit{A Practical Guide
to Splines}. Springer-Verlag, New York.

\vskip.075in\noindent Dempster, A. P., Laird, N. M. and Rubin, D. B.
(1977) Maximum likelihood from incomplete data via the EM algorithm
(with discussion). \textit{Journal of the Royal Statistical Society,
Series B} {\bf 39}, 1-38.

\vskip.075in\noindent Diggle, P. J., Heagerty, P., Liang, K.-Y., and
Zeger, S. L. (2002) \textit{Analysis of Longitudinal Data, 2nd.
Edition}. Oxford University Press.

\vskip.075in\noindent Edelman, A., Arias, T. A. and Smith, S. T.
(1999) The geometry of algorithms with orthogonality constraints,
\textit{SIAM Journal on Matrix Analysis and Applications} {\bf 20},
303-353.


\vskip.075in\noindent Fan, J. and Zhang, J. T. (2000) Two-step
estimation of functional linear models with applications to
longitudinal data. \textit{Journal of Royal Statistical Society,
Series B} {\bf 62}, 303-322.

\vskip.075in\noindent Ferraty, F. and Vieu, P. (2006)
\textit{Nonparametric Functional Data Analysis : Theory and
Practice}. Springer.


\vskip.075in\noindent Green, P. J. and Silverman, B. W. (1994)
\textit{Nonparametric Regression and Generalized Linear Models : A
Roughness Penalty Approach}. Chapman \& Hall/CRC.

\vskip.075in\noindent Hall, P. and Horowitz, J. L. (2007)
Methodology and convergence rates for functional linear regression.

\noindent\texttt{http://www.faculty.econ.northwestern.edu/faculty/horowitz/papers/hhor-final.pdf}

\vskip.075in\noindent Hall, P., M\"{u}ller, H.-G. and Wang, J.-L.
(2006) Properties of principal component methods for functional and
longitudinal data analysis. \textit{Annals of Statistics} {\bf 34},
1493-1517.


\vskip.075in\noindent James, G. M., Hastie, T. J. and Sugar, C. A.
(2000) Principal component models for sparse functional data.
\textit{Biometrika}, {\bf 87}, 587-602.

\vskip.075in\noindent Kaslow R. A., Ostrow D. G., Detels R., Phair
J. P., Polk B. F., Rinaldo C. R. (1987) The Multicenter AIDS Cohort
Study: rationale, organization, and selected characteristics of the
participants. \textit{American Journal of Epidemiology} {\bf
126}(2), 310-318.

\vskip.075in\noindent Lee, J. M. (1997) \textit{Riemannian
Manifolds: An Introduction to Curvature}, Springer.

\vskip.075in\noindent Marron, S. J., M\"{u}ller, H.-G., Rice, J.,
Wang, J.-L., Wang, N. and Wang, Y. (2004) Discussion of
nonparametric and semiparametric regression. \textit{Statistica
Sinica} {\bf 14}, 615-629.

\vskip.075in\noindent Meyer, K. and Kirkpatrick, M. (2005)
Restricted maximum likelihood estimation of genetic principal
components and smoothed covariance matrices. \textit{Genetic
Selection Evolution} {\bf 37}, 1-30.

\vskip.075in\noindent Muirhead, R. J. (1982) \textit{Aspects of
Multivariate Statistical Theory}, John Wiley \& Sons.


\vskip.075in\noindent Paul, D. (2005) \textit{Nonparametric
Estimation of Principal Components}. Ph.D. Thesis,
\textit{Stanford University}.



\vskip.075in\noindent Rice, J. A. and Wu, C. O. (2001) Nonparametric
mixed effects models for unequally sampled noisy curves.
\textit{Biometrics} {\bf 57}, 253-259.

\vskip.075in\noindent Ramsay, J. and Silverman, B. W. (2005)
\textit{Functional Data Analysis, 2nd Edition}. Springer.


\vskip.075in\noindent Smith, S. T. (2005) Covariance, subspace and
intrinsic Cram\'{e}r-Rao bounds. \textit{IEEE Transactions on Signal
Processing} {\bf 53}, 1610-1630.

\vskip.075in\noindent Wu, C. and Chiang, C. (2000) Kernel smoothing
on varying coefficient models with longitudinal dependent variables.
\textit{Statistica Sinica} {\bf 10}, 433-456.

\vskip.075in\noindent Yao, F., M\"{u}ller, H.-G. and Wang, J.-L.
(2005) Functional data analysis for sparse longitudinal data.
\textit{Journal of the American Statistical Association} {\bf 100},
577-590.

\vskip.075in\noindent Yao, F., M\"{u}ller, H.-G. and Wang, J.-L.
(2006) Functional linear regression for longitudinal data.
\textit{Annals of Statistics} {\bf 33}, 2873-2903.


\section*{Appendix A : Review of some Riemannian geometric concepts}


Let $({\cal M},g)$ be a smooth manifold with Riemannian metric $g$.
We shall denote the tangent space of ${\cal M}$ at $p \in {\cal M}$
by ${\cal T}_p {\cal M}$. We shall first give some basic definitions
related to the work we present in this article. A good reference is
Lee (1997).

\subsection*{Gradient and Hessian of a function}

\begin{itemize}
\item {\bf Gradient :} Let $f:{\cal M} \to \mathbb{R}$ be a smooth
function. Then $\nabla f$, the \textit{gradient} of $f$, is a
\textit{vector field} on $\mathcal{M}$ defined by the following:
\begin{itemize}
\item[]
for any $X \in {\cal T}{\cal M}$, (i.e., a vector field on ${\cal
M}$), $\langle \nabla f, X \rangle_g = X(f)$, where $X(f)$ is the
\textit{directional derivative} of $f$ w.r.t. $X$ : $
X(f)\left|_p\right. = \frac{df(\gamma(t))}{dt}\left|_{t=0}\right. $
for any differentiable curve $\gamma$ on ${\cal M}$ with $\gamma(0)
= p$, $\overset{.}{\gamma}(0) = X(p)$.
\end{itemize}
Note that $X(f): p \rightarrow X(f)\left|_p\right.$ is a function
that maps ${\cal M}$ to $\mathbb{R}$.

\item {\bf Covariant derivative :} (also known as \textit{Riemannian connection}) : Let
$X, Y \in {\cal T M}$ be two vector fields on ${\cal M}$. Then the
vector field $\overline{\nabla}_Y X \in {\cal TM}$ is called the
\textit{covariant derivative of $X$ in the direction of $Y$} if the
operator $\overline{\nabla}$ satisfies the following properties:
\begin{itemize}
\item[(a)] (Bi-linearity) : For $\lambda_1,\lambda_2 \in
\mathbb{R}$,
$$
\overline{\nabla}_Y (\lambda_1 X_1 + \lambda_2 X_2) = \lambda_1
\overline{\nabla}_Y X_1 + \lambda_2 \overline{\nabla}_Y X_2
$$
and
$$\overline{\nabla}_{\lambda_1 Y_1 + \lambda_2 Y_2} X = \lambda_1
\overline{\nabla}_{Y_1} X + \lambda_2 \overline{\nabla}_{Y_2} X.
$$

\item[(b)] (Leibniz) : for a smooth function $f : {\cal M} \to
\mathbb{R}$ ,
$$
\overline{\nabla}_Y (f \cdot X) = Y(f) \cdot X + f \cdot
\overline{\nabla}_Y X.
$$

\item[(c)] (Preserving metric) : for $X,Y,Z \in {\cal TM}$,
$$
Z(\langle X, Y\rangle_g) = \langle \overline{\nabla}_Z X, Y\rangle_g
+ \langle X, \overline{\nabla}_Z Y\rangle_g.
$$

\item[(d)] (Symmetry) : $\overline{\nabla}_X Y - \overline{\nabla}_Y X = [X,Y]$
where $[X,Y] := X(Y) - Y(X) \in {\cal TM}$, i.e., for a smooth $f :
{\cal M} \to \mathbb{R}$, $[X,Y](f) = X(Y(f)) - Y(X(f))$.

\end{itemize}

\item {\bf Hessian operator:} For a smooth function $f : {\cal M}
\to \mathbb{R}$, $H_f : {\cal TM}\times {\cal TM} \to \mathbb{R}$ is
the bi-linear form defined as
$$
H_f(Y,X) = \langle \overline{\nabla}_Y (\nabla f), X\rangle_g, \ X,Y
\in {\cal TM}.
$$
Note that, by definition, $H_f$ is bi-linear and symmetric (i.e.,
$H_f(Y,X) = H_f(X,Y)$). For national simplicity, sometimes we also
write $\overline{\nabla}_Y (\nabla f)$ as $H_f(Y)$. $H_f$ can be
calculated in the following manner. Note that, for a smooth curve
$\gamma(\cdot)$ on the manifold ${\cal M}$, with $X =
\overset{.}{\gamma}(t)$, it follows that
\begin{eqnarray*}
\frac{d^2 f(\gamma(t))}{dt^2} = \frac{d}{dt} \left(
\frac{df(\gamma(t))}{dt} \right) &=&  \frac{d}{dt} (\langle \nabla
f, X \rangle_g \left|_{\gamma(t)}\right.)
\\
&=& X(\langle \nabla f, X \rangle_g) = \langle \overline{\nabla}_X
(\nabla f), X\rangle_g + \langle \nabla f,\overline{\nabla}_X X
\rangle_g,
\end{eqnarray*}
where the last step follows by applying the \textit{Leibniz rule}
for the covariant derivative. Since $\gamma(\cdot)$ is a geodesic if
and only if $\overline{\nabla}_X X = 0$ (self-parallel), this
implies that, for such a $\gamma(\cdot)$,
$$
\frac{d^2 f(\gamma(t))}{dt^2} = \langle \overline{\nabla}_X(\nabla
f), X\rangle_g = H_f(X,X).
$$
From this, we can derive the Hessian of $f$:
$$
H_f(X,Y) = \frac{1}{2}(H_f(X+Y,X+Y) - H_f(X,X) - H_f(Y,Y)).
$$

\item {\bf Inverse of Hessian :} For $X \in {\cal TM}$, and a
smooth function $f:{\cal M}\to \mathbb{R}$, $H_f^{-1}(X) \in {\cal
TM}$ is defined as the vector field satisfying: for
$\forall~\mathit{\Delta} \in {\cal TM}$,
$$
\ H_f(H_f^{-1}(X), \mathit{\Delta}) = \langle X,
\mathit{\Delta}\rangle_g.
$$
To understand the definition of $H_f^{-1}$, note that if $H : {\cal
TM} \times {\cal TM}\to \mathbb{R}$ is bi-linear and $\langle \cdot,
\cdot \rangle_g$ is an inner product on ${\cal TM}$ (i.e., a
Riemannian metric on ${\cal M}$), then by the \textit{Riesz
representation theorem}, $\exists ! ~A :{\cal TM}\to {\cal TM}$,
that is 1-1 and linear such that $H(X,Y) = \langle A(X),
Y\rangle_g$. This implies that $H(A^{-1}(X),Y) = \langle X,
Y\rangle_g$, so that $H_f^{-1} = A^{-1}$ when $H = H_f$.

\end{itemize}

\subsection*{Some facts about Stiefel manifold}\label{subsec:stiefel}

The manifold ${\cal M} = \{ B \in \mathbb{R}^{M\times r} : B^T B =
I_r\}$ is known as the \textit{Steifel manifold} in $\mathbb{R}^{M
\times r}$. Here we present some basic facts about this manifold
which are necessary for implementing the proposed method. A more
detailed description is given in Edelman \textit{et al.} (1999).
\begin{itemize}
\item
{\bf Tangent space :} ${\cal T}_B {\cal M} = \{ \mathit{\Delta} \in
 \mathbb{R}^{M\times r} : B^T \mathit{\Delta}$ is skew-symmetric $\}$.

\item {\bf Canonical metric :} For $\mathit{\Delta}_1,\mathit{\Delta}_2 \in {\cal
T}_B {\cal M}$ with $B \in {\cal M}$, the \textit{canonical metric}
(a Riemannian metric on ${\cal M}$) is defined as
$$
\langle \mathit{\Delta}_1, \mathit{\Delta}_2 \rangle_c =
Tr(\mathit{\Delta}_1^T (I - \frac{1}{2} BB^T) \mathit{\Delta}_2).
$$

\item {\bf Gradient :} For a smooth
function $f:{\cal M}\to \mathbb{R}$,
$$
\nabla f\left|_{B}\right. = f_B - B f_B^T B,
$$
where $f_B$ is the usual Euclidean gradient of $f$ defined through
$(f_B)_{ij} = \frac{\partial f}{\partial B_{ij}}$.

\item {\bf Hessian operator :} (derived from the geodesic
equation): For $\mathit{\Delta}_1,\mathit{\Delta}_2 \in {\cal T}_B
{\cal M}$, {\small
$$
H_f(\mathit{\Delta}_1,\mathit{\Delta}_2)\left|_{B}\right. =
f_{BB}(\mathit{\Delta}_1,\mathit{\Delta}_2) + \frac{1}{2}
Tr\left[(f_B^T \mathit{\Delta}_1 B^T + B^T \mathit{\Delta}_1
f_B^T)\mathit{\Delta}_2\right] - \frac{1}{2} Tr\left[ (B^T f_B +
f_B^T B) \mathit{\Delta}_1^T\mathit{\Pi} \mathit{\Delta}_2\right],
$$
} where $\mathit{\Pi} = I - BB^T$.

\item
{\bf Inverse of Hessian :} For $\mathit{\Delta}, G \in {\cal T}_B
{\cal M}$, the equation $\mathit{\Delta} = H_f^{-1}(G)$ means that
$\mathit{\Delta}$ is the solution of
$$
f_{BB}(\mathit{\Delta}) - B ~\mbox{skew}(f_B^T \mathit{\Delta}) -
~\mbox{skew}(\mathit{\Delta} f_B^T) B - \frac{1}{2} \mathit{\Pi}
\mathit{\Delta} B^T f_B = G,
$$
subject to the condition that $B^T \mathit{\Delta}$ is
skew-symmetric, i.e., $B^T \mathit{\Delta} + \mathit{\Delta}^T B =
0$, where $f_{BB}(\mathit{\Delta})\in {\cal T}_B {\cal M}$ such that
$$
\langle f_{BB}(\mathit{\Delta}),X\rangle_c =
f_{BB}(\mathit{\Delta},X) = Tr(\mathit{\Delta}^T f_{BB} X) \qquad
\forall~X \in {\cal T}_B {\cal M}.
$$
This implies that $f_{BB}(\mathit{\Delta}) = H(\mathit{\Delta}) - B
H^T(\mathit{\Delta}) B$, where $H(\mathit{\Delta}) = f_{BB}^T
\mathit{\Delta}$. Here skew$(X) = \frac{1}{2}(X-X^T)$.
\end{itemize}

\subsection*{Exponential of skew-symmetric matrices}\label{subsec:exp_skew}

Let $X = - X^T$ be a $p\times p$ matrix. Want to compute $\exp(tX)
:= \sum_{k=0}^\infty \frac{t^k}{k!} X^k$ for $t \in \mathbb{R}$,
where $X^0 = I$. Let the SVD of $X$ be given by $X = U D V^T$, where
$U^T U = V^T V = I_p$, and $D$ is diagonal. So, $X^2 = X X = - XX^T
= - UDV^T V DU^T = - UD^2 U^T$. This also shows that all the
eigenvalues of $X$ are purely imaginary. Using the facts that $D^0 =
I_p$; $ X^{2k} = (X^2)^k = (-1)^k (U D^2 U^T)^k = (-1)^k U D^{2k}
U^T$; and $X^{2k+1} = (-1)^k U D^{2k} U^T U D V^T = (-1)^k U
D^{2k+1} V^T$, we have
\begin{eqnarray*}
\exp(tX) &=& U \left[\sum_{k=0}^\infty \frac{(-t)^k}{(2k)!}
D^{2k}\right]U^T
+ U \left[\sum_{k=0}^\infty \frac{(-t)^k}{(2k+1)!}  D^{2k+1} \right]V^T \\
&=& U \cos(tD) U^T + U \sin(tD) V^T,
\end{eqnarray*}
where $\cos(tD) = diag((\cos(td_{jj}))_{j=1}^p)$ and $\sin(tD) =
diag((\sin(td_{jj}))_{j=1}^p)$, if $D = ~diag((d_{jj})_{j=1}^p)$.

\subsection*{Vectorization of matrix equations}\label{subsec:vectorization}

A general form of the equation in the $M\times r$ matrix
$\mathit{\Delta}$ is given by
$$
L = A\mathit{\Delta} + \mathit{\Delta} K + C\mathit{\Delta} D + E
\mathit{\Delta}^T F,
$$
where $L$ is $M \times r$, $A$ is $M\times M$, $K$ is $r\times r$,
$C$ is $M \times M$, $D$ is $r \times r$, $E$ is $M \times r$, and
$F$ is $M \times r$. Vectorization of this equation using the vec
operation means that vec$(L)$ is given by
\begin{eqnarray}\label{eq:vec_general}
&& \mbox{vec}(A\mathit{\Delta}) + \mbox{vec}(\mathit{\Delta} K) +
\mbox{vec}(C\mathit{\Delta} D) + \mbox{vec}(E\mathit{\Delta}^T
F)\nonumber\\
&=& \left[(I_r \otimes A) + (K^T \otimes I_M) + (D^T \otimes C) +
(F^T \otimes E) P_{M,r}\right]
\mbox{vec}(\mathit{\Delta}),\nonumber\\
&&
\end{eqnarray}
where, $\otimes$ denotes the Kronecker product, and we have used the
following properties of the vec operator (Muirhead (1982)): (i)
$\mbox{vec}(K X C) = (C^T \otimes K) \mbox{vec}(X)$; (ii)
$\mbox{vec}(X^T) = P_{m,n} \mbox{vec}(X)$. Here $X$ is $m\times n$,
$K$ is $r \times m$, $C$ is $n \times s$, and $P_{m,n}$ is an
appropriate $mn \times mn$ permutation matrix.

\section*{Appendix B : Gradients and Hessians with respect to $B$}


We use the following lemmas (cf. Muirhead (1982)) repeatedly in our
computations in this subsection.

\noindent{\bf Lemma 1 :} \textit{Let $P = I_p + AE$ where $A$ is
$p\times q$, $E$ is $q\times p$. Then
$$
\det(P) = |I_p+AE| = |I_q + EA|.
$$
}

\vskip.1in\noindent{\bf Lemma 2 :} \textit{Let $A$ be $p\times p$
and $E$ be $q\times q$, both nonsingular. If $P = A + CED$, for any
$p\times q$ matrix $C$ and any $q\times p$ matrix $D$, then
$$
P^{-1} = (A + CED)^{-1} = A^{-1}[A - CQ^{-1}D]A^{-1},
~~\mbox{where},~~ Q = E^{-1} + D A^{-1}C
$$
is $q \times q$.}

\subsection*{Application to the likelihood setting}

Let $P_i = \sigma^2 I_{m_i} + \mathit{\Phi}_i^T B \mathit{\Lambda}
B^T \mathit{\Phi}_i$ (is an $m_i\times m_i$ matrix), where
$\mathit{\Phi}_i$ is $M \times m_i$, $B$ is $M\times r$ and
$\mathit{\Lambda}$ is $r\times r$ matrices. Then, by \textit{Lemma
1},
\begin{equation}\label{eq:det_P_i}
|P_i| = \sigma^{2m_i} |I_r + \sigma^{-2} \mathit{\Lambda} B^T
\mathit{\Phi}_i \mathit{\Phi}_i^T B| = \sigma^{2(m_i-r)}
|\mathit{\Lambda}| |\sigma^2\mathit{\Lambda}^{-1} + B^T
\mathit{\Phi}_i \mathit{\Phi}_i^T B| = \sigma^{2(m_i-r)}
|\mathit{\Lambda}| |Q_i|,
\end{equation}
where
$$
Q_i = \sigma^2 \mathit{\Lambda}^{-1} + B^T \mathit{\Phi}_i
\mathit{\Phi}_i^T B
$$
is an $r\times r$ positive definite matrix.  Also, by \textit{Lemma
2}
\begin{equation}\label{eq:inv_P_i}
P_i^{-1} = \sigma^{-2} I_{m_i} - \sigma^{-4} \mathit{\Phi}_i^T B
(\mathit{\Lambda}^{-1} + \sigma^{-2} B^T \mathit{\Phi}_i
\mathit{\Phi}_i^T B)^{-1} B^T \mathit{\Phi}_i =
\sigma^{-2}\left[I_{m_i} - \mathit{\Phi}_i^T B Q_i^{-1} B^T
\mathit{\Phi}_i\right].
\end{equation}
In our problem, we consider the $i$-th term in the expression for
the log-likelihood and recall that $\widetilde{\mathbf{Y}}_i  =
\mathbf{Y}_i - \boldsymbol{\mu}_i$. Then
\begin{eqnarray}\label{eq:F_12_alt}
F_i^1 &=& Tr[(\sigma^2 I_{m_i} + \mathit{\Phi}_i^T B
\mathit{\Lambda} B^T \mathit{\Phi}_i)^{-1} \widetilde{\mathbf{Y}}_i
\widetilde{\mathbf{Y}}_i^T] =
Tr(P_i^{-1}\widetilde{\mathbf{Y}}_i \widetilde{\mathbf{Y}}_i^T) \nonumber\\
F_i^2 &=& \log |\sigma^2 I_{m_i} + \mathit{\Phi}_i^T B
\mathit{\Lambda} B^T \mathit{\Phi}_i| = \log |P_i|.
\end{eqnarray}
For simplifying notations we shall drop the subscript $i$ from
these functions. We view $F^1 = F^1(B)$ as a function of $B$.
Since $F^1 = Tr(P_i^{-1}
\widetilde{\mathbf{Y}}_i\widetilde{\mathbf{Y}}_i^T)$, using
(\ref{eq:inv_P_i}) we have
\begin{eqnarray}\label{eq:F_1_alt}
F^1(B) = Tr(P_i^{-1} \widetilde Y_i\widetilde{\mathbf{Y}}_i^T) &=&
\sigma^{-2} Tr(\widetilde{\mathbf{Y}}_i\widetilde{\mathbf{Y}}_i^T) -
\sigma^{-2} Tr(\mathit{\Phi}_i^T B Q_i^{-1} B^T
\mathit{\Phi}_i \widetilde{\mathbf{Y}}_i \widetilde{\mathbf{Y}}_i^T)\nonumber\\
&=& \sigma^{-2}
Tr(\widetilde{\mathbf{Y}}_i\widetilde{\mathbf{Y}}_i^T) - \sigma^{-2}
Tr(BQ_i^{-1} B^T \mathit{\Phi}_i \widetilde{\mathbf{Y}}_i
\widetilde{\mathbf{Y}}_i^T \mathit{\Phi}_i^T).
\end{eqnarray}
Similarly,
\begin{equation}
F^2 = F^2(B) = \log |P_i| = \log (\sigma^{2(m_i-r)}
|\mathit{\Lambda}|) + \log |Q_i|.
\end{equation}

\subsection*{Gradient of $F^1$}

Let $B(t) = B + t\mathit{\Delta}$. Then
$$
\frac{dQ_i(t)}{dt}\left|_{t=0}\right. = \mathit{\Delta}^T
\mathit{\Phi}_i \mathit{\Phi}_i^T B + B^T \mathit{\Phi}_i
\mathit{\Phi}_i^T \mathit{\Delta},
$$
so that
\begin{eqnarray}\label{eq:F_1_grad_Delta_alt}
\langle F_B^1,\mathit{\Delta}\rangle &=&
\frac{dF^1(B(t))}{dt}\left|_{t=0}\right.\nonumber\\
&=& -\sigma^{-2} Tr\left[(\mathit{\Delta} Q_i^{-1} B^T + B Q_i^{-1}
\mathit{\Delta}^T - BQ_i^{-1} \frac{dQ_i}{dt}\left|_{t=0}\right.
Q_i^{-1} B^T)
\mathit{\Phi}_i \widetilde{\mathbf{Y}}_i \widetilde{\mathbf{Y}}_i^T \mathit{\Phi}_i^T\right]\nonumber\\
&=& - 2\sigma^{-2} Tr\left[(\mathit{\Phi}_i \widetilde{\mathbf{Y}}_i
\widetilde{\mathbf{Y}}_i^T \mathit{\Phi}_i^T B Q_i^{-1} -
\mathit{\Phi}_i \mathit{\Phi}_i^T B Q_i^{-1} B^T \mathit{\Phi}_i
\widetilde{\mathbf{Y}}_i \widetilde{\mathbf{Y}}_i^T
\mathit{\Phi}_i^T B Q_i^{-1} ) \mathit{\Delta}^T \right].
\end{eqnarray}
Thus the Euclidean gradient of $F^1$ is
\begin{eqnarray*}
F_B^1 &=& - 2\sigma^{-2} \left[\mathit{\Phi}_i
\widetilde{\mathbf{Y}}_i \widetilde{\mathbf{Y}}_i^T
\mathit{\Phi}_i^T B Q_i^{-1} - \mathit{\Phi}_i \mathit{\Phi}_i^T B
Q_i^{-1} B^T \mathit{\Phi}_i \widetilde{\mathbf{Y}}_i
\widetilde{\mathbf{Y}}_i^T \mathit{\Phi}_i^T B Q_i^{-1} \right]
\nonumber\\
&=& 2\sigma^{-2} \left[\mathit{\Phi}_i \mathit{\Phi}_i^T B
Q_i^{-1} B^T - I_M\right] \mathit{\Phi}_i \widetilde{\mathbf{Y}}_i
\widetilde{\mathbf{Y}}_i^T \mathit{\Phi}_i^T B Q_i^{-1},\nonumber
\end{eqnarray*}
and the gradient of $F^1$ with respect to $B$ is
\begin{equation}\label{eq:F_1_grad_alt}
\nabla F^1 = F_B^1 - B (F_B^1)^T B.
\end{equation}

\subsection*{Hessian of $F^1$}

Let $B(t,s) = B + t\mathit{\Delta} + s X$. Then using
(\ref{eq:F_1_grad_Delta_alt}),
\begin{eqnarray*}
F_{BB}^1(\mathit{\Delta},X)&=&\langle F^1_{BB}(\Delta),X
\rangle_c\nonumber \\
 &=&\langle H^1_{BB}(\Delta),X \rangle = \frac{\partial}{\partial t}
\frac{\partial}
{\partial s} F^1(B(t,s)) \left|_{s,t=0}\right.\nonumber\\
&=& 2\sigma^{-2} Tr\left[\frac{\partial}{\partial t}
(\mathit{\Phi}_i \mathit{\Phi}_i^T B(t,0) Q_i(t)^{-1} B(t,0)^T -
I_M)\left|_{t=0}\right. \mathit{\Phi}_i \widetilde{\mathbf{Y}}_i
\widetilde{\mathbf{Y}}_i^T
\mathit{\Phi}_i^T B Q_i^{-1} X^T\right] \nonumber\\
&& ~+ 2\sigma^{-2} Tr\left[(\mathit{\Phi}_i \mathit{\Phi}_i^T
BQ_i^{-1} B^T - I_M) \mathit{\Phi}_i \widetilde{\mathbf{Y}}_i
\widetilde{\mathbf{Y}}_i^T \mathit{\Phi}_i^T
\frac{\partial}{\partial t}(B(t,0) Q_i(t)^{-1}) \left|_{t=0}\right.
X^T\right] .
\end{eqnarray*}
Note that
$$
\frac{\partial}{\partial t}(B(t,0) Q_i(t)^{-1})\left|_{t=0}\right. =
\mathit{\Delta} Q_i^{-1} - B Q_i^{-1} (\mathit{\Delta}^T
\mathit{\Phi}_i \mathit{\Phi}_i^T B + B^T \mathit{\Phi}_i
\mathit{\Phi}_i^T \mathit{\Delta}) Q_i^{-1},
$$
and
$$
\frac{\partial}{\partial t}(B(t,0)
Q_i(t)^{-1}B(t,0)^T)\left|_{t=0}\right. = \mathit{\Delta}
Q_i^{-1}B^T + B Q_i^{-1} \mathit{\Delta}^T - B Q_i^{-1}
(\mathit{\Delta}^T \mathit{\Phi}_i \mathit{\Phi}_i^T B + B^T
\mathit{\Phi}_i \mathit{\Phi}_i^T \mathit{\Delta}) Q_i^{-1} B^T.
$$
Thus $H_{BB}^1 (\Delta)$ is given by,
\begin{eqnarray*}
&&H_{BB}^1(\Delta) \\
&=& 2\sigma^{-2}
\mathit{\Phi}_i\mathit{\Phi}_i^T \left[\mathit{\Delta} Q_i^{-1}
B^T + B Q_i^{-1} \mathit{\Delta}^T - B Q_i^{-1} (\mathit{\Delta}^T
\mathit{\Phi}_i \mathit{\Phi}_i^T B + B^T \mathit{\Phi}_i
\mathit{\Phi}_i^T \mathit{\Delta})Q_i^{-1} B^T \right]
\mathit{\Phi}_i \widetilde{\mathbf{Y}}_i
\widetilde{\mathbf{Y}}_i^T \mathit{\Phi}_i^T
B  Q_i^{-1}\nonumber\\
&& ~+ 2\sigma^{-2} \left[(\mathit{\Phi}_i \mathit{\Phi}_i^T B
Q_i^{-1} B^T - I_M) \mathit{\Phi}_i \widetilde{\mathbf{Y}}_i
\widetilde{\mathbf{Y}}_i^T \mathit{\Phi}_i^T (\mathit{\Delta}
Q_i^{-1} - B Q_i^{-1} (\mathit{\Delta}^T \mathit{\Phi}_i
\mathit{\Phi}_i^T B + B^T \mathit{\Phi}_i \mathit{\Phi}_i^T
\mathit{\Delta}) Q_i^{-1})\right],
\end{eqnarray*}
and
\begin{equation}\label{eq:F_1_hessian_alt}
F_{BB}^1(\mathit{\Delta}) = H_{BB}^1(\Delta) - B (H_{BB}^1
(\Delta))^T B.
\end{equation}

\subsection*{Gradient of $F^2$}

Let $B(t) = B + t\mathit{\Delta}$. Then
\begin{eqnarray}\label{eq:F_2_grad_Delta_alt}
\langle F_B^2,\mathit{\Delta}\rangle =
\frac{dF^2(B(t))}{dt}\left|_{t=0}\right.
&=& Tr\left(Q_i^{-1} \frac{dQ_i}{dt}\left|_{t=0}\right.\right)\nonumber\\
&=& Tr(Q_i^{-1}(\mathit{\Delta}^T \mathit{\Phi}_i \mathit{\Phi}_i^T
B + B^T \mathit{\Phi}_i \mathit{\Phi}_i^T \mathit{\Delta}))
\nonumber\\
&=& 2 Tr(Q_i^{-1} B^T \mathit{\Phi}_i \mathit{\Phi}_i^T
\mathit{\Delta}).
\end{eqnarray}
Thus
\begin{equation}\label{eq:F_2_grad_alt}
\nabla F^2 = F_B^2 - B (F_B^2)^T B, ~~~\mbox{where}~~F_B^2 =
2\mathit{\Phi}_i\mathit{\Phi}_i^T B Q_i^{-1}.
\end{equation}

\subsection*{Hessian of $F^2$}

Let $B(t,s) = B + t\mathit{\Delta} + sX$. Then using
(\ref{eq:F_2_grad_Delta_alt}),
\begin{eqnarray*}
F_{BB}^2(\mathit{\Delta},X) &=& \langle
F_{BB}^2(\mathit{\Delta}),X\rangle_c \nonumber \\
&=& \langle H_{BB}^2(\mathit{\Delta}),X\rangle=
\frac{\partial}{\partial t} \frac{\partial}
{\partial s} F^2(B(t,s)) \left|_{s,t=0}\right. \nonumber\\
&=& \frac{\partial}{\partial t}  [2Tr(Q_i(t)^{-1} B(t,0)^T
\mathit{\Phi}_i \mathit{\Phi}_i^T X)]\left|_{t=0}\right. \nonumber\\
&=& 2 Tr\left[(-Q_i^{-1} \frac{dQ_i(t)}{dt}\left|_{t=0}\right.
Q_i^{-1} B^T + Q_i^{-1} \mathit{\Delta}^T) \mathit{\Phi}_i \mathit{\Phi}_i^T X\right]\nonumber\\
&=& 2 Tr\left[(-Q_i^{-1}(\mathit{\Delta}^T
\mathit{\Phi}_i\mathit{\Phi}_i^T B + B^T \mathit{\Phi}_i
\mathit{\Phi}_i^T
\mathit{\Delta})Q_i^{-1} B^T + Q_i^{-1} \mathit{\Delta}^T)\mathit{\Phi}_i\mathit{\Phi}_i^T X\right]. \nonumber\\
\end{eqnarray*}
From this  $H_{BB}^2(\Delta)$ is
\begin{eqnarray}
H_{BB}^2(\Delta) &=& 2\left[-Q_i^{-1} (\mathit{\Delta}^T
\mathit{\Phi}_i\mathit{\Phi}_i^T B + B^T \mathit{\Phi}_i
\mathit{\Phi}_i^T
\mathit{\Delta})Q_i^{-1} B^T + Q_i^{-1} \mathit{\Delta}^T)\mathit{\Phi}_i\mathit{\Phi}_i^T\right]^T\nonumber\\
&=& 2\mathit{\Phi}_i \mathit{\Phi}_i^T \left[\mathit{\Delta} - B
Q_i^{-1}(\mathit{\Delta}^T \mathit{\Phi}_i\mathit{\Phi}_i^T B +
B^T \mathit{\Phi}_i \mathit{\Phi}_i^T \mathit{\Delta})\right]
Q_i^{-1}, \nonumber
\end{eqnarray}
and
\begin{equation}\label{eq:F_2_hessian_alt}
F_{BB}^2(\mathit{\Delta}) = H_{BB}^2(\Delta) -
B(H_{BB}^2(\Delta))^T B,
\end{equation}

\section*{Appendix C : Derivation of $\widetilde{CV}$ (\ref{eq:cv_MLE_approx_final})}\label{sec:cv_derivation}

For now, in (\ref{eq:Psi_hat_i_eqn}), considering only the part
corresponding to the gradient w.r.t. $B$ and expanding it around
$\widehat\Psi$, while approximating
$(\widehat\tau^{(-i)},\widehat{\boldsymbol{\zeta}}^{(-i)})$ by
$(\widehat\tau,\widehat{\boldsymbol{\zeta}})$, we have (for
notational simplicity, write $\ell_j(\widehat{B})$ to denote
$\ell_j(\widehat{\Psi})$)
\begin{equation}\label{eq:score_eqn_expansion}
0 = \sum_{j\neq i} \nabla_B\ell_j(\widehat\Psi^{(-i)}) \approx
\sum_{j\neq i} \nabla_B\ell_j(\widehat B) + \sum_{j\neq i}
\overline{\nabla}_{\mathit{\Delta}_i} (\nabla_B\ell_j(\widehat
B)),
\end{equation}
where $\overline{\nabla}_{\mathit{\Delta}_i} (\nabla_B \ell_j)$ is
the \textit{covariant derivative} of $\nabla_B \ell_j$ in the
direction of $\mathit{\Delta}_i$. Now, substituting
(\ref{eq:Psi_hat_eqn}) in (\ref{eq:score_eqn_expansion}), we get
\begin{equation}\label{eq:score_eqn_expansion2}
0 \approx -\nabla_B\ell_i(\widehat B) +
\overline{\nabla}_{\mathit{\Delta}_i}[\sum_{j \neq i} \nabla_B
\ell_j(\widehat B)].
\end{equation}
Then for any $X \in {\cal T}_{\widehat B}{\cal M}$,
$$
\langle \overline{\nabla}_{\mathit{\Delta}_i} (\sum_{j \neq i}
\nabla_B \ell_j (\widehat B)), X\rangle_c =[\sum_{j\neq
i}\nabla_B^2 \ell_j(\widehat B) ](\mathit{\Delta}_i,X) \approx
\langle \nabla_B \ell_i(\widehat B), X\rangle_c.
$$
Thus by the definition of the Hessian inverse operator,
$$
\mathit{\Delta}_i \approx [\sum_{j\neq i}\nabla_B^2
\ell_j(\widehat B) ]^{-1} (\nabla_B \ell_i(\widehat B)).
$$
This, together with (\ref{eq:score_eqn_expansion2}), leads to the
approximation of $\mathit{\Delta}_i$,
\begin{eqnarray}\label{eq:Psi_hat_i_expand}
\mathit{\Delta}_i &\approx& [\sum_{j \neq i} \nabla_B^2
\ell_j(\widehat B)]^{-1} \nabla_B\ell_i(\widehat B)= [\sum_j
\nabla_B^2\ell_j(\widehat B)- \nabla_B^2 \ell_i(\widehat B)]^{-1}
(\nabla_B\ell_i(\widehat B))\nonumber\\
&\approx&  \left[ I + [\sum_j \nabla_B^2\ell_j(\widehat B)]^{-1}
\nabla_B^2 \ell_i(\widehat B)\right]
[\sum_j\nabla_B^2\ell_j(\widehat B)]^{-1} (\nabla_B\ell_i(\widehat
B))\nonumber\\
&=& \left[ I + [\mathbf{H}_B(\widehat B)]^{-1}\nabla_B^2
\ell_i(\widehat B)\right] [\mathbf{H}_B(\widehat B)]^{-1}
(\nabla_B\ell_i(\widehat B)),
\end{eqnarray}
where $\mathbf{H}_B=\sum_j \nabla_B^2\ell_j$. Note that the last
approximation is because, for linear operators $A$ and $C$ such
that $A$ is invertible and $\parallel A^{-1} C\parallel$ is small,
we have
$$
(A-C)^{-1} = (A(I - A^{-1} C))^{-1} = (I - A^{-1} C)^{-1} A^{-1}
\approx (I+A^{-1}C)A^{-1}.
$$
The interpretation of the linear operator  $[\mathbf{H}_B(\widehat
B)]^{-1} \nabla_B^2 \ell_i(\widehat B)$ in
(\ref{eq:Psi_hat_i_expand}) is as follows : for $\gamma \in {\cal
T}_{\widehat B} {\cal M}$,
$$
\left[ [\mathbf{H}_B(\widehat B)]^{-1} \nabla_B^2 (\ell_i(\widehat
B))\right](\gamma) =[\mathbf{H}_B(\widehat B)]^{-1}(\nabla_B^2
(\ell_i(\widehat B))(\gamma)).
$$
Now, for $X \in {\cal T}_{\widehat B}{\cal M}$, by definition of
Hessian,
\begin{eqnarray}\label{eq:covariant_Delta_i_ell_sum}
&& \langle \sum_{j=1}^n \overline{\nabla}_{\mathit{\Delta}_i}
(\nabla_B \ell_j(\widehat B)), X\rangle_c = [\nabla_B^2
(\sum_{j=1}^n \ell_j(\widehat B))]
(\mathit{\Delta}_i,X)\nonumber\\
&\approx& \mathbf{H}_B(\widehat B)\left( [\mathbf{H}_B(\widehat
B)]^{-1} (\nabla_B \ell_i(\widehat B)) + [(\mathbf{H}_B(\widehat
B)]^{-1} \nabla_B^2 \ell_i(\widehat B) [\mathbf{H}_B(\widehat
B)]^{-1} (\nabla_B \ell_i(\widehat B)), X \right)\nonumber\\
&=& \langle \nabla_B \ell_i(\widehat B), X\rangle_c + \langle
\nabla_B^2(\ell_i(\widehat B))[\mathbf{H}_B(\widehat B)]^{-1}
(\nabla_B \ell_i(\widehat B)), X\rangle_c,
\end{eqnarray}
where, by definition,
$$
\nabla_B^2 (\ell_i(\widehat B))(\gamma) = \overline{\nabla}_\gamma
(\nabla_B \ell_i(\widehat B))
$$
for $\gamma \in {\cal T}_{\widehat B} {\cal M}$. In the first
approximation of (\ref{eq:covariant_Delta_i_ell_sum}), we have
used the approximation (\ref{eq:Psi_hat_i_expand}), and the last
step follows from the definition of Hessian inverse and linearity
of the Hessian. From (\ref{eq:score_eqn_expansion}) we also have,
\begin{equation}\label{eq:covariant_Delta_i_ell_i}
\overline{\nabla}_{\mathit{\Delta}_i} (\nabla_B \ell_i(\widehat
B)) \approx - \nabla_B \ell_i(\widehat B) + \sum_{j=1}^n
\overline{\nabla}_{\mathit{\Delta}_i} (\nabla_B \ell_j (\widehat
B)).
\end{equation}
Substituting (\ref{eq:covariant_Delta_i_ell_sum}) in
(\ref{eq:covariant_Delta_i_ell_i}), we then have the approximation
\begin{eqnarray}\label{eq:Hess_ell_i_delta_approx}
\nabla_B^2 \ell_i(\widehat B)(\mathit{\Delta}_i) =
\overline{\nabla}_{\mathit{\Delta}_i} (\nabla_B \ell_i(\widehat
B))
&\approx& \nabla_B^2 \ell_i(\widehat B) [\mathbf{H}_B(\widehat
B)]^{-1} (\nabla_B\ell_i(\widehat B)).
\end{eqnarray}

Using (\ref{eq:Psi_hat_i_expand}) and
(\ref{eq:Hess_ell_i_delta_approx}),  and ignoring terms higher
than the second order, we have the approximation {\small
\begin{eqnarray}\label{eq:cv_MLE_approx_second}
&&  \sum_{i=1}^n \langle \nabla_B \ell_i(\widehat B),
\mathit{\Delta}_i\rangle_c + \frac{1}{2}\sum_{i=1}^n \nabla_B^2
\ell_i(\widehat B)(\mathit{\Delta}_i,\mathit{\Delta}_i)\nonumber\\
&\approx& \left[\sum_{i=1}^n \langle \nabla_B\ell_i(\widehat B),
[\mathbf{H}_B(\widehat B)]^{-1} (\nabla_B\ell_i(\widehat
B))\rangle_c + \sum_{i=1}^n \langle \nabla_B\ell_i(\widehat
B),[\mathbf{H}_B(\widehat B)]^{-1} \nabla_B^2 \ell_i(\widehat B)
[\mathbf{H}_B(\widehat B)]^{-1} (\nabla_B\ell_i(\widehat
B))\rangle_c \right]
\nonumber\\
&& + \frac{1}{2} \sum_{i=1}^n \langle [\mathbf{H}_B(\widehat
B)]^{-1} (\nabla_B\ell_i(\widehat B)), \nabla_B^2 \ell_i(\widehat
B) [\mathbf{H}_B(\widehat B)]^{-1} (\nabla_B\ell_i(\widehat
B))\rangle_c
\nonumber\\
&=&  \sum_{i=1}^n \langle \nabla_B\ell_i(\widehat B),
[\mathbf{H}_B(\widehat B)]^{-1} \nabla_B\ell_i(\widehat
B)\rangle_c+ \frac{3}{2} \sum_{i=1}^n \nabla_B^2 \ell_i(\widehat
B) ( [\mathbf{H}_B(\widehat B)]^{-1} \nabla_B\ell_i(\widehat B),
[\mathbf{H}_B(\widehat B)]^{-1} \nabla_B\ell_i(\widehat B)).
\end{eqnarray}
} Here  we give brief justifications  for the steps in
(\ref{eq:cv_MLE_approx_second}). The first approximation follows
from the definition of Hessian, and the approximation of
$\mathit{\Delta}_i$ by (\ref{eq:Psi_hat_i_expand}). The last
equation follows from: (i) by definition of Hessian, applied to
$\nabla_B^2 \ell_i(\widehat B)$, the term on the third line equals
$$
\frac{1}{2}\sum_{i=1}^n \nabla_B^2 \ell_i(\widehat B) (
[\mathbf{H}_B(\widehat B)]^{-1} (\nabla_B\ell_i(\widehat B)),
[\mathbf{H}_B(\widehat B)]^{-1} (\nabla_B\ell_i(\widehat B)));
$$
and (ii) the second term on the second line equals the same term
as above, except for the factor $\frac{1}{2}$, by definition of
{Hessian}$^{-1}$, now applied to $[\mathbf{H}_B(\widehat
B)]^{-1}$:
\begin{eqnarray*}
&&\langle \nabla_B\ell_i(\widehat B),[\mathbf{H}_B(\widehat
B)]^{-1} \nabla_B^2 \ell_i(\widehat B) [\mathbf{H}_B(\widehat
B)]^{-1} (\nabla_B\ell_i(\widehat B))\rangle_c\\
&=&\mathbf{H}_B(\widehat B)([\mathbf{H}_B(\widehat B)]^{-1}
(\nabla_B\ell_i(\widehat B)),[\mathbf{H}_B(\widehat B)]^{-1}
\nabla_B^2 \ell_i(\widehat B) [\mathbf{H}_B(\widehat B)]^{-1}
(\nabla_B\ell_i(\widehat B)))\\ &=&\langle [\mathbf{H}_B(\widehat
B)]^{-1}(\nabla_B\ell_i(\widehat B)), \nabla_B^2 \ell_i(\widehat
B) [\mathbf{H}_B(\widehat B)]^{-1} (\nabla_B\ell_i(\widehat
B))\rangle_c.
\end{eqnarray*}

Using very similar (but conceptually much simpler) arguments, we
also have the second order approximation
{\small
\begin{eqnarray}\label{eq:cv_MLE_approx_third}
&&  \sum_{i=1}^n \langle \nabla_{(\tau,\zeta)}
\ell_i(\widehat\tau,\widehat{\boldsymbol{\zeta}}),
(\delta_\tau^i,\boldsymbol{\delta}_\zeta^i)^T\rangle + \frac{1}{2}
\sum_{i=1}^n \langle [\nabla_{(\tau,\zeta)}^2
\ell_i(\widehat\tau,\widehat{\boldsymbol{\zeta}})](\delta_\tau^i,\boldsymbol{\delta}_\zeta^i)^T
,(\delta_\tau^i,\boldsymbol{\delta}_\zeta^i)^T\rangle \nonumber\\
&\approx& \sum_{i=1}^n \langle
\nabla_{(\tau,\zeta)}\ell_i(\widehat\tau,\widehat{\boldsymbol{\zeta}}),
[\mathbf{H}_{(\tau,\zeta)}(\widehat\tau,\widehat{\boldsymbol{\zeta}})]^{-1}
\nabla_{(\tau,\zeta)}\ell_i(\widehat\tau,\widehat{\boldsymbol{\zeta}})\rangle\nonumber\\
&+& \frac{3}{2} \sum_{i=1}^n \langle \nabla_{(\tau,\zeta)}^2
\ell_i(\widehat\tau,\widehat{\boldsymbol{\zeta}})
[\mathbf{H}_{(\tau,\zeta)}(\widehat\tau,\widehat{\boldsymbol{\zeta}})]^{-1}
\nabla_{(\tau,\zeta)}\ell_i(\widehat\tau,\widehat{\boldsymbol{\zeta}}),
[\mathbf{H}_{(\tau,\zeta)}(\widehat\tau,\widehat{\boldsymbol{\zeta}})]^{-1}
\nabla_{(\tau,\zeta)}\ell_i(\widehat\tau,\widehat{\boldsymbol{\zeta}})\rangle.
\end{eqnarray}
} Combining (\ref{eq:cv_MLE_approx_first}),
(\ref{eq:cv_MLE_approx_second}) and
(\ref{eq:cv_MLE_approx_third}), we have the approximate CV score
given by (\ref{eq:cv_MLE_approx_final}).


\section*{Appendix D : Gradients and Hessians
with respect to $\boldsymbol{\zeta}$ and $\tau$}

Define $H_i = \mathit{\Phi}_i^T B$, $i=1,\ldots,n$. In the
following we shall use $\exp(\boldsymbol{\zeta})$ and $\exp(\tau)
$to denote the $r\times r$ diagonal matrix $\mathit{\Lambda}$ and
$\sigma^2$, respectively. Then, as defined in Appendix A,
$$
P_i = \sigma^2 I_{m_i} + \mathit{\Phi}_i^T B \mathit{\Lambda} B^T
\mathit{\Phi}_i = \sigma^2 I_{m_i} + H_i\mathit{\Lambda} H_i^T =
e^\tau I_{m_i} + H_i \exp(\boldsymbol{\zeta}) H_i^T,
$$
$$
Q_i = \sigma^2 \mathit{\Lambda}^{-1} + B^T \mathit{\Phi}_i
\mathit{\Phi}_i^T B = \sigma^2 \mathit{\Lambda}^{-1} + H_i^T H_i =
e^\tau \exp(-\boldsymbol{\zeta}) + H_i^T H_i.
$$
and re-writing (\ref{eq:inv_P_i}):
$$
P_i^{-1} = \sigma^{-2} I_{m_i} - \sigma^{-4}
H_i(\mathit{\Lambda}^{-1} +\sigma^{-2} H_i^T H_i)^{-1} H_i^T =
e^{-\tau} \left[I_{m_i} - H_i Q_i^{-1} H_i^T\right].
$$
For future uses, we calculate the following derivatives.
\begin{equation}\label{eq:del_tau_P_i}
\frac{\partial P_i}{\partial \tau} = \frac{\partial}{\partial \tau}
[e^\tau I_{m_i} + H_i\exp(\boldsymbol{\zeta}) H_i^T] = e^\tau
I_{m_i}.
\end{equation}
Let $H_{ik} = \mathit{\Phi}_i^T B_k$, $k=1,\ldots,r$ where $B_k$ is
the $k$-th column of $B$. We shall use the following fact
\begin{equation}\label{eq:del_zeta_P_i}
\frac{\partial P_i}{\partial \zeta_k} = \frac{\partial}{\partial
\zeta_k} [e^\tau I_{m_i} + \sum_{k=1}^r e^{\zeta_k} H_{ik}H_{ik}^T]
= e^{\zeta_k} H_{ik} H_{ik}^T.
\end{equation}
In the following, we shall drop the subscript $i$ from the functions
$F_i^1$ and $F_i^2$, and treat the latter as functions of
$(\tau,\zeta)$.

\subsection*{Gradient of $F^1$ and $F^2$}

By direct computations we have,
\begin{eqnarray}\label{eq:grad_tau_F_1_tilde}
\frac{\partial F^1}{\partial\tau} &=&
\frac{\partial}{\partial\tau} Tr[P_i^{-1} \widetilde{\mathbf{Y}}_i
\widetilde{\mathbf{Y}}_i^T] = - Tr\left[P_i^{-1}
\left(\frac{\partial P_i}{\partial \tau}\right) P_i^{-1}
\widetilde{\mathbf{Y}}_i \widetilde{\mathbf{Y}}_i^T\right]\nonumber\\
&=& - e^\tau Tr[P_i^{-2}\widetilde{\mathbf{Y}}_i
\widetilde{\mathbf{Y}}_i^T] = -e^\tau \widetilde{\mathbf{Y}}_i^T
P_i^{-2} \widetilde{\mathbf{Y}}_i, ~~(\mbox{by}~
(\ref{eq:del_tau_P_i})),
\end{eqnarray}
and
\begin{eqnarray}\label{eq:grad_zeta_F_1_tilde}
\frac{\partial F^1}{\partial\zeta_k} &=&
\frac{\partial}{\partial\zeta_k} Tr[P_i^{-1}
\widetilde{\mathbf{Y}}_i \widetilde{\mathbf{Y}}_i^T] = -
Tr\left[P_i^{-1} \left(\frac{\partial P_i}{\partial \zeta_k}\right)
P_i^{-1}\widetilde{\mathbf{Y}}_i \widetilde{\mathbf{Y}}_i^T\right]\nonumber\\
&=& - e^{\zeta_k} Tr[P_i^{-1}H_{ik}H_{ik}^T P_i^{-1}
\widetilde{\mathbf{Y}}_i \widetilde{\mathbf{Y}}_i^T] = -e^{\zeta_k}
(H_{ik}^T P_i^{-1} \widetilde{\mathbf{Y}}_i)^2,
~~(\mbox{by}~(\ref{eq:del_zeta_P_i})).
\end{eqnarray}
Also,
\begin{eqnarray}
\frac{\partial F^2}{\partial\tau} =
\frac{\partial}{\partial\tau}\log |P_i| &=& Tr\left[P_i^{-1}
\left(\frac{\partial P_i}{\partial \tau}\right)\right] = e^\tau
Tr(P_i^{-1}), ~~(\mbox{by}~(\ref{eq:del_tau_P_i})),
\label{eq:grad_tau_F_2_tilde}\\
\frac{\partial F^2}{\partial\zeta_k}
=\frac{\partial}{\partial\zeta_k} \log |P_i| &=& Tr\left[P_i^{-1}
\left(\frac{\partial P_i}{\partial \zeta_k}\right)\right] =
e^{\zeta_k} H_{ik}^T P_i^{-1} H_{ik},
~~(\mbox{by}~(\ref{eq:del_zeta_P_i})).
\label{eq:grad_zeta_F_2_tilde}
\end{eqnarray}

\subsection*{Hessian of $F^1$}

From (\ref{eq:grad_tau_F_1_tilde}),
\begin{eqnarray}\label{eq:hess_tau_tau_F_1_tilde}
\frac{\partial^2 F^1}{\partial \tau^2} &=& \frac{\partial}{\partial
\tau} \left[-e^\tau \widetilde{\mathbf{Y}}_i^T
P_i^{-2} \widetilde{\mathbf{Y}}_i\right]\nonumber\\
&=& - e^\tau \widetilde{\mathbf{Y}}_i P_i^{-2}
\widetilde{\mathbf{Y}}_i + e^\tau \widetilde{\mathbf{Y}}_i^T
P_i^{-1} \left(\frac{\partial P_i} {\partial \tau}\right) P_i^{-2}
\widetilde{\mathbf{Y}}_i + \widetilde{\mathbf{Y}}_i^T
P_i^{-2}\left(\frac{\partial P_i}{\partial \tau}\right)
P_i^{-1}   \widetilde{\mathbf{Y}}_i \nonumber\\
&=& e^\tau \widetilde{\mathbf{Y}}_i^T [2e^\tau P_i^{-3} -
P_i^{-2}] \widetilde{\mathbf{Y}}_i,
~~(\mbox{by}~(\ref{eq:del_tau_P_i})).
\end{eqnarray}
From (\ref{eq:grad_zeta_F_1_tilde}),
\begin{eqnarray}\label{eq:hess_tau_zeta_F_1_tilde}
\frac{\partial^2 F^1}{\partial \tau\partial \zeta_k} &=&
\frac{\partial}{\partial \tau} \left[-e^{\zeta_k} (H_{ik}^T P_i^{-1}
\widetilde{\mathbf{Y}}_i)^2\right] \nonumber\\
&=& 2e^{\zeta_k} (\widetilde{\mathbf{Y}}_i^T P_i^{-1} H_{ik})
\left[H_{ik}^T P_i^{-1}\left(\frac{\partial P_i}{\partial
\tau}\right) P_i^{-1}
\widetilde{\mathbf{Y}}_i\right]\nonumber\\
&=& 2 e^{\zeta_k + \tau} \widetilde{\mathbf{Y}}_i^T P_i^{-1}
H_{ik} H_{ik}^T P_i^{-2} \widetilde{\mathbf{Y}}_i,
~~(\mbox{by}~(\ref{eq:del_tau_P_i})).
\end{eqnarray}
Again using (\ref{eq:grad_zeta_F_1_tilde}), and denoting by
$\delta_{kl}$ the indicator of $\{k=l\}$,
\begin{eqnarray}\label{eq:hess_zeta_zeta_F_1_tilde}
\frac{\partial^2 F^1}{\partial \zeta_l\partial \zeta_k} &=&
\frac{\partial}{\partial \zeta_l} \left[-e^{\zeta_k} (H_{ik}^T
P_i^{-1}
\widetilde{\mathbf{Y}}_i)^2\right] \nonumber\\
&=& -\delta_{kl} e^{\zeta_k} (H_{ik}^T P_i^{-1}
\widetilde{\mathbf{Y}}_i)^2 + 2e^{\zeta_k}
(\widetilde{\mathbf{Y}}_i^T P_i^{-1} H_{ik}) \left[H_{ik}^T
P_i^{-1}\left(\frac{\partial P_i}{\partial \zeta_l}\right)
P_i^{-1} \widetilde{\mathbf{Y}}_i\right]\nonumber\\
&=& -\delta_{kl} e^{\zeta_k} (H_{ik}^T P_i^{-1}
\widetilde{\mathbf{Y}}_i)^2 + 2e^{\zeta_k+\zeta_l}
\widetilde{\mathbf{Y}}_i^T P_i^{-1} H_{ik} H_{ik}^T P_i^{-1} H_{il}
H_{il}^T P_i^{-1}\widetilde{\mathbf{Y}}_i,
~~(\mbox{by}~(\ref{eq:del_zeta_P_i}))
\nonumber\\
&=& \begin{cases} 2e^{\zeta_k+\zeta_l} (H_{ik}^T P_i^{-1}
\widetilde{\mathbf{Y}}_i) (H_{il}^T P_i^{-1}
\widetilde{\mathbf{Y}}_i)(H_{ik}^T P_i^{-1} H_{il}) & \mbox{if}~
k\neq l\\
e^{\zeta_k}  (H_{ik}^T P_i^{-1}
\widetilde{\mathbf{Y}}_i)^2\left[2e^{\zeta_k} (H_{ik}^T P_i^{-1}
H_{ik}) - 1\right] & \mbox{if}~k = l.
\end{cases}
\end{eqnarray}

\subsection*{Hessian of $F^2$}

From (\ref{eq:grad_tau_F_2_tilde}),
\begin{equation}\label{eq:hess_tau_tau_F_2_tilde}
\frac{\partial^2 F^1}{\partial \tau^2} = \frac{\partial}{\partial
\tau}\left[e^\tau Tr(P_i^{-1})\right] = e^\tau Tr(P_i^{-1}) -
e^\tau Tr\left[P_i^{-1} \left(\frac{\partial P_i}{\partial
\tau}\right)P_i^{-1}\right] = e^\tau [Tr(P_i^{-1}) - e^\tau
Tr(P_i^{-2})].
\end{equation}
From (\ref{eq:grad_zeta_F_2_tilde}),
\begin{equation}\label{eq:hess_tau_zeta_F_2_tilde}
\frac{\partial^2 F^2}{\partial\tau \partial \zeta_k} =
\frac{\partial }{\partial \tau} [e^{\zeta_k} H_{ik}^T P_i^{-1}
H_{ik}] = - e^{\zeta_k} H_{ik}^T P_i^{-1} \left(\frac{\partial
P_i}{\partial \tau}\right) P_i^{-1} H_{ik} = -e^{\zeta_k + \tau}
H_{ik}^T P_i^{-2} H_{ik}.
\end{equation}
Finally,
\begin{eqnarray}\label{eq:hess_zeta_zeta_F_2_tilde}
\frac{\partial^2 F^2}{\partial \zeta_l \zeta_k} &=&
\frac{\partial}{\partial \zeta_l} \left[e^{\zeta_k} H_{ik}^T
P_i^{-1} H_{ik}\right] \nonumber\\
&=& \delta_{kl} e^{\zeta_k} H_{ik}^T P_i^{-1} H_{ik} - e^{\zeta_k}
H_{ik}^T P_i^{-1} \left(\frac{\partial P_i}{\partial \zeta_l}\right)
P_i^{-1} H_{ik}
\nonumber\\
&=& \delta_{kl} e^{\zeta_k} H_{ik}^T P_i^{-1} H_{ik} - e^{\zeta_k +
\zeta_l} (H_{ik}^T P_i^{-1} H_{il})^2,
~~(\mbox{by}~(\ref{eq:del_zeta_P_i}))
\nonumber\\
&=& \begin{cases}
- e^{\zeta_k +\zeta_l} (H_{ik}^T P_i^{-1} H_{il})^2 & \mbox{if}~k \neq l\\
e^{\zeta_k} H_{ik} P_i^{-1} H_{ik}[1- e^{\zeta_k} H_{ik}^T P_i^{-1}
H_{ik}] & \mbox{if}~ k = l.
\end{cases}
\end{eqnarray}

\clearpage 
\section*{Tables}
{
\begin{table}[h]
\scriptsize

 \centering \caption{\texttt{Easy}, $n=200$,
$\sigma^2=1/16$, Gaussian noise}\label{table:easy_200}
\begin{tabular}{ccccccccc}\\\hline\hline
${\bf M=5}$&
&$\psi_1$&$\psi_2$&$\psi_3$&$\lambda_1$&$\lambda_2$&$\lambda_3$&$\sigma^2$\\\hline
\texttt{New.loc}&MISE/MSE& 0.060&   0.214&   0.217& 1.36&1.07& 3.75&88.23\\
&(Sd)&   (0.066)&   (0.273)&   (0.337)\\
&Reduction ($\%$)&  81.8 &  64.3 &  65.4 & 73.8 & 65.8 & 36.0\\
\cline{2-9} \texttt{loc}&MISE/MSE& 0.329&   0.599&
0.628&5.19&3.13&
5.86&261.82 \\
&(Sd)&(0.333)&   (0.577)& (0.617)&\\\hline
\texttt{New.EM}&MISE/MSE& 0.058&0.191& 0.169& 1.27& 1.04& 1.65&0.96\\
\hline\hline
\end{tabular}
\end{table}
}

{
\begin{table}
\scriptsize
 \centering \caption{\texttt{Practical}, $n=500$,
$\sigma^2=1/16$, Gaussian noise}\label{table:prac_500}
\begin{tabular}{cccccccccccc}\\\hline\hline
${\bf M=10}$&
&$\psi_1$&$\psi_2$&$\psi_3$&$\psi_4$&$\psi_5$&$\lambda_1$&$\lambda_2$&$\lambda_3$&$\lambda_4$&$\lambda_5$\\\hline
\texttt{New.loc}&MISE/MSE& 0.035 & 0.195 & 0.463 & 0.556 & 0.343 & 0.69 & 0.54 & 0.62 & 0.54 & 0.68 \\
&(Sd)&   (0.025)  &  (0.347) & (0.532) & (0.531) & (0.404)\\
&Reduction ($\%$)&  91.9 & 81.6 & 59.5 & 54.1 & 69.6 & 88.6 & 80.0 & 74.2 & 85.5 & 90.5\\
\cline{2-12}
\texttt{loc}&MISE/MSE& 0.434 &  1.059 &  1.143 & 1.211 & 1.127 & 6.04 & 2.70 & 2.40 & 3.73 & 7.19  \\
&(Sd)&(0.387) & (0.502) & (0.514) & (0.536) &  (0.523)&\\\hline
\texttt{New.EM}&MISE/MSE& 0.036 & 0.172 & 0.396 & 0.498 & 0.332 & 0.62 & 0.54 & 0.52 & 0.43 & 0.80 \\
\hline\hline

\end{tabular}
\end{table}
}

{
\begin{table}
\scriptsize
 \centering \caption{\texttt{Practical}, $n=1000$,
$\sigma^2=1/16$, Gaussian noise}\label{table:prac_1000}
\begin{tabular}{cccccccccccc}\\\hline\hline
${\bf M=10}$&
&$\psi_1$&$\psi_2$&$\psi_3$&$\psi_4$&$\psi_5$&$\lambda_1$&$\lambda_2$&$\lambda_3$&$\lambda_4$&$\lambda_5$\\\hline
\texttt{New.loc}&MISE/MSE& 0.015 & 0.067 & 0.169 & 0.228 & 0.146 & 0.26 & 0.34 & 0.25 & 0.35 & 0.41 \\
&(Sd)&   (0.014) & (0.134) & (0.226)& (0.259) & (0.203)\\
&Reduction ($\%$)& 93.3 & 91.8 & 84.2 & 78.0 & 84.3 & 94.6 &  82.7 &  80.8 & 82.0 & 91.3\\
\cline{2-12}
\texttt{loc}&MISE/MSE& 0.224 & 0.813 & 1.069 & 1.035 & 0.930 & 4.85 & 1.96 & 1.30 & 1.94 & 4.72 \\
&(Sd)& (0.137) & (0.523) & (0.466) & (0.580) &  (0.541)&\\\hline
\texttt{New.EM}&MISE/MSE& 0.016 & 0.063 & 0.145 & 0.232 & 0.172 & 0.25 & 0.29 & 0.24 & 0.33 & 0.41\\
\hline\hline

\end{tabular}
\end{table}
}

{
\begin{table}
\scriptsize
 \centering \caption{\texttt{Challenging}, $n=500$,
$\sigma^2=1/16$, Gaussian noise}\label{table:spike_500}
\begin{tabular}{ccccccccc}\\\hline\hline
${\bf M=30}$&
&$\psi_1$&$\psi_2$&$\psi_3$&$\lambda_1$&$\lambda_2$&$\lambda_3$&$\sigma^2$\\\hline
\texttt{New.loc}&MISE/MSE& 0.162 & 0.196 &  0.135 & 2.67 & 2.06 & 2.27  & 2103.76\\
&(Sd)&   (0.254) & (0.330) & (0.285)\\
&Reduction ($\%$)&  57.5 & 62.1 & 60.2 & 37.5 & 6.4 & 0.9  \\
\cline{2-9} \texttt{loc}&MISE/MSE& 0.381 & 0.517 & 0.339 & 4.27 & 2.20 & 2.29 & 2469.33 \\
&(Sd)& (0.307) & (0.453)  & (0.373)\\\hline
\texttt{New.EM}&MISE/MSE& 0.124 & 0.130 & 0.082 & 1.43 & 0.53 & 0.67  & 20.82\\
\hline\hline
\end{tabular}
\end{table}
}


{
\begin{table}
\scriptsize
 \begin{center} \caption{Model selection : \texttt{Easy} ($n=200$), \texttt{Practical}
 ($n=500, 1000$), and \texttt{Challenging} ($n=500$),
$\sigma^2=1/16$, Gaussian noise}\label{table:modsel_all}
\begin{tabular}{c|c|cccc|cccc}\\\hline\hline
Model & Method & \multicolumn{4}{c}{Number of converged
replicates}
&\multicolumn{4}{c}{Frequency of models selected}\\
\hline\hline
& & $M=4$&${\bf M=5}$&$M=6$&$M=9$&$M=4$&${\bf M=5}$&$M=6$&$M=9$\\
\cline{3-10}
    &\texttt{New.loc}& 91 & 82 & 86 & 82 & 3 & 80 & 1& 9\\
    &\texttt{New.EM}&99& 99& 99 & 98 & 0 & 99 & 0 & 0\\
\texttt{Easy} &\texttt{Combine$^{1}$}& 100 & 100 & 100 & 100 & 0 & 100 & 0 & 0\\
 ($n=200$)   &\texttt{New.EM.ns}& 99 & 99 & 99 & 99 & 0 & 99 & 0 & 0\\
    &\texttt{Combine.ns$^{2}$}& 99 & 99 & 99 & 99 & 0 & 99 & 0 & 0\\
\hline\hline
& & $M=5$&${\bf M=10}$&$M=15$&$M=20$&$M=5$&${\bf M=10}$&$M=15$&$M=20$\\
\cline{3-10}
     &\texttt{New.loc}& 53& 46& 21& 7 & 25 & 46 & 8 & 1\\
     &\texttt{New.EM}& 96& 93& 94& 88 & 0 & 93&  2&  5\\
\texttt{Practical} & \texttt{Combine}& 100 & 97 & 95 & 88 & 2& 97 & 1 & 2\\
($n=500$) &\texttt{New.EM.ns}&  72 & 60 & 78 & 86 & 3 &60& 17& 20 \\
     &\texttt{Combine.ns} & 90 & 82 & 81 & 87 & 2 & 82 & 7 & 9\\
\hline\hline
& & $M=5$&${\bf M=10}$&$M=15$&$M=20$&$M=5$&${\bf M=10}$&$M=15$&$M=20$\\
\cline{3-10}
     &\texttt{New.loc}& 53 & 77 & 43 & 28 & 7 & 77 & 4 &  5\\
     &\texttt{New.EM}& 98 & 97 & 98 & 92 & 0 & 97 & 0 & 3\\
\texttt{Practical} & \texttt{Combine}& 99 & 100 & 99 & 96 & 0  & 100  & 0  & 0 \\
($n=1000$) &\texttt{New.EM.ns}&   83& 62& 84& 91 & 0 & 59 & 6 & 35\\
     &\texttt{Combine.ns} & 93& 94& 89& 93 & 0 & 91 & 1 &  8\\
\hline\hline
& & $M=20$&$M=25$&${\bf M=30}$&$M=35$&$M=20$&$M=25$&${\bf M=30}$&$M=35$\\
\cline{3-10}
     &\texttt{New.loc}& 71 & 71 & 58 & 46 & 5& 16& 56& 16 \\
     &\texttt{New.EM} & 93 & 91 & 79 & 70 & 1 & 7 & 79 & 11\\
\texttt{Challenging} &\texttt{Combine} & 95 & 96 & 88 & 79 & 1 & 4  & 88 & 6  \\
 ($n=500$)    &\texttt{New.EM.ns}& 65& 76& 62& 62 & 2& 10& 62& 25\\
     &\texttt{Combine.ns}& 94 & 92 & 89 & 77 & 2 & 2& 87&  9\\
\hline\hline
\end{tabular}
\end{center}
$^1$ combine the results of \texttt{New.loc} with \texttt{New.EM}:
replace one by another if the first one fails to converge; $^2$
combine the results of \texttt{New.loc} with \texttt{New.EM.ns}.
\end{table}
}



{
\begin{table}
\scriptsize

\centering\caption{Model selection : \texttt{Hybrid}, $n=500$,
$\sigma^2=1/16$, Gaussian noise}\label{table:hybrid_modsel_500}
\begin{tabular}{c|cccc|cccc}\\
& \multicolumn{4}{c}{\texttt{New.loc}} &
 \multicolumn{4}{c}{\texttt{New.EM}}\\\hline\hline
\multicolumn{9}{c}{{\bf I. number of converged replicates (total=100)}}\\
\hline\hline  $r$ & ${\bf M=10}$ & $M=15$ & $M=20$ & $M=25$ & ${\bf
M=10}$ & $M=15$ & $M=20$ & $M=25$\\\hline\hline
2 & 75 & 63 & 60 & 54 & 100 & 98 & 99 & 85 \\
3 & 81 & 78 & 60 & 61 &  99 & 99 & 97 & 96 \\
4 & 36 & 11 & 3 & 1 & 99 & 94 & 94 & 93 \\
5 & 6 & 0 & 0 &  0 & 95 & 85 & 60 & 38 \\
6 & 1 & 0 & 0 & 0 & 61 & 60 & 29 & 8 \\
7 & 4 & 0 & 0 & 0 & 7 & 22 & 7 & 3 \\
\hline\hline
\multicolumn{9}{c}{{\bf II. frequencies of models selected}}\\
\hline\hline  $r$ & ${\bf M=10}$ & $M=15$ & $M=20$ & $M=25$ & ${\bf
M=10}$ & $M=15$ & $M=20$ & $M=25$\\\hline\hline
1 & 0 ~(1,1,1)& 0 ~(0,0,1)& 0 ~(0,0,0)& 0 ~(0,0,1)& 0 ~(0,0,0)& 0 ~(0,0,0)& 0 ~(0,0,0)& 0 ~(0,0,0)\\
2 & 2 ~(4,6,6)& 2 ~(3,3,2)&  2 ~(2,2,2)& 2 ~(2,2,1)& 0 ~(0,0,0)& 0 ~(0,0,0)& 0 ~(0,0,0)& 0 ~(0,0,0)\\
3 & 47 ~(61,63,79)& 2 ~(1,1,1)& 1 ~(1,1,1)& 0 ~(0,0,0)& 0 ~(0,0,98)& 0 ~(0,0,1)& 0 ~(0,0,1)& 0 ~(0,0,0)\\
4 & 30 ~(20,16,0)& 0 ~(0,0,0)& 0 ~(0,0,0)& 0 ~(0,0,0)& 2 ~(43,97,0) & 0 ~(0,1,0)& 0 ~(1,1,0)& 0 ~(0,0,0)\\
5 & 3 ~(0,0,0)& 0 ~(0,0,0)& 0 ~(0,0,0)& 0 ~(0,0,0)& 56 ~(55,1,0)& 0 ~(1,0,0)& 1 ~(0,0,0)& 0 ~(0,0,0)\\
6 & 1 ~(0,0,0)& 0 ~(0,0,0)& 0 ~(0,0,0)& 0 ~(0,0,0)& 33 ~(0,0,0)& 0 ~(0,0,0)& 0 ~(0,0,0)& 0 ~(0,0,0)\\
7 & 3 ~(0,0,0)& 0 ~(0,0,0)& 0 ~(0,0,0)& 0 ~(0,0,0)& 7 ~(0,0,0)& 1 ~(0,0,0)& 0 ~(0,0,0)& 0 ~(0,0,0)\\
\hline
\end{tabular}
\end{table}
}


{
\begin{table}
\small
 \begin{center} \caption{Mean running time for one replicate in seconds: \texttt{Easy} ($n=200$), \texttt{Practical}
 ($n=500$),
$\sigma^2=1/16$, Gaussian noise}\label{table:comp}
\begin{tabular}{c|c|cccc}\\\hline\hline
Model & Method &\multicolumn{4}{c}{{\bf Mean running time
(standard
deviation) $^{\ast}$}}\\\hline\hline & & $M=4$&${\bf M=5}$&$M=6$&$M=9$\\
\cline{3-6} &\texttt{New.loc}& 14.3 (4.4)&
14.4 (4)&15.4 (4.9)&16.7 (5.2)\\
 \texttt{Easy}&\texttt{loc}&19 (1.9)&19 (1.9)&19 (1.9)&19 (1.9)\\
 ($n=200$)&\texttt{New.EM}&14.7 (0.48)&14.4 (0.52)&14.8 (0.42)&16.3 (0.48)\\
   &\texttt{EM}&9.8 (0.79)& 9.7 (0.48)& 10.1 (0.32)& 10.7 (1.1)
\\
\hline\hline
& & $M=5$&${\bf M=10}$&$M=15$&$M=20$\\
\cline{3-6} &\texttt{New.loc}& 63.8 (27.9)& 80.9 (45.1)& 87.4 (35.8)&92.7 (31.2)\\
 \texttt{Practical}&\texttt{loc}& 28.4 (3.4)&28.4 (3.4)&28.4 (3.4)&28.4 (3.4)\\
 ($n=500$)&\texttt{New.EM}& 60.2 (9.5)&59.4 (3.1)& 70.6 (17.9)& 91.9 (30.2)\\
   &\texttt{EM}& 54.1 (6.7)&47.6 (2.2)&53.7 (6.7)& 61.2 (11.9)\\
\hline\hline

\end{tabular}
\end{center}
{\small $\ast$ for \texttt{New.loc} and \texttt{New.EM}, this
means the additional computational cost after obtaining the
initial estimates.}
\end{table}

}


{
\begin{table}
\small

\centering\caption{\texttt{CD4 counts} data: estimated error
variance and eigenvalues}\label{table:real_data}
\begin{tabular}{c|ccccc}\\
\hline\hline {\bf Model : $M=10$, $r=4$} &
$\widehat{\sigma}^2$&$\widehat\lambda_1$ & $\widehat\lambda_2$ &
$\widehat\lambda_3$ & $\widehat\lambda_4$
 \\
\hline\hline
\texttt{loc} & 42,359.3&615,735.6& 94,188.6& 47,012.6& 37,687.1\\
\texttt{New.EM} &38,411.0 & 473,416.8& 208,201.4& 53,253.9&
24,582.0\\
\texttt{EM} & 38,132.2& 469,784.3& 207,961.1 &54,007.2 &24,344.5\\
\hline\hline
\end{tabular}
\end{table}
}



 \clearpage
\section*{Figures}


 \clearpage

\clearpage
\section*{Supplementary material}
\section*{Tables}
In the following tables, \texttt{New.loc}: \texttt{Newton} with
\texttt{loc} as initial estimate; \texttt{New.EM}: \texttt{Newton}
with \texttt{EM} (B-spline basis) as initial estimate;
\texttt{New.EM.ns}: \texttt{Newton} with \texttt{EM.ns} (natural
spline basis) as initial estimate; \texttt{Combine}: result after
replacing \texttt{New.EM} with \texttt{New.loc} if the former
fails to converge; \texttt{Combine.ns}: result after replacing
\texttt{New.EM.ns} with \texttt{EM.loc} if the former fails to
converge; in block (IV), number within parentheses after
\texttt{loc} is the number of replicates with $\widehat\sigma^2 <
0$ by \texttt{loc}.

{
\begin{table}
\scriptsize

\centering \caption{\texttt{Easy}, $n=100$; $\sigma^2=1/16$; noise
distribution : $N(0,1)$.}
\begin{tabular}{ccccc}\\\hline\hline
&$M=4$& ${\bf M=5}$& $M=6$& $M=9$\\\hline\hline
\multicolumn{5}{c}{{\bf I. number of converged
replicates}}\\
\hline\hline
\texttt{New.loc}& 91  & 91 & 89 & 72\\
\texttt{New.EM}& 99 &100& 100& 96\\
 \texttt{Combine}& 99& 100 &100& 99\\
\texttt{New.EM.ns}&96 &100&  99&  98\\
\texttt{Combine.ns} & 98 & 100 & 100 & 100 \\
\hline\hline

\end{tabular}
\end{table}
}

\newpage

{
\begin{table}
\scriptsize

\centering \caption{\texttt{Easy}; $n=200$; $\sigma^2=1/16$; noise
distribution : $N(0,1)$}
\begin{tabular}{ccccc}\\\hline\hline
&$M=4$& ${\bf M=5}$& $M=6$& $M=9$\\\hline\hline
\multicolumn{5}{c}{{\bf
I. number of converged replicates}}\\
\hline\hline
\texttt{New.loc}& 91 &82& 86& 82\\
\texttt{New.EM}&99&  99&  99&  98\\
\texttt{Combine}&100&100& 100& 100\\
\texttt{New.EM.ns}& 99 &99& 99& 99\\
\texttt{Combine.ns}& 99 &99& 99& 99\\
\hline\hline

\end{tabular}
\end{table}
}

\newpage

{
\begin{table}
\scriptsize

\centering \caption{\texttt{Easy}; $n=500$; $\sigma^2=1/16$; noise
distribution : $N(0,1)$}
\begin{tabular}{ccccc}\\\hline\hline
&$M=4$& ${\bf M=5}$& $M=6$& $M=9$\\\hline\hline
\multicolumn{5}{c}{{\bf I. number of converged
replicates}}\\
\hline\hline
\texttt{New.loc}&  83& 84& 87& 80\\
\texttt{New.EM}&       99 &100& 100& 100\\
\texttt{Combine}& 100& 100& 100& 100\\
\texttt{New.EM.ns}&97& 100& 100& 100\\\hline\hline
 \multicolumn{5}{c}{{\bf II. mean integrated squared error of estimated
eigenfunctions ($\psi_\nu$)}}\\
\hline\hline
\end{tabular}
\end{table}
}

\newpage


{

\begin{table}
\scriptsize

\centering \caption{\texttt{Easy}; $n=200$; $\sigma^2=1/8$; noise
distribution : $N(0,1)$}
\begin{tabular}{ccccc}\\\hline\hline
&$M=4$& ${\bf M=5}$& $M=6$& $M=9$\\\hline\hline
\multicolumn{5}{c}{{\bf I. number of converged
replicates}}\\\hline\hline
\texttt{New.loc}& 86 & 86 & 86 & 85\\
\texttt{New.EM}&  99 & 99 & 99 & 99 \\
\texttt{Combine}& 100 & 100 & 100 & 100 \\
\hline\hline \multicolumn{5}{c}{{\bf II. mean integrated squared
error of estimated eigenfunctions ($\psi_\nu$)}}\\
\hline\hline \texttt{New.loc}
&  0.345, 0.369, 0.463 &  0.064, 0.186, 0.158 &  0.065, 0.177, 0.170 & 0.074, 0.181, 0.153 \\
\hline\hline
\end{tabular}
\end{table}
}

\newpage


{
\begin{table}
\scriptsize

\centering \caption{\texttt{Easy}; $n=100$; $\sigma^2=1/16$; noise
distribution : $\frac{1}{\sqrt{2}}t_4$}
\begin{tabular}{ccccc}\\\hline\hline
&$M=4$& ${\bf M=5}$& $M=6$& $M=9$\\\hline\hline
\multicolumn{5}{c}{{\bf I. number of converged
replicates}}\\\hline\hline
\texttt{New.loc}& 85 & 79 & 79 & 66\\
\texttt{New.EM}& 100 & 100 & 99 & 99 \\
 \texttt{Combine}& 100 & 100 & 100 & 99 \\
\hline\hline
 \multicolumn{5}{c}{{\bf II. mean integrated squared error of estimated
eigenfunctions ($\psi_\nu$)}}\\\hline\hline
\texttt{New.loc}& 0.366, 0.613, 0.682 & 0.120, 0.393, 0.404 & 0.129, 0.360, 0.370 & 0.165, 0.400, 0.343 \\
\hline\hline

\end{tabular}
\end{table}
}

\newpage


{
\begin{table}
\scriptsize

\centering \caption{\texttt{Easy}; $n=200$; $\sigma^2=1/16$; noise
distribution : $\frac{1}{\sqrt{2}} t_4$}
\begin{tabular}{ccccc}\\\hline\hline
&$M=4$& ${\bf M=5}$& $M=6$& $M=9$\\\hline\hline
\multicolumn{5}{c}{{\bf I. number of converged
replicates}}\\\hline\hline
\texttt{New.loc}& 91 & 90 & 88 & 81 \\
\texttt{New.EM}& 98 & 98 & 99 & 99 \\
\texttt{Combine}&  99 & 100 & 100 & 99 \\
\hline\hline
 \multicolumn{5}{c}{{\bf II. mean integrated squared error of estimated
eigenfunctions ($\psi_\nu$)}}\\\hline\hline
\texttt{New.loc}& 0.343, 0.472, 0.587 &  0.047, 0.189, 0.173 &  0.062, 0.195, 0.183 & 0.054, 0.183, 0.163 \\
\hline\hline
\end{tabular}
\end{table}
}

\newpage


{
\begin{table}
\scriptsize

\centering \caption{\texttt{Easy}; $n=100$; $\sigma^2=1/16$; noise
distribution : Exp$(1) - 1$}
\begin{tabular}{ccccc}\\\hline\hline
&$M=4$& ${\bf M=5}$& $M=6$& $M=9$\\\hline\hline
\multicolumn{5}{c}{{\bf I. number of converged
replicates}}\\
\hline\hline \texttt{New.loc} & 83 & 84 & 84 & 67\\
\texttt{New.EM} & 97 & 100 & 100 & 99 \\
\texttt{Combine}& 98 & 100 & 100 & 100\\
\hline\hline
 \multicolumn{5}{c}{{\bf II. mean integrated squared error of estimated
eigenfunctions ($\psi_\nu$)}}\\\hline\hline
\texttt{New.loc}& 0.403, 0.548, 0.652 &  0.128, 0.358, 0.340 & 0.136, 0.369, 0.412 & 0.150, 0.377, 0.320\\
\hline\hline
\end{tabular}
\end{table}
}

\newpage


{
\begin{table}
\scriptsize

\centering \caption{\texttt{Easy}; $n=200$; $\sigma^2=1/16$; noise
distribution : Exp$(1)-1$}
\begin{tabular}{ccccc}\\\hline\hline
&$M=4$& ${\bf M=5}$& $M=6$& $M=9$\\\hline\hline
\multicolumn{5}{c}{{\bf I. number of converged
replicates}}\\
\hline\hline \texttt{New.loc}& 90 & 91 & 87 & 79\\
\texttt{New.EM}& 99 & 100 & 100 & 99\\
 \texttt{Combine}& 100 & 100 & 100 & 100 \\
\hline\hline
 \multicolumn{5}{c}{{\bf II. mean integrated squared error of estimated
eigenfunctions ($\psi_\nu$)}}\\
\hline\hline\texttt{New.loc}
& 0.337, 0.463, 0.527  & 0.061, 0.248, 0.221  & 0.057, 0.208, 0.184  & 0.073, 0.233, 0.191 \\
\hline\hline
\end{tabular}
\end{table}
}

\newpage

{
\begin{table}

\tiny

\centering \caption{\texttt{Practical}; $n=300$; $\sigma^2=1/16$;
noise distribution : $N(0,1)$}
\begin{tabular}{ccccc}\\\hline\hline
&$M=5$& ${\bf M=10}$& $M=15$& $M=20$\\\hline\hline
\multicolumn{5}{c}{{\bf I. number of converged
replicates}}\\
\hline\hline
\texttt{New.loc}& 60 & 17 &  4 & 0\\
\texttt{New.EM}& 96 & 98 & 87 & 73\\
\texttt{Combine}& 99 & 98 & 89 & 73\\
\texttt{New.EM.ns}& 81 & 48 & 67 & 80\\\hline\hline
 \multicolumn{5}{c}{{\bf II. mean  integrated squared error of estimated
eigenfunctions ($\psi_\nu$)}}\\
\hline\hline
\end{tabular}
\end{table}
}

\newpage

{
\begin{table}
\tiny

\centering \caption{\texttt{Practical}; $n=500$; $\sigma^2=1/16$;
noise distribution : $N(0,1)$}
\begin{tabular}{ccccc}\\\hline\hline
&$M=5$& ${\bf M=10}$& $M=15$& $M=20$\\\hline\hline
\multicolumn{5}{c}{{\bf I. number of converged
replicates}}\\
\hline\hline
\texttt{New.loc}& 53& 46& 21& 7\\
\texttt{New.EM}&       96& 93& 94& 88\\
\texttt{Combine}& 100 &97& 95& 88\\
\texttt{New.EM.ns}&  72& 60& 78& 86\\
\texttt{Combine.ns} & 90 & 82 & 81 & 87 \\
\hline\hline
\end{tabular}
\end{table}
}

\newpage
{
\begin{table}
\tiny

\centering \caption{\texttt{Practical}; $n=500$; $\sigma^2=1/8$;
noise distribution : $N(0,1)$}
\begin{tabular}{ccccc}\\\hline\hline
&$M=5$& ${\bf M=10}$& $M=15$& $M=20$\\\hline\hline
\multicolumn{5}{c}{{\bf I. number of converged
replicates}}\\\hline\hline
\texttt{New.loc}& 51 & 51 & 26 & 6 \\
\texttt{New.EM}& 96 & 97 & 98 & 91 \\
\texttt{Combine}&  96 & 98 & 99 & 92 \\
\hline\hline
 \multicolumn{5}{c}{{\bf II. mean integrated squared error of estimated
eigenfunctions ($\psi_\nu$)}}\\
\hline\hline \texttt{New.loc} &  1.612, 1.782, 1.383, 1.807, 1.537 &
0.037, 0.183, 0.495, 0.595, 0.340 & 0.101, 0.189, 0.581, 0.793, 0.537 & 0.057, 0.301,  0.579, 0.711, 0.439\\
\hline\hline
\end{tabular}
\end{table}
}

\newpage

{
\begin{table}
\tiny

\centering \caption{\texttt{Practical}; $n=1000$; $\sigma^2=1/16$;
noise distribution : $N(0,1)$}
\begin{tabular}{ccccc}\\\hline\hline
&$M=5$& ${\bf M=10}$& $M=15$& $M=20$\\\hline\hline
\multicolumn{5}{c}{{\bf I. number of converged
replicates}}\\
\hline\hline
\texttt{New.loc}&53 & 77 & 43 & 28\\
\texttt{New.EM} &  98 & 97 & 98 & 92\\
\texttt{Combine} & 99 & 100 & 99 & 96 \\
\texttt{New.EM.ns}&  83& 62& 84& 91\\
\texttt{Combine.ns}&93& 94& 89& 93
\\
\hline\hline
\end{tabular}
\end{table}
}

\newpage

{
\begin{table}
\tiny

\centering \caption{\texttt{Practical}; $n=300$; $\sigma^2=1/16$;
noise distribution : $\frac{1}{\sqrt{2}}
t_4$}
\begin{tabular}{ccccc}\\\hline\hline
&$M=5$& ${\bf M=10}$& $M=15$& $M=20$\\\hline\hline
\multicolumn{5}{c}{{\bf I. number of converged
replicates}}\\
\hline\hline
\texttt{New.loc}   & 55 & 23 & 0 & 0\\
\texttt{New.EM}& 99 & 94 & 95 & 66 \\
\texttt{Combine}& 100 & 94 & 95 & 66\\
\\\hline\hline
 \multicolumn{5}{c}{{\bf II. mean integrated squared error of estimated
eigenfunctions ($\psi_\nu$)}}\\
\hline\hline
\end{tabular}
\end{table}
}

\newpage


{
\begin{table}
\tiny

\centering\caption{\texttt{Practical}; $n=500$; $\sigma^2=1/16$;
noise distribution :
$\frac{1}{\sqrt{2}}t_4$}
\begin{tabular}{ccccc}\\\hline\hline
&$M=5$& ${\bf M=10}$& $M=15$& $M=20$\\\hline\hline
\multicolumn{5}{c}{{\bf I. number of converged
replicates}}\\
\hline\hline
\texttt{New.loc}&  44 & 41 & 27 & 6\\
\texttt{New.EM}& 98 & 95 & 93 & 93 \\
\texttt{Combine}& 99 & 99 & 96 & 93 \\
\hline\hline
 \multicolumn{5}{c}{{\bf II. mean integrated squared error of estimated
eigenfunctions ($\psi_\nu$)}}\\\hline\hline \texttt{New.loc} &
1.621, 1.820, 1.382, 1.789, 1.494 &
0.046, 0.182, 0.290, 0.431, 0.376 & 0.081, 0.144, 0.300, 0.406, 0.425 &  0.113, 0.280, 0.358, 0.537, 0.469\\
\hline\hline
\end{tabular}
\end{table}
}

\newpage


{
\begin{table}
\tiny

\centering \caption{\texttt{Practical}; $n=500$; $\sigma^2=1/16$;
noise distribution : Exp$(1)-1$}
\begin{tabular}{ccccc}\\\hline\hline
&$M=5$& ${\bf M=10}$& $M=15$& $M=20$\\\hline\hline
\multicolumn{5}{c}{{\bf I. number of converged
replicates}}\\
\hline\hline
\texttt{New.loc}& 56 & 51 & 20 & 6\\
\texttt{New.EM}& 98 & 95 & 96 & 88 \\
\texttt{Combine}& 100 & 95 & 96 &  88 \\
\hline\hline
 \multicolumn{5}{c}{{\bf II. mean integrated squared error of estimated
eigenfunctions ($\psi_\nu$)}}\\\hline\hline \texttt{New.loc} &
1.626, 1.839, 1.355, 1.794, 1.501 & 0.042, 0.186,
0.402, 0.438, 0.365 & 0.013, 0.012, 0.004, 0.001, 0.007 &  0.058, 0.272, 0.346,  0.285, 0.168 \\
\hline\hline
\end{tabular}
\end{table}
}

\newpage

{
\begin{table}
\scriptsize

\centering \caption{\texttt{Challenging}; $n=300$; $\sigma^2=1/16$;
noise distribution : $N(0,1)$}
\begin{tabular}{ccccc}\\\hline\hline
&$M=20$& $M=25$& ${\bf M=30}$& $M=35$\\\hline\hline
\multicolumn{5}{c}{{\bf I. number of converged
replicates}}\\
\hline\hline
\texttt{New.loc}&56& 51& 45& 21\\
\texttt{New.EM}& 92& 92& 90& 64\\
\texttt{Combine}& 96& 96& 93& 72\\
\hline\hline
 \multicolumn{5}{c}{{\bf II. mean integrated squared error of estimated
eigenfunctions ($\psi_\nu$)}}\\\hline\hline
 \texttt{Bias}& 0.169, 0.061, 0.026& 0.142, 0.045, 0.020& 0.053, 0.009, 0.005& 0.084,
 0.019, 0.009\\
\hline\hline
\end{tabular}
\end{table}
}

\newpage

{
\begin{table}
\scriptsize

\centering \caption{\texttt{Challenging}; $n=500$; $\sigma^2=1/16$;
noise distribution : $N(0,1)$}
\begin{tabular}{ccccc}\\\hline\hline
&$M=20$& $M=25$& ${\bf M=30}$& $M=35$\\\hline\hline
\multicolumn{5}{c}{{\bf I. number of converged
replicates}}\\
\hline\hline
\texttt{New.loc}& 71 & 71 & 58 & 46 \\
\texttt{New.EM} & 93 & 91 & 79 & 70\\
\texttt{Combine} & 95 & 96 & 88 & 79 \\
\texttt{New.EM.ns}& 65& 76& 62& 62\\
\texttt{Combine.ns}& 94 & 92 & 89 & 77\\
\hline\hline
\end{tabular}
\end{table}
}

\newpage

{
\begin{table}
\scriptsize

\centering \caption{\texttt{Challenging}; $n=500$; $\sigma^2=1/8$;
noise distribution : $N(0,1)$}
\begin{tabular}{ccccc}\\\hline\hline
&$M=20$& $M=25$& ${\bf M=30}$& $M=35$\\\hline\hline
\multicolumn{5}{c}{{\bf I. number of converged
replicates}}\\
\hline\hline
\texttt{New.loc}& 76 & 60 & 54 & 39 \\
\texttt{New.EM}&  90 & 81 & 79 & 82 \\
\texttt{Combine}& 96 & 86 & 87 & 85
\\\hline\hline
 \multicolumn{5}{c}{{\bf II. mean integrated squared error of estimated
eigenfunctions ($\psi_\nu$)}}\\\hline\hline
 \texttt{Bias} & 0.169, 0.061, 0.026& 0.142, 0.045, 0.020& 0.053, 0.009, 0.005& 0.084,
 0.019, 0.009 \\
\hline\hline
\end{tabular}
\end{table}
}
\clearpage

\clearpage 

{
\begin{table}
\scriptsize

\centering \caption{\texttt{Challenging}; $n=1000$; $\sigma^2=1/16$;
noise distribution : $N(0,1)$}
\begin{tabular}{ccccc}\\\hline\hline
&$M=20$& $M=25$& ${\bf M=30}$& $M=35$\\\hline\hline
\multicolumn{5}{c}{{\bf I. number of converged
replicates}}\\\hline\hline \texttt{New.loc}& 76 & 59 & 49 & 45 \\
\texttt{New.EM}&  95 & 71 & 79 & 88 \\
\texttt{Combine} & 95 & 81 & 90 & 93 \\\hline\hline
 \multicolumn{5}{c}{{\bf II. mean integrated squared error of estimated
eigenfunctions ($\psi_\nu$)}}\\\hline\hline
 \texttt{Bias}& 0.169, 0.061, 0.026& 0.142, 0.045, 0.020& 0.053, 0.009, 0.005& 0.084,
 0.019, 0.009\\
\hline\hline
\end{tabular}
\end{table}
}

\newpage


{
\begin{table}
\scriptsize
 \centering \caption{Model selection : \texttt{Easy} ($n=200$), \texttt{Practical}
 ($n=500, 1000$), and \texttt{Challenging} ($n=500$),
$\sigma^2=1/16$,  $N(0,1)$ noise}
\begin{tabular}{c|c|cccc|cccc}\\\hline\hline
Model & Method & \multicolumn{4}{c}{Number of converged replicates}
&\multicolumn{4}{c}{Frequency of models selected}\\
\hline\hline
& & $M=4$&${\bf M=5}$&$M=6$&$M=9$&$M=4$&${\bf M=5}$&$M=6$&$M=9$\\
\cline{3-10}
    &\texttt{New.loc}& 91 & 82 & 86 & 82 & 3 & 80 & 1& 9\\
    &\texttt{New.EM}&99& 99& 99 & 98 & 0 & 99 & 0 & 0\\
\texttt{Easy} &\texttt{Combine}& 100 & 100 & 100 & 100 & 0 & 100 & 0 & 0\\
 ($n=200$)   &\texttt{New.EM.ns}& 99 & 99 & 99 & 99 & 0 & 99 & 0 & 0\\
    &\texttt{Combine.ns}& 99 & 99 & 99 & 99 & 0 & 99 & 0 & 0\\
\hline\hline
& & $M=5$&${\bf M=10}$&$M=15$&$M=20$&$M=5$&${\bf M=10}$&$M=15$&$M=20$\\
\cline{3-10}
     &\texttt{New.loc}& 53& 46& 21& 7 & 25 & 46 & 8 & 1\\
     &\texttt{New.EM}& 96& 93& 94& 88 & 0 & 93&  2&  5\\
\texttt{Practical} & \texttt{Combine}& 100 & 97 & 95 & 88 & 2& 97 & 1 & 2\\
($n=500$) &\texttt{New.EM.ns}&  72 & 60 & 78 & 86 & 3 &60& 17& 20 \\
     &\texttt{Combine.ns} & 90 & 82 & 81 & 87 & 2 & 82 & 7 & 9\\
\hline\hline
& & $M=5$&${\bf M=10}$&$M=15$&$M=20$&$M=5$&${\bf M=10}$&$M=15$&$M=20$\\
\cline{3-10}
     &\texttt{New.loc}& 53 & 77 & 43 & 28 & 7 & 77 & 4 &  5\\
     &\texttt{New.EM}& 98 & 97 & 98 & 92 & 0 & 97 & 0 & 3\\
\texttt{Practical} & \texttt{Combine}& 99 & 100 & 99 & 96 & 0  & 100  & 0  & 0 \\
($n=1000$) &\texttt{New.EM.ns}&   83& 62& 84& 91 & 0 & 59 & 6 & 35\\
     &\texttt{Combine.ns} & 93& 94& 89& 93 & 0 & 91 & 1 &  8\\
\hline\hline
& & $M=20$&$M=25$&${\bf M=30}$&$M=35$&$M=20$&$M=25$&${\bf M=30}$&$M=35$\\
\cline{3-10}
     &\texttt{New.loc}& 71 & 71 & 58 & 46 & 5& 16& 56& 16 \\
     &\texttt{New.EM} & 93 & 91 & 79 & 70 & 1 & 7 & 79 & 11\\
\texttt{Challenging} &\texttt{Combine} & 95 & 96 & 88 & 79 & 1 & 4  & 88 & 6  \\
     &\texttt{New.EM.ns}& 65& 76& 62& 62 & 2& 10& 62& 25\\
     &\texttt{Combine.ns}& 94 & 92 & 89 & 77 & 2 & 2& 87&  9\\
\hline\hline
\end{tabular}
\end{table}
}

{
\begin{table}
\scriptsize

\centering\caption{Model selection : \texttt{Hybrid}, $n=300$,
$\sigma^2=1/16$,  $N(0,1)$ noise}
\begin{tabular}{c|cccc|cccc}\\
& \multicolumn{4}{c}{\texttt{New.loc}} &
 \multicolumn{4}{c}{\texttt{New.EM}}\\\hline\hline
\multicolumn{9}{c}{{\bf I. number of converged replicates (total=100)}}\\
\hline\hline  \hline $r$ & ${\bf M=10}$ & $M=15$ & $M=20$ & $M=25$
& ${\bf M=10}$  & $M=15$ & $M=20$ & $M=25$\\\hline\hline
2 & 76 & 62 & 43 & 29 & 99 & 98 & 94 & 90 \\
3 & 65 & 42 & 30 & 26 & 97 & 99 & 95 & 94 \\
4 & 19 & 7  & 0 & 0 & 97 & 89 & 87 & 79 \\
5 & 8 & 0 & 0 & 0 & 92 & 78 & 32 & 18 \\
6 & 0 & 0 & 0 & 0 & 70 & 44 & 13 & 4 \\
7 & 2 & 0 & 0 & 0 & 11 & 12 & 6  & 0\\
\hline\hline
\multicolumn{9}{c}{{\bf II. frequencies of models selected}}\\
\hline\hline  $r$ & ${\bf M=10}$ & $M=15$ & $M=20$ & $M=25$ & ${\bf
M=10}$ & $M=15$ & $M=20$ & $M=25$\\\hline\hline
1 & 0 ~(2,2,2)& 0 ~(0,0,0)& 0 ~(0,0,0) & 0 ~(1,1,1) & 0 ~(0,0,0)& 0 ~(0,0,0)& 0~(0,0,0)& 0 ~(0,0,0)\\
2 & 8 ~(8,8,9)& 7 ~(8,8,8)& 4 ~(4,4,4)& 1 ~(0,0,0)& 0 ~(0,0,0)& 0 ~(0,0,0)& 0 ~(0,0,0)& 0 ~(0,0,0)\\
3 & 46 ~(52,54,62)& 2 ~(1,2,2)& 1 ~(1,1,1)& 0 ~(0,0,0)& 0 ~(0,0,98)& 0 ~(0,0,1)& 0 ~(0,0,0)& 0 ~(0,0,0)\\
4 & 12 ~(11,9,0)& 1 ~(1,0,0)& 0 ~(0,0,0)& 0 ~(0,0,0)& 2 ~(32,90,1)& 1 ~(1,1,0)& 0 ~(0,0,0)& 0 ~(0,0,0)\\
5 & 5 ~(0,0,0)& 0 ~(0,0,0)& 0 ~(0,0,0)& 0 ~(0,0,0)& 36 ~(64,9,0)& 0 ~(0,0,0)& 0 ~(0,0,0)& 0 ~(0,0,0)\\
6 & 0 ~(0,0,0)& 0 ~(0,0,0)& 0 ~(0,0,0)& 0 ~(0,0,0)& 50 ~(2,0,0)& 0 ~(0,0,0)& 0 ~(0,0,0)& 0 ~(0,0,0)\\
7 & 2 ~(0,0,0)& 0 ~(0,0,0)& 0 ~(0,0,0)& 0 ~(0,0,0)& 11 ~(0,0,0)& 0 ~(0,0,0)& 0 ~(0,0,0)& 0 ~(0,0,0)\\
\hline
\end{tabular}
\end{table}
}

\newpage

{
\begin{table}
\scriptsize

\centering\caption{Model selection : \texttt{Hybrid}, $n=500$,
$\sigma^2=1/16$,  $N(0,1)$ noise}
\begin{tabular}{c|cccc|cccc}\\
& \multicolumn{4}{c}{\texttt{New.loc}} &
 \multicolumn{4}{c}{\texttt{New.EM}}\\\hline\hline
\multicolumn{9}{c}{{\bf I. number of converged replicates (total=100)}}\\
\hline\hline  $r$ & ${\bf M=10}$ & $M=15$ & $M=20$ & $M=25$ &
${\bf M=10}$ & $M=15$ & $M=20$ & $M=25$\\\hline\hline
2 & 75 & 63 & 60 & 54 & 100 & 98 & 99 & 85 \\
3 & 81 & 78 & 60 & 61 &  99 & 99 & 97 & 96 \\
4 & 36 & 11 & 3 & 1 & 99 & 94 & 94 & 93 \\
5 & 6 & 0 & 0 &  0 & 95 & 85 & 60 & 38 \\
6 & 1 & 0 & 0 & 0 & 61 & 60 & 29 & 8 \\
7 & 4 & 0 & 0 & 0 & 7 & 22 & 7 & 3 \\
\hline\hline
\multicolumn{9}{c}{{\bf II. frequencies of models selected}}\\
\hline\hline  $r$ & ${\bf M=10}$ & $M=15$ & $M=20$ & $M=25$ & ${\bf
M=10}$ & $M=15$ & $M=20$ & $M=25$\\\hline\hline
1 & 0 ~(1,1,1)& 0 ~(0,0,1)& 0 ~(0,0,0)& 0 ~(0,0,1)& 0 ~(0,0,0)& 0 ~(0,0,0)& 0 ~(0,0,0)& 0 ~(0,0,0)\\
2 & 2 ~(4,6,6)& 2 ~(3,3,2)&  2 ~(2,2,2)& 2 ~(2,2,1)& 0 ~(0,0,0)& 0 ~(0,0,0)& 0 ~(0,0,0)& 0 ~(0,0,0)\\
3 & 47 ~(61,63,79)& 2 ~(1,1,1)& 1 ~(1,1,1)& 0 ~(0,0,0)& 0 ~(0,0,98)& 0 ~(0,0,1)& 0 ~(0,0,1)& 0 ~(0,0,0)\\
4 & 30 ~(20,16,0)& 0 ~(0,0,0)& 0 ~(0,0,0)& 0 ~(0,0,0)& 2 ~(43,97,0) & 0 ~(0,1,0)& 0 ~(1,1,0)& 0 ~(0,0,0)\\
5 & 3 ~(0,0,0)& 0 ~(0,0,0)& 0 ~(0,0,0)& 0 ~(0,0,0)& 56 ~(55,1,0)& 0 ~(1,0,0)& 1 ~(0,0,0)& 0 ~(0,0,0)\\
6 & 1 ~(0,0,0)& 0 ~(0,0,0)& 0 ~(0,0,0)& 0 ~(0,0,0)& 33 ~(0,0,0)& 0 ~(0,0,0)& 0 ~(0,0,0)& 0 ~(0,0,0)\\
7 & 3 ~(0,0,0)& 0 ~(0,0,0)& 0 ~(0,0,0)& 0 ~(0,0,0)& 7 ~(0,0,0)& 1 ~(0,0,0)& 0 ~(0,0,0)& 0 ~(0,0,0)\\
\hline
\end{tabular}
\end{table}
}

\newpage

{
\begin{table}
\scriptsize

\centering\caption{Model selection : \texttt{Hybrid}, $n=1000$,
$\sigma^2=1/16$, $N(0,1)$ noise}
\begin{tabular}{c|cccc|cccc}\\
& \multicolumn{4}{c}{\texttt{New.loc}} &
 \multicolumn{4}{c}{\texttt{New.EM}}\\\hline\hline
\multicolumn{9}{c}{{\bf I. number of converged replicates (total=50)}}\\
\hline\hline  $r$ & ${\bf M=10}$ & $M=15$ & $M=20$ & $M=25$ &
${\bf M=10}$ & $M=15$ & $M=20$ & $M=25$\\\hline\hline
2 & 41 & 39 & 34 & 31 & 48 & 49 & 49 & 46 \\
3 & 42 & 37 & 35 & 35 & 49 & 49 & 45 & 43 \\
4 & 26 & 13 & 8 & 11 & 49 & 45 & 43 & 46 \\
5 & 9 & 3 & 0 & 0 & 49 & 45 & 44 & 32 \\
6 & 5 & 0 & 0 & 0 & 29 & 32 & 24 & 12 \\
7 & 1 & 0 & 0 & 0 & 6 & 11 &  9 & 3 \\
\hline\hline
\multicolumn{9}{c}{{\bf II. frequencies of models selected}}\\
\hline\hline  $r$ & ${\bf M=10}$ & $M=15$ & $M=20$ & $M=25$ & ${\bf
M=10}$ & $M=15$ & $M=20$ & $M=25$\\\hline\hline
1 & 0 ~(0,0,0)& 0 ~(0,0,1)& 0 ~(0,0,1)& 0 ~(0,0,1)& 0 ~(0,0,0)& 0 ~(0,0,0)& 0 ~(0,0,0)& 0 ~(0,0,0)\\
2 & 0 ~(1,1,1)& 1 ~(1,1,0)& 1 ~(1,1,0)& 1 ~(1,1,0)& 0 ~(0,0,0)& 0 ~(0,0,0)& 0 ~(0,0,0)& 0 ~(0,0,0)\\
3 & 11 ~(17,21,41)& 0 ~(0,0,0)& 0 ~(0,0,0)& 0 ~(0,0,1)& 0 ~(0,0,49)& 0 ~(0,0,0)& 0 ~(0,0,1)& 0 ~(0,0,0)\\
4 & 19 ~(22,20,0)& 0 ~(0,0,0)& 0 ~(0,0,0)& 1 ~(1,1,0)& 0 ~(27,49,0)& 0 ~(0,0,0)& 0 ~(0,1,0)& 0 ~(0,0,0)\\
5 & 7 ~(2,0,0)& 0 ~(0,0,0)& 0 ~(0,0,0)& 0 ~(0,0,0)& 26 ~(22,0,0)& 0 ~(0,0,0)& 0 ~(1,0,0)& 0 ~(0,0,0)\\
6 & 5 ~(0,0,0)& 0 ~(0,0,0)& 0 ~(0,0,0)& 0 ~(0,0,0)& 18 ~(0,0,0)& 0 ~(0,0,0)& 1 ~(0,0,0)& 0 ~(0,0,0)\\
7 & 0 ~(0,0,0)& 0 ~(0,0,0)& 0 ~(0,0,0)& 0 ~(0,0,0)& 5 ~(0,0,0)& 0 ~(0,0,0)& 0 ~(0,0,0)& 0 ~(0,0,0)\\
\hline
\end{tabular}
\end{table}
}

\newpage

{
\begin{table}
\scriptsize

\centering\caption{Risk for most frequently selected models:
\texttt{Hybrid}, $\sigma^2=1/16$, $N(0,1)$ noise (based on
\textit{all converged replicates})}
\begin{tabular}{c|c|c|c}\\\hline\hline
\multicolumn{4}{c}{{\bf I. mean integrated squared error of
estimated eigenfunctions ($\psi_\nu$)}}\\\hline\hline
 & &  ${\bf M=10}$, $r=5$    & ${\bf M=10}$, $r=6$ \\\hline
$n=300$ & \texttt{New.EM} &   0.042,  0.138, 0.118,  0.023, 0.513 &   0.042, 0.153, 0.136, 0.022, 0.474, 1.248 \\
 & (sd) & 0.054,   0.168,    0.157,    0.013,    0.528 & 0.054, 0.195, 0.185, 0.011, 0.532, 0.508 \\
 &(\% reduction) & 0.0, 34.3, 38.9, 11.5, 6.4 & 2.3, 36.5, 39.6, 18.5, 17.0, -8.6\\
\cline{2-4} &  \texttt{EM} & 0.042,  0.210, 0.193, 0.026, 0.548 & 0.043, 0.241, 0.225, 0.027, 0.571, 1.149 \\
 & (sd) & 0.048, 0.291, 0.289, 0.016, 0.465 & 0.047, 0.312, 0.310, 0.016, 0.507, 0.497 \\
\hline\hline
\end{tabular}
\end{table}
}

\newpage

{
\begin{table}
\scriptsize

\centering\caption{Risk for most frequently selected models:
\texttt{Hybrid}, $\sigma^2=1/16$, $N(0,1)$ noise (based on
\textit{selected replicates
only})}
\begin{tabular}{c|c|c|c}\\\hline\hline
\multicolumn{4}{c}{{\bf I. mean integrated squared error of
estimated eigenfunctions ($\psi_\nu$)}}\\\hline\hline
 & &  ${\bf M=10}$, $r=5$   & ${\bf M=10}$, $r=6$ \\\hline
$n=300$ & \texttt{New.EM} &  0.044, 0.125, 0.103, 0.022, 0.406 &   0.041, 0.140, 0.122, 0.023, 0.495, 1.212 \\
 & (sd) & 0.057, 0.149, 0.118, 0.013, 0.430 & 0.055, 0.165, 0.163, 0.011, 0.572, 0.529 \\
 &(\% reduction) &   2.2, 36.2, 42.1, 12.0, -4.1 & -5.1, 36.4, 40.8, 14.8, 14.8, -4.3\\
\cline{2-4} &  \texttt{EM} & 0.045, 0.196, 0.178, 0.025, 0.390 &  0.039, 0.220, 0.206, 0.027, 0.581, 1.162 \\
 & (sd) & 0.052, 0.289, 0.273, 0.014, 0.350 & 0.045, 0.278, 0.283, 0.016, 0.515, 0.505 \\
\hline\hline
\end{tabular}
\end{table}
}


{
\begin{table}
\small
 \begin{center} \caption{Mean running time for one replicate in seconds: \texttt{Easy} ($n=200$), \texttt{Practical}
 ($n=500$),
$\sigma^2=1/16$, Gaussian noise}
\begin{tabular}{c|c|cccc}\\\hline\hline
Model & Method &\multicolumn{4}{c}{{\bf Mean running time (standard
deviation) $^{\ast}$}}\\\hline\hline & & $M=4$&${\bf M=5}$&$M=6$&$M=9$\\
\cline{3-6} &\texttt{New.loc}& 14.3 (4.4)&
14.4 (4)&15.4 (4.9)&16.7 (5.2)\\
 \texttt{Easy}&\texttt{loc}&19(1.9)&19(1.9)&19(1.9)&19(1.9)\\
 ($n=200$)&\texttt{New.EM}&14.7(0.48)&14.4 (0.52)&14.8 (0.42)&16.3 (0.48)\\
   &\texttt{EM}&9.8 (0.79)& 9.7 (0.48)& 10.1 (0.32)& 10.7 (1.1)
\\
\hline\hline
& & $M=5$&${\bf M=10}$&$M=15$&$M=20$\\
\cline{3-6} &\texttt{New.loc}& 63.8 (27.9)& 80.9 (45.1)& 87.4 (35.8)&92.7 (31.2)\\
 \texttt{Practical}&\texttt{loc}& 28.4 (3.4)&28.4 (3.4)&28.4 (3.4)&28.4 (3.4)\\
 ($n=500$)&\texttt{New.EM}& 60.2 (9.5)&59.4 (3.1)& 70.6 (17.9)& 91.9 (30.2)\\
   &\texttt{EM}& 54.1 (6.7)&47.6 (2.2)&53.7 (6.7)& 61.2 (11.9)\\
\hline\hline

\end{tabular}
\end{center}
{\small $\ast$ for \texttt{New.loc} and \texttt{New.EM}, this means
the additional computational cost after obtaining the initial
estimates.}
\end{table}

}


{
\begin{table}
\small

\centering\caption{\texttt{CD4 counts} data: estimated error
variance and eigenvalues}
\begin{tabular}{c|ccccc}\\
\hline\hline {\bf Model : $M=10$, $r=4$} &
$\widehat{\sigma}^2$&$\widehat\lambda_1$ & $\widehat\lambda_2$ &
$\widehat\lambda_3$ & $\widehat\lambda_4$
 \\
\hline\hline
\texttt{loc} & 42,359.3&615,735.6& 94,188.6& 47,012.6& 37,687.1\\
\texttt{New.EM} &38,411.0 & 473,416.8& 208,201.4& 53,253.9&
24,582.0\\
\texttt{EM} & 38,132.2& 469,784.3& 207,961.1 &54,007.2 &24,344.5\\
\hline\hline
\end{tabular}
\end{table}
}






\end{document}